\documentclass[12pt]{article}

\ifx\pdfoutput\undefined
\usepackage[dvips,bookmarks]{hyperref}
\else
\usepackage{hyperref}
\fi
\hypersetup{colorlinks=false,bookmarksopen,bookmarksnumbered,citecolor=blue,
   pdfstartview=FitH}

\usepackage[pdftex]{graphicx}	
\usepackage{latexsym}

\usepackage{amssymb,amsfonts,amsmath}
\usepackage{graphicx} 
\usepackage{indentfirst}

\usepackage{bbm}

\parskip=\medskipamount

\newcommand {\cA}{{\cal A}}

\newcommand {\cC}{{\cal C}}
\newcommand {\cD}{{\cal D}}
\newcommand {\cE}{{\cal E}}
\newcommand {\cF}{{\cal F}}
\newcommand {\cG}{{\cal G}}
\newcommand {\cH}{{\cal H}}

\newcommand {\cJ}{{\cal J}}
\newcommand {\cK}{{\cal K}}
\newcommand {\cL}{{\cal L}}
\newcommand {\cM}{{\cal M}}
\newcommand {\cN}{{\cal N}}

\newcommand {\cQ}{{\cal Q}}
\newcommand {\cR}{{\cal R}}
\newcommand {\cS}{{\cal S}}

\newcommand {\cW}{{\cal W}}
\newcommand {\cX}{{\cal X}}
\newcommand {\cY}{{\cal Y}}

\def\a{\alpha}

\def\b{\beta}
\def\c{\chi}
\def\d{\delta}
\def\e{\epsilon}
\def\f{\phi}
\def\g{\gamma}
\def\G{\Gamma}

\def\k{\kappa}
\def\l{\lambda}
\def\m{\mu}
\def\n{\nu}
\def\o{\omega}
\def\p{\pi}
\def\q{\theta}
\def\r{\rho}
\def\s{\sigma}

\def\x{\xi}
\def\z{\zeta}
\def\D{\Delta}
\def\F{\Phi}
\def\J{\Psi}
\def\L{\Lambda}
\def\O{\Omega}

\def\S{\Sigma}
\def\U{\Upsilon}
\def\X{\Xi}

\def\rd{{\rm d}}
\def\ri{{\rm i}}
\def\re{{\rm e}}

\def\rb{{\rm b}}
\def\rc{{\rm c}}
\def\ra{{\rm a}}

\newcommand{\ad}{{\dot{\alpha}}}                           %new
\newcommand{\bd}{{\dot{\beta}}}                            %new
\newcommand{\ve}{\varepsilon}                            %new
\newcommand{\cDB}{{\bar\cD}}                            %new
\newcommand{\DB}{\bar{D}}

\newcommand{\pa}{\partial}                           %new
\newcommand{\hf}{\frac12}
\newcommand{\vf}{\varphi}
\newcommand{\be}{\begin{equation}}
\newcommand{\ee}{\end{equation}}
\newcommand{\bea}{\begin{eqnarray}}
\newcommand{\eea}{\end{eqnarray}}
\newcommand{\non}{\nonumber}

\newcommand{\1}{{\underline{1}}}
\newcommand{\2}{{\underline{2}}}

\newcommand{\bm}[1]{\mbox{\boldmath$#1$}}

\def\double #1{#1{\hbox{\kern-2pt $#1$}}}

\newcommand{\ha}{{\hat{a}}}
\newcommand{\hb}{{\hat{b}}}
\newcommand{\hc}{{\hat{c}}}
\newcommand{\hd}{{\hat{d}}}

\newcommand{\gd}{{\dot\g}}
\newcommand{\dd}{{\dot\d}}

\newcommand{\ts}{{\tilde{\s}}}

\newcommand{\sba}{{\bar{\s}}}

\newcommand{\qb}{{\bar{\theta}}}

\newif\ifdtup

\def\de{{\nabla}}                                         % del
\newcommand{\bsubeq}{\begin{subequations}}
\newcommand{\esubeq}{\end{subequations}}

\newcommand{\bms}{{{\bm s}}}

\newcommand{\mub}{{\bar{\mu}}}

\renewcommand{\(}{\left(}
\renewcommand{\)}{\right)}

\newcommand{\bmq}{{\bm \q}}
\newcommand{\bmqb}{{\bar{{\bm \q}}}}

\newcommand{\eol}{\non\\}

\numberwithin{equation}{section}

\begin{document}
%%%%%%%%%%%%%%%%
%%%%%%%%%%%%%%%%
\begin{titlepage}
\begin{flushright}
UUITP-07/12\\
March, 2012\\
\end{flushright}
\vspace{5mm}

\begin{center}
{\Large \bf Extended supersymmetric sigma models in AdS$_4$
from projective superspace}\\ 
\end{center}

\begin{center}

{\bf 
Daniel Butter and Sergei M. Kuzenko
}

{\footnotesize{
{\it School of Physics M013, The University of Western Australia\\
35 Stirling Highway, Crawley W.A. 6009, Australia}} ~\\
\texttt{daniel.butter@uwa.edu.au},
\texttt{sergei.kuzenko@uwa.edu.au}}\\
\vspace{3mm}

{\bf Ulf Lindstr\"om and Gabriele Tartaglino-Mazzucchelli
} 

{\footnotesize{
{\it Theoretical Physics, Department of Physics and Astronomy,
Uppsala University \\ 
Box 516, SE-751 20 Uppsala, Sweden}}\\
\texttt{ulf.lindstrom@physics.uu.se},
\texttt{gabriele.tartaglino-mazzucchelli@physics.uu.se}}
\vspace{3mm}

\end{center}
%\vspace{5mm}

\begin{abstract}
\baselineskip=12pt
There exist two superspace approaches to describe $\cN=2$ supersymmetric nonlinear $\s$-models 
in four-dimensional anti-de Sitter ($\rm AdS_4$) space:  (i) in terms of $\cN=1$ AdS chiral superfields, 
as developed in arXiv:1105.3111 and arXiv:1108.5290; and (ii) in terms of $\cN=2$ 
polar supermultiplets  using the $\rm AdS$ projective-superspace techniques developed in arXiv:0807.3368. 
The virtue of the approach (i) is that it makes manifest the geometric properties of 
the $\cN=2$ supersymmetric  $\s$-models in $\rm AdS_4$.
The target space must be a non-compact hyperk\"ahler manifold 
endowed with a Killing vector field which generates an
SO(2) group of rotations on the two-sphere of complex structures.  The power of the approach (ii) is that it 
allows us, in principle, to generate hyperk\"ahler metrics as well as to address the problem 
of deformations of such metrics. 

Here we show how to relate the formulation (ii) to (i) by integrating out 
an infinite number of $\cN=1$ $\rm AdS$ auxiliary superfields and performing a superfield duality transformation.
We also develop a novel description of the most general $\cN=2$ supersymmetric nonlinear $\s$-model
in $\rm AdS_4$ in terms of chiral superfields on three-dimensional $\cN=2$ flat superspace 
without central charge.  
This superspace naturally originates from a conformally flat realization 
for the four-dimensional $\cN=2$ $\rm AdS$ superspace that makes use of Poincar\'e coordinates 
for $\rm AdS_4$. This novel formulation allows us to uncover several interesting geometric results. 

\end{abstract}
\vspace{1cm}

\vfill
\end{titlepage}

\newpage
\renewcommand{\thefootnote}{\arabic{footnote}}
\setcounter{footnote}{0}

%%%%%%%%%%%%%%%%%%%%%%%%%%%%%%%%%%%%%%%%%%%
%%%%%%%%%%%%%%%%%%%%%%%%%%%%%%%%%%%%%%%%%%%
%%%%%%%%%%%%%%%%%%%%%%%%%%%%%%%%%%%%%%%%%%%

\tableofcontents

%%%%%%%%%%%%%%%%%%%%%%%%%%%%%%%%%%%%%%%%%%%
%%%%%%%%%%%%%%%%%%%%%%%%%%%%%%%%%%%%%%%%%%%
%%%%%%%%%%%%%%%%%%%%%%%%%%%%%%%%%%%%%%%%%%%

\section{Introduction}
\setcounter{equation}{0}

Recently, two of us  (DB and SMK) 
have constructed  the most general $\cN=2$ supersymmetric 
$\s$-model in four-dimensional anti-de Sitter space 
($\rm AdS_4$) using a formulation in terms of $\cN=1$ covariantly chiral superfields \cite{BKsigma1,BKsigma2}.
The target space of such a $\s$-model proves to be a non-compact hyperk\"ahler manifold 
restricted to possess a special Killing vector field which generates an
SO(2) group of rotations on the two-sphere of complex structures and necessarily
leaves one of them, $\mathbb J$,  invariant; each of the complex structures that are orthogonal to $\mathbb J$
is characterized by  an exact K\"ahler two-form.\footnote{In the case of 4D $\cN=2$ Poincar\'e supersymmetry, 
the target-space geometry of general $\sigma$-models is only required to be hyperk\"ahler \cite{A-GF}.}
The existence of such hyperk\"ahler spaces was pointed out
twenty five years ago in \cite{HitchinKLR}.  One of the main virtues of the $\cN=1$ formulation 
\cite{BKsigma1,BKsigma2} is its geometric character. The superfield Lagrangian proves to be a globally defined function on the target space, which is simultaneously the K\"ahler potential (with respect to each 
complex structure orthogonal to $\mathbb J$) and
the Killing potential of the SO(2) isometry group (with respect to $\mathbb J$). 
Another remarkable property of the $\cN=2$ supersymmetric $\s$-model 
constructed in \cite{BKsigma1,BKsigma2} is that the algebra of ${\rm OSp(2|4)}$ transformations 
closes off the mass shell. 
The only disadvantage of the $\cN=1$ formulation \cite{BKsigma1,BKsigma2} is that it cannot be used 
to generate hyperk\"ahler metrics (a hyperk\"ahler space has to be given in order to 
define the $\s$-model action). The latter goal can be achieved by resorting to 
powerful $\cN=2$ superspace techniques
such as harmonic superspace \cite{GIKOS,GIOS} and projective superspace \cite{KLR,LR88,LR90}.

A few years ago, general off-shell $\cN=2$ supersymmetric $\sigma$-models in $\rm AdS_4$
were constructed in $\cN=2$ AdS superspace \cite{KT-M-4D-conf-flat} building on the projective-superspace
formulations for  $\cN=2$ matter-coupled supergravity in four dimensions  \cite{KLRT-M1,KLRT-M2}
and $\cN=1$ matter-coupled supergravity in five dimensions \cite{KT-M5DSUGRA1,KT-M5DSUGRA2}.
The work of \cite{KT-M-4D-conf-flat} is a natural extension of the earlier 5D AdS approach
developed in  \cite{KT-M,KT-M_5D_conf-flat}.
The  powerful property of the construction given in \cite{KT-M-4D-conf-flat} is that $\cN=2$ 
supersymmetric $\sigma$-models in $\rm AdS_4$ 
can be generated from a Lagrangian that is an arbitrary real analytic 
function of $2n $ real variables, where $4n$ is the dimension of the target space. 
Its technical disadvantage is that the hyperk\"ahler geometry of the target space is hidden, 
unlike in the $\cN=1$ formulation \cite{BKsigma1,BKsigma2}. To uncover the explicit structure of the 
target space, the $\cN=2$ formulation of  \cite{KT-M-4D-conf-flat} has to be related to that given 
\cite{BKsigma1,BKsigma2}.
For the series of $\cN=2$ supersymmetric $\s$-models presented in  \cite{KT-M-4D-conf-flat},  
one can in principle derive their reformulation in terms of $\cN=1$ chiral superfields by: 

(i) eliminating the (infinitely many) $\cN=1$ auxiliary superfields; 
and 

(ii) performing appropriate $\cN=1$ superfield duality transformations.\\
These are nontrivial technical problems which are more difficult to address than in Minkowski space, 
due to non-zero curvature of the AdS space-time.  These problems will be dealt with in the present paper. 
Before turning to the description of our novel approach, it is appropriate to recall the salient points of 
\cite{BKsigma1,BKsigma2} and  \cite{KT-M-4D-conf-flat}. 

\subsection{$\cN=2$ anti-de Sitter superspace}
\label{N=2AdSgeneral}
In order to make use of the power of projective superspace techniques \cite{KLR,LR88,LR90}, 
one has to pick an $\cN=1$ subspace of a given $\cN=2$ superspace, 
which in our case is $\cN=2$ AdS superspace. It is pertinent here to recall its definition. 
The four-dimensional $\cN=2$ AdS superspace 
$$
{\rm AdS^{4|8} } := \frac{{\rm OSp}(2|4)}{{\rm SO}(3,1) \times {\rm SO} (2)}
$$
is a maximally symmetric geometry that originates within 
the superspace formulation of $\cN=2$ conformal supergravity developed in  \cite{KLRT-M1}.
The corresponding covariant derivatives\footnote{The SU(2) generators, $J_{kl}$, 
act on the spinor  derivatives 
by the rule: $[J_{kl} , { \cD}_\a^i ] = -\hf ( \d^i_k  { \cD}_{\a l} +\d^i_l  { \cD}_{\a k})$.} 
\bea
{\cD}_{\cA} =({ \cD}_{a}, { \cD}_{{\a}}^i, { \cDB}^\ad_i)
= E_{\cA}{}^\cM \pa_\cM + \hf  {\O}_{\cA}{}^{ bc} M_{ bc} 
+  \F_{\cA}{}^{ij} J_{ij}~, \qquad i,j =\1 , \2
\label{1.1}
\eea
obey the algebra  \cite{KT-M-4D-conf-flat, KLRT-M1}
\begin{subequations}\label{1.2}
\bea
&\{\cD_\a^i,\cD_\b^j\}=
4{\cS}^{ij}M_{\a\b}
+2 \ve_{\a\b}\ve^{ij}\cS^{kl}J_{kl}~,
\qquad
\{\cD_\a^i,\cDB^\bd_j\}=
-2\ri\d^i_j(\s^c)_\a{}^\bd\cD_c
~,~~~
\\
&{[}\cD_a,\cD_\b^j{]}=
\frac{\ri}{2} ({\s}_a)_{\b\gd}\cS^{jk}\cDB^\gd_k~, \qquad
{[}\cD_a,\cDB^\bd_j{]}=
\frac{\ri}{ 2} (\tilde{\s}_a)^{\bd\g}\cS_{jk}\cD_\g^k
~,~~~~ \\
&
[\cD_a,\cD_b]= - \cS^2
M_{ab}~,
~~~~~~
\label{AdS-N2-2}
\eea
\end{subequations} 
where ${ \cS}^{ij} $ is a {\it covariantly constant} 
 real isotriplet, $\cD_\cA \cS^{ij}=0$, with the algebraic properties
 ${ \cS}^{ji} = {\cS}^{ij}$, 
 $\overline{ { \cS}^{ij}} = { \cS}_{ij} =\ve_{ik}\ve_{jl}{ \cS}^{kl}$, 
and  ${ \cS}^2 :=  \frac{1}{2} { \cS}^{ij} { \cS}_{ij} = \text{const}$.
The constant $\cS^2$ is positive and so \eqref{AdS-N2-2} gives the
algebra of covariant derivatives in AdS.
This superspace is conformally flat  \cite{KT-M-4D-conf-flat}
and proves to be a  solution to the equations of motion 
for $\cN=2$ supergravity with a cosmological term \cite{BK_AdS_supercurrent}.

Due to \eqref{1.2}, the SU(2) gauge freedom can be used to choose the SU(2) connection
$\F_{\cA}{}^{ij} $ in 
\eqref{1.1}
to look like $ \F_{\cA}{}^{ i j } =  \F_{\cA}\cS^{ij}$, for some one-form $\F_\cA$
describing the residual U(1) connection associated with the generator $\cS^{ij} J_{ij}$.  
Then $\cS^{ij}$ becomes a constant isotriplet, $\cS^{ij} = s^{ij}= {\rm const}$.
The remaining global SU(2) rotations
can take $s^{ij}$ to any position on the two-sphere of radius 
$s =\sqrt{\hf s^{ij}s_{ij}} \equiv \cS$.\footnote{In what follows, we do not distinguish between 
$s$ and $\cS$.} 
There are two natural options for how to choose $\cS^{ij}$: 
\begin{subequations} 
\bea
 s^{\1 \2} &=&0~;  
 \label{1.3a} \\
 s^{\1\1}&=& s^{\2\2} =0~.
 \label{1.3b}
 \eea
 \end{subequations}
 Of course, these options are physically equivalent. However, choosing one or the other 
 may be more preferable to achieve certain technical simplifications.
It turns out that the choice \eqref{1.3a} must be used in order to embed an $\cN=1$ AdS superspace,
${\rm AdS^{4|4} }$, into the full
$\cN=2$ AdS superspace \cite{KT-M-4D-conf-flat}. 
As to the second choice, eq. \eqref{1.3b}, it will be shown in this paper that it corresponds to choosing 
Poincar\'e coordinates\footnote{The Poincar\'e patch covers half of  AdS${}_4$. It is sufficient to restrict 
our analysis to this coordinate patch when considering infinitesimal  isometry transformations.} 
for AdS${}_4$  in which the space-time metric takes the form 
\bea
{\rm d}s^2 = \Big(\frac{1}{s z}\Big)^2 \Big( 
\eta_{mn}{\rm d}x^m   {\rm d}x^n +  {\rm d}z^2 \Big)~.
\label{1.4}
\eea
The slices $z = {\rm const} $ foliate AdS${}_4$ into a family of three-dimensional Minkowski spaces.  
We will show, closely following the 5D AdS construction of \cite{KT-M_5D_conf-flat}, 
 that the choice \eqref{1.3b} allows us\footnote{More precisely, we will use a conformally flat representation 
 for the covariant derivatives \eqref{1.1} such that $\cS^{ij} $ is not constant but instead 
$\cS^{ij}  = s^{ij} +O(\q) $. Then, the choice \eqref{1.3b} leads to the required 3D foliation.}
 to choose a different $\cN=1$ subspace of ${\rm AdS^{4|8} }$, specifically
$  {\mathbb R}^{3|4} \times {\mathbb R}_+$, where ${\mathbb R}^{3|4}$ denotes
three-dimensional $\cN=2$ Minkowski superspace {\it without} central charge, 
and ${\mathbb R}_+:=\big\{ z\in {\mathbb R},~z>0 \big\} $.
Using such a setting, 
the problem of reformulating the off-shell $\cN=2$ supersymmetric $\sigma$-models in AdS${}_4$ 
\cite{KT-M-4D-conf-flat} 
in terms of $\cN=1$ chiral superfields proves to become almost identical to that appearing in the case of
 off-shell $\cN=2$ supersymmetric $\s$-models in 4D $\cN=2$ Minkowski space. The latter problem has been addressed in a number of publications \cite{GK1,GK2,AN,AKL1,AKL2,KN} (see \cite{K-Srni} for a review), 
and here we can make use of the results obtained in these papers.

\subsection{Formulation of $\cN=2$ supersymmetric $\s$-models in AdS${}^{\bf 4|4} $}\label{Review4DAdS}
As already mentioned, the choice \eqref{1.3a} is required
for embedding ${\rm AdS^{4|4} }$ into ${\rm AdS^{4|8} }$. 
We assume that ${\rm AdS^{4|8} }$  is parametrized  by local bosonic ($x$) and fermionic ($\q, \bar \q$) 
coordinates  ${\bm z}^{\cM}=(x^{m},\q^{\mu}_{\imath},{\bar \q}_{\dot{\mu}}^{\imath})$
(where $m=0,1,\cdots,3$, $\mu=1,2$, $\dot{\mu}=1,2$ and  $\imath=\1,\2$).
By applying certain general coordinate and local U(1) transformations in ${\rm AdS^{4|4} }$, 
it is possible to identify  ${\rm AdS^{4|4} }$ with the surface $\q^{\mu}_{\2} = 0 $ and 
${\bar \q}_{\dot{\mu}}^{\2} =0$. The covariant derivatives for ${\rm AdS^{4|4} }$,
\be
\cD_{A}=(\cD_a,\cD_\a,{\bar \cD}^\ad)
= E_A{}^M\pa_M+\hf\O_A{}^{bc}M_{bc}~,
\ee 
are related to \eqref{1.1} as follows 
\be
\cD_\a := \cD_\a^{\1}\big|~, \qquad
\bar \cD^\ad := \bar \cD^\ad_{\1} \big|~,
\ee
and similarly for the vector covariant derivative.
Here the 
bar-projection is defined by 
\be
U | :=  U(x,\q_{\imath},\bar \q^{\imath})|_{\q_\2={\bar \q}^\2=0}~,
\label{bar-projection}
\ee 
for any $\cN=2$  tensor superfield  $ U(x,\q_{\imath},\bar \q^{\imath})$. It follows from \eqref{1.2} 
that the $\cN=1$ covariant derivatives obey the algebra 
\bsubeq \label{1.5}
\bea
&\{\cD_\a,\cD_\b\}=-4\bar{\mu}M_{\a\b}~,~~~
\{\cDB_\ad,\cDB_\bd\}=4\mu\bar{M}_{\ad\bd}~,~~~
\{\cD_\a,\cDB_\bd\}=-2\ri\cD_{\a\bd}
~,
\\
&{[}\cD_a,\cD_\b{]}=-\frac{\ri}{2}\bar{\mu}(\s_a)_{\b\gd}\cDB^{\gd}~,~~~
{[}\cD_a,\cDB_\bd{]}=\frac{\ri}{2}\mu(\s_a)_{\g\bd}\cD^{\g}~, \\
&{[}\cD_a,\cD_b{]}=-|\mu|^2 M_{ab}
~,
\eea
\esubeq
where $\m = -s_{\1\1} = -s^{\2\2}$. As a result, each $\cN=2$ supersymmetric field theory in 
${\rm AdS^{4|8} }$ can be reformulated as some theory in ${\rm AdS^{4|4} }$. 

Any $\cN=2$ supersymmetric nonlinear $\s$-model describes a self-interaction of hypermultiplets.
When formulated in $\cN=1$ AdS superspace, a single hypermultiplet can be realized in terms of two covariantly  chiral scalar superfields. As shown in \cite{BKsigma1,BKsigma2}, the most general 
$\cN=2$ supersymmetric nonlinear $\s$-model in AdS can be described by 
an action in  ${\rm AdS^{4|4} }$ of the form\footnote{The target space of the most general $\cN=1$ supersymmetric 
$\s$-model in AdS \eqref{4DN1action} is characterized by an exact K\"ahler two-form
 \cite{Adams:2011vw,FS,BKsigma1},  and therefore this manifold is non-compact.} 
\begin{align}\label{4DN1action}
S = \int \rd^4x\, \rd^2\q\, \rd^2\bar\q \, E\, \cK(\vf^a, \bar\vf{}^{\bar b})~, 
\qquad
E^{-1}={\rm Ber}(E_A{}^M)
\end{align}
where $\vf^a$ is a chiral scalar, $\bar \cD_\ad \vf^a =0$. 
Here $\cK(\vf, \bar\vf  ) $ is a globally defined real function over the  
target space $\cM$ which is a hyperk\"ahler manifold. 
In terms of $\cK(\vf, \bar \vf )$, the target space metric is 
$g_{a \bar b} = \pa_a \pa_{\bar b} \cK$, and hence   the K\"ahler two-form is exact. This implies that
the target space is non-compact. 
The variables $\vf^a$ are local complex coordinates with respect to one of the complex structures on $\cM$, 
\begin{align}\label{complex_structure1} 
J_3 = \left(\begin{array}{cc}
\ri \,\delta^a{}_b & 0 \\
0 & -\ri \,\delta^{\bar a}{}_{\bar b}
\end{array}\right)~.
\end{align}
Two other complex structures can be chosen as
\begin{align}
\label{complex_structure2}
J_1 = \left(\begin{array}{cc}
0 & \omega^a{}_{\bar b} \\
\omega^{\bar a}{}_b & 0
\end{array}\right)~, \qquad
J_2 = \left(\begin{array}{cc}
0 & \ri\, \omega^a{}_{\bar b} \\
-\ri\, \omega^{\bar a}{}_b & 0
\end{array}\right)~,
\end{align}
and $\cM$ is K\"ahler with respect to each of them.
Here $\o_{ab}:= g_{a \bar c} \o^{\bar c}{}_b = -\o_{ba}$ is  a covariantly constant 
(2,0) form with respect to $J_3$, 
\be
\nabla_c \o_{ab} = \nabla_{\bar c} \o_{ab}=0~,
\ee
and hence it is holomorphic, $\o_{a b}= \o_{ab}(\vf)$.
The operators $J_A = (J_1, J_2, J_3)$ obey the quaternionic algebra
$
J_A J_B = -\delta_{A B} {\mathbbm 1} + \ve_{ABC} J_C.
$

As shown in \cite{BKsigma1,BKsigma2}, 
the $\s$-model  \eqref{4DN1action} is $\cN=2$ supersymmetric provided
the following vector field
\begin{align}\label{eq_VAds}
V^\n = (V^a, V^{\bar a}) 
 = \Big( \frac{\mu}{2|\mu|} \omega^{ab} \cK_b\, , 
 \frac{\bar\mu}{2|\mu|} \omega^{\bar a \bar b} \cK_{\bar b}\Big)
\end{align}
obeys the Killing equations\footnote{The equation $\nabla_a V_b + \nabla_b V_a =0$
trivially follows from the definition \eqref{eq_VAds}.}
\begin{align}\label{eq_VKilling}
\nabla_a V_b + \nabla_b V_a = \nabla_a V_{\bar b} + \nabla_{\bar b} V_a = 0~.
\end{align}
It can be shown that this Killing vector field \emph{rotates} the complex structures:
\begin{align}\label{eq_CSrotate}
\cL_V J_1 = J_3 \,\sin\theta ~,\quad
\cL_V J_2  = -J_3 \,\cos\theta ~,\quad
\cL_V J_3  = J_2 \,\cos\theta  - J_1\, \sin\theta ~,
\end{align}
where $\theta := \arg\mu$. There is a preferred complex structure
\begin{align}\label{eq_JAdS}
{\mathbb J} := J_1 \cos \theta  +  J_2 \sin \theta
	= \frac{1}{|\mu|} \left(\begin{array}{cc}
	0 & \mu\, \omega^a{}_{\bar b} \\
	\bar\mu\, \omega^{\bar a}{}_b & 0
	\end{array}\right)
\end{align}
with respect to which $V^\n$ is \emph{holomorphic},
\begin{align}\label{eq_JAdSInv}
\cL_V {\mathbb J} = 0~.
\end{align}

\subsection{Formulation of $\cN=2$ supersymmetric $\s$-models in  AdS${}^{\bf 4|8} $ }
\label{subsection1.3} 

General supersymmetric field theories in  AdS${}^{4|8}$
can be formulated in terms of covariant projective supermultiplets \cite{KT-M-4D-conf-flat}.
The covariant projective supermultiplets in four-dimensional $\cN=2$ supergravity were
introduced in \cite{KLRT-M1}. 
The definition given in \cite{KLRT-M1} was then specialized to the case 
of $\cN=2$ AdS supersymmetry in \cite{KT-M-4D-conf-flat}.
A  projective supermultiplet of weight $n$,
$\cQ^{(n)}(v^i)$, 
is defined to be a scalar superfield that
lives on  AdS$^{4|8}$,
is holomorphic with respect to
the isotwistor variables $v^i $ on an open domain of
${\mathbb C}^2 \setminus  \{0\}$,
and is characterized by the following conditions:\\
${}\quad$(1) it obeys the covariant analyticity constraints
\bea
\cD^{(1)}_{\a} \cQ^{(n)}  = {\bar \cD}^{(1)}_{\ad} \cQ^{(n)}  =0~, \qquad 
\cD^{(1)}_\a := v_i \cD^{i}_\a ~, \quad 
{\bar \cD}^{(1)}_\ad := v_i {\bar \cD}^{i}_\ad 
~;
\label{ana}
\eea
${}\quad$(2) it is  a homogeneous function of $v^i$
of degree $n$, that is,
\be
\cQ^{(n)}(c\, v )\,=\,c^n\,\cQ^{(n)}( v)~, \qquad c\in \mathbb{C}\setminus \{0\}~;
\label{weight}
\ee
${}\quad$(3)  the OSp$(2|4)$ transformation law of  $\cQ^{(n)}$
is as follows:
\bea
\d_\x \cQ^{(n)}
&=& -\Big( { \x}
+2{\ve}  {\cS}^{ij} J_{ij} \Big)\cQ^{(n)} ~,
\non \\
{\cS}^{ij} J_{ij}  \cQ^{(n)}&:=& -
 \Big( {\cS}^{{(2)}} { \pa}^{(-2)}
-n \, {\cS}^{(0)}\Big) \cQ^{(n)} ~, \qquad
{ \pa}^{(-2)} := \frac{1}{(v,u)}u^{i} \frac{\pa}{\pa v^{i}}~,
\label{harmult1}
\eea
where 
$$
{ \x} := { \x}^{\rm a}{ \cD}_{\rm a} + { \x}^\a_i \cD_\a^i + \bar { \x}_\ad^i \bar {\cD}^\ad_i
$$
is  an $\cN=2$ AdS Killing vector field, see section \ref{IntrinsicVH} for the definition.
In (\ref{harmult1}) we have introduced 
\bea
{\cS}^{(2)}:=v_i v_j {\cS}^{ij}~,\qquad
{\cS}^{(0)}:= \frac{1}{(v,u)} v_i u_j {\cS}^{ij}~.
\eea
The transformation law (\ref{harmult1}) involves an additional isotwistor,  $u_i$, 
which is only subject 
to the condition $(v,u) := v^{i}u_i \neq 0$, and is otherwise completely arbitrary.
Both  $\cQ^{(n)}$ and $\d_\x  \cQ^{(n)}$ are independent of $u_i$.
It is  seen that the projective 
supermultiplets live in the AdS projective superspace  ${\rm AdS^{4|8} } \times {\mathbb C}P^1$.

In the family of projective multiplets,  a generalized conjugation, 
$\cQ^{(n)} (v^i) \to \breve{\cQ}^{(n)} (v^i)$, is defined as follows:
\be
\breve{\cQ}^{(n)} (v) := \bar{\cQ}^{(n)}\big(\overline{v} \to 
{\rm i}\, \s_2\, v
 \big)~, 
\ee
with $\bar{\cQ}^{(n)}(\overline{v}) $ the complex conjugate of $\cQ^{(n)}(v)$ 
and $\s_2$ the second Pauli matrix. 
One can check that $\breve{\cQ}^{ (n) } (v)$ is a projective multiplet of weight $n$.
One can also see that
$\breve{\breve{\cQ}}{}^{(n)}=(-1)^n \cQ^{(n)}$,
and therefore real supermultiplets can be consistently defined when 
$n$ is even.
The $\breve{\cQ}^{(n)}$ is called the smile-conjugate of 
${\cQ}^{(n)}$.

To describe the dynamics of supersymmetric field theories in AdS$^{4|8}$, 
the following supersymmetric action principle can be used
\bea
S&=&
\frac{1}{2\pi} \oint_C (v, \rd v)
\int \rd^4 x \, {\rm d}^4\q \, {\rm d}^4{\bar \q}\,{\mathbb E}\, \frac{\cL^{(2)}}{(\cS^{(2)})^2}~,
\qquad
{\mathbb E}^{-1}={\rm Ber}({E}_\cA{}^\cM)
~,
\label{InvarAc1}
\eea
with $(v, \rd v) := v^i \rd v_i$.
Here the Lagrangian is a real weight-two projective multiplet in AdS$^{4|8}$.
 The first integral in \eqref{InvarAc1}
 is along a contour in ${\mathbb C}P^1$ parametrized by
complex homogeneous  coordinates $v^i$.
The second integral is over AdS$^{4|8}$. 
 
In this paper, we mostly concentrate on studying a certain class of 
$\cN=2$ supersymmetric $\s$-models in $\rm AdS_4$
introduced in \cite{KT-M-4D-conf-flat}. Such a theory  is a system
of interacting {\it covariant arctic weight-zero} multiplets 
\bea
{ \U}^I (v) = \sum_{n=0}^{\infty}  \, \z^n \U_n^I  ~, \qquad \z:= \frac{v^{\2}}{v^{\1} }
\label{1.23}
\eea
 and their smile-conjugates
\be
\breve{ \U}^{\bar I} (v) =\sum_{n=0}^{\infty}  \,  (-\z)^{-n}\,
{\bar \U}_n^{\bar I}~.
\label{1.24}
\ee
  described by the Lagrangian
\bea
\cL^{(2)} = \frac{1}{2s} \cS^{(2)}\, {K}({ \U}, \breve{ \U})~,
\label{1.25}
\eea
with $ s = \sqrt{ \hf \cS^{ij} \cS_{ij} }$.
Here ${K}(\F^I, {\bar \F}^{\bar J}) $ is the K\"ahler potential of a real analytic K\"ahler manifold $\cX$.
The interpretation of $K$ as a K\"ahler potential is consistent, since the action 
generated by \eqref{1.25} turns out to be invariant under K\"ahler transformations of the form
\be
{ K}({ \U}, \breve{\U})~\to ~{ K}({ \U}, \breve{ \U})
+{ \L}({\U}) +{\bar { \L}} (\breve{\U} )~,
\label{1.26}
\ee
with ${ \L}(\F^I)$ a holomorphic function.
The target space $\cM$ of this $\s$-model proves to be an open domain of the zero section 
of the cotangent bundle of $\cX$, $\cM \subset T^*\cX$. This can be shown by generalizing 
the flat-superspace considerations of \cite{K98,GK2}. 
 
The $\cN=2$ supersymmetric $\s$-models  defined by \eqref{4DN1action}
 and  \eqref{InvarAc1}, \eqref{1.25} are off-shell. 
This is a built-in property of the latter theory formulated in the $\cN=2$ AdS superspace. 
The off-shell nature of the former theory is a non-trivial result established in  \cite{BKsigma1,BKsigma2}.
Each hypermultiplet in the model \eqref{4DN1action} is described in terms of $8+8$ degrees
of freedom  which are packaged into two $\cN=1$ chiral superfields and their conjugates.
On the other hand, each arctic multiplet ${ \U}^I (v) $, eq. \eqref{1.23}, 
contains an infinite number of ordinary fields, most of which are auxiliary. 
One of the main virtues of the $\s$-model   \eqref{InvarAc1}, \eqref{1.25} 
is that its Lagrangian  \eqref{1.25} is given in terms of an arbitrary function 
${K}(\F^I, {\bar \F}^{\bar J}) $. Therefore, this $\s$-model formulation allows us, 
in principle, to generate hyperk\"ahler manifolds as well as to address the problem 
of deformations of such manifolds. To achieve these goals, however, 
we have to develop techniques to eliminate the infinite number of auxiliary fields. 
In particular, we have to understand how to relate the $\s$-model  \eqref{InvarAc1}, \eqref{1.25} 
to the $\cN=1$ formulation \eqref{4DN1action}. 
 
Due to the analyticity constraints  $\cD^{(1)}_{\a} \U^I  = {\bar \cD}^{(1)}_{\ad} \U^I  =0$, 
the Taylor coefficients $\U^I_n$ in \eqref{1.23} are constrained $\cN=2$ superfields. 
Once restricted to an $\cN=1$ subspace of the $\cN=2$ superspace  ${\rm AdS^{4|8} }$, 
the coefficients $\U^I_2, \U^I_3, \dots$,  can be shown to be unconstrained $\cN=1$ superfields. 
Upon reducing the superspace integral in  \eqref{InvarAc1}
with Lagrangian \eqref{1.25} to that over the 
$\cN=1$ subspace chosen, it can be shown that the superfields $\U^I_2, \U^I_3, \dots$, appear in the action
without derivatives, and therefore they are purely auxiliary and can be eliminated algebraically 
using their equations of motion. A natural option for how to define this $\cN=1$ subspace of  ${\rm AdS^{4|8} }$
is to choose the condition \eqref{1.3a} and 
embed ${\rm AdS^{4|4} }$ into ${\rm AdS^{4|8} }$
using the procedure described above.
However, such a set-up does not allow us to make use of the methods
which have been developed for the general $\cN=2$ supersymmetric $\s$-models in Minkowski space
\cite{GK1}. In other words, 
some conceptually new techniques are required if ${\rm AdS^{4|4} }$ is chosen
as the desired $\cN=1$ subspace of  ${\rm AdS^{4|8} }$. Such techniques 
have not yet been developed. 
On the other hand, the problem of eliminating the auxiliary superfields can be reduced to that 
studied in \cite{GK1,GK2,AN,AKL1,AKL2,KN,K-duality,K-comments} if we choose
\eqref{1.3a} and follow the five-dimensional construction of \cite{KT-M_5D_conf-flat}
to foliate ${\rm AdS^{4|8} }$ into a family of three-dimensional $\cN=2$ Minkowski superspaces. 

We will show that  the choice \eqref{1.3b} 
 leads, upon elimination of the auxiliary superfields in the $\s$-model 
 defined by eqs.  \eqref{InvarAc1} and \eqref{1.25},   to an action
in Poincar\'e coordinates\footnote{The definition of the Grassmann coordinates
$\bm\q$ and $\bar{\bm \q}$ will be given later.}
\begin{align}\label{3DN2action}
S &= \int \frac{\rd z}{ (sz)^2} \Bigg\{    \int \rd^3 x\, {\rm d}^2\bmq \,{\rm d}^2\bmqb\,
	 \mathbb K(\phi, \bar\phi)
	+ \Big( \ri \int \rd^3 x \, {\rm d}^2\bmq \,
	H_\ra  (\f)  \partial_z \phi^\ra +\text{c.c.}\Big)
 \Bigg\} ~.
\end{align}
The coordinates $(x, \bm\q, \bar{\bm\q})$ parametrize the 3D $\cN=2$
Minkowski superspace $\mathbb R^{3|4}$ lying at constant values of $z$.
The real function ${\mathbb K }(\f, \bar \f)$ 
 is a K\"ahler potential of the hyperk\"ahler target space $\cM$.
The three-dimensional  $\cN=2$ chiral superfields $\phi^\ra$ are complex coordinates 
with respect to the complex structure $\mathbb J$ defined by \eqref{eq_JAdS}. 
Finally, 
$H = H_\ra (\f ) \rd\phi^\ra$ is a globally defined holomorphic (1,0) form on $\cM$. 
Several additional
geometric requirements are imposed, which we will discuss.

This paper is organized as follows. In section 2 we introduce two invariant tensors of the $\cN=2$ 
AdS supergroup $\rm OSp(2|4)$: the intrinsic vector multiplet and the intrinsic hypermultiplet. 
The latter is then used to realize general $\cN=2$ superconformal $\s$-models 
as a subclass of the $\s$-model family \eqref{1.25}. The main thrust of section 3 is to show
how the off-shell supersymmetric $\s$-models in AdS described by \eqref{1.25}
can be reformulated in terms of $\cN=1$ chiral superfields in AdS, that is in the form 
\eqref{4DN1action}. We also describe gauged 
 $\cN=2$ supersymmetric $\s$-models 
 in the AdS projective superspace  ${\rm AdS^{4|8} } \times {\mathbb C}P^1$ and their reformulation 
in terms of $\cN=1$ chiral superfields on  ${\rm AdS^{4|4} }$. 
Section 4 provides a new conformally flat realization for 
${\rm AdS^{4|8} }$ with the key property that this superspace becomes foliated
into a union of 3D $\cN=4$ flat superspaces with a real central charge 
(to be called 3D $\cN=4$ central charge superspace) 
corresponding to a derivative in the fourth dimension. 
In section 5 we introduce a new set of Grassmann variables for 3D $\cN=4$
central charge superspace which provides the simplest embedding of 
3D $\cN=2$ Minkowski superspace without central charge.  This technical construction 
(to be referred to as  the 3D foliated frame) allows us 
to reformulate general supersymmetric theories in  ${\rm AdS^{4|8} } \times {\mathbb C}P^1$
in terms of flat projective supermultiplets. In section 6 we repeat most of the analysis of section 3 
using the 3D foliated frame. One of the advantages of this frame, as compared to the AdS frame 
used in section 3, is that we can explicitly construct the hyperk\"ahler potential for a large class 
of $\cN=2$ supersymmetric $\s$-models in AdS. 
Section 7 is devoted to the analysis of the most general $\cN=2$ supersymmetric $\sigma$-model 
in AdS using the 3D foliation. In section 8 we describe the general geometric features 
of the hyperk\"ahler target spaces of $\cN=2$ supersymmetric $\s$-models in AdS.
Section 9 is concerned with the $\cN=2$ AdS supersymmetric  $\s$-model on $T^*{\mathbb C}P^n$.
This is the only nontrivial example of a nonlinear $\s$-model in AdS
in which we have been able to explicitly eliminate 
the auxiliary superfields in the AdS frame. Our main findings are summarized in section 10. 
The main body of this paper is accompanied by two technical appendices. 
Appendix A describes the explicit form of the Killing vector fields of ${\rm AdS^{4|8} } $ in the 
3D foliated frame. Appendix B describes the tropical prepotential for the intrinsic 
vector multiplet in the 3D foliated frame.

%%%%%%%%%%%%%%%%%%%%%%%%%%%%%%%%%%%%%%%
%%%%%%%%%%%%%%%%%%%%%%%%%%%%%%%%%%%%%%%

\section{Intrinsic vector multiplet and hypermultiplet}\label{IntrinsicVH}
\setcounter{equation}{0}

In four-dimensional $\cN=2$ Minkowski superspace, the standard mechanism to make a charged
off-shell hypermultiplet massive \cite{vev,projective3} consists in coupling the hypermultiplet 
to a frozen U(1) vector multiplet such that its chiral field strength $W$ is constant. 
The same procedure also works in five-dimensional $\cN=1$ Poincar\'e supersymmetry 
where the field strength of a vector multiplet,  $W$,  is real \cite{Kuzenko:Superpotentials}.
Applying this mechanism to an off-shell 4D $\cN=2$ and 5D $\cN=1$ supersymmetric $\s$-model
(for this the target space has to possess a tri-holomorphic isometry) 
generates a superpotential  \cite{Kuzenko:Superpotentials}.
In this section we first discuss an AdS analogue of the frozen vector multiplet
-- the intrinsic vector multiplet \cite{KT-M-4D-conf-flat}, which will be used 
in subsequent sections. We also introduce a covariantly constant hypermultiplet which proves 
to be closely related to the geometry of AdS$^{4|8}$. Making use of this hypermultiplet
allows us to realize general $\cN=2$ superconformal $\s$-models 
as a subclass of the models \eqref{1.25}. 
\subsection{Intrinsic vector multiplet}
Consider an Abelian vector multiplet in AdS$^{4|8}$. It can be described by gauge-covariant 
derivatives 
\bea
{\bm \cD}_\cA =\cD_\cA +\ri  V_\cA  \hat e~,
\label{g-cd}
\eea
with $V_\cA$ the gauge one-form, and $\hat e$ the generator of the U(1) gauge group.
The gauge-covariant  derivatives are subject to the anti-commutation relations
\begin{subequations}\label{VM-al}
\bea
\{{\bm \cD}_\a^i, {\bm \cD}_\b^j\}&=&\phantom{-}
4{\cS}^{ij}M_{\a\b}
+2 \ve_{\a\b}\ve^{ij}\Big( \cS^{kl}J_{kl} +\ri \bar \cW \hat e\Big) ~, \\
\{\bar{\bm  \cD}_{\ad i}, \bar{\bm \cD}_{\bd j}\}&=&
-4{\cS}_{ij}\bar M_{\ad\bd}
+2 \ve_{\ad\bd}\ve_{ij}\Big( \cS^{kl}J_{kl} +\ri  \cW \hat e\Big) ~, \\
\{{\bm \cD}_\a^i, \bar{\bm \cD}_{\bd j} \}&=&
-2{\ri} \d^i_j ({\s}^c)_{\a\bd} {\bm \cD}_c~
\eea
\end{subequations}
which are obtained by combining the AdS algebra of covariant derivatives, \eqref{1.2}, 
with that describing the U(1) vector multiplet in Minkowski superspace \cite{GSW}.
Here the field strength $\cW$  is covariantly chiral, 
\be
\bar{ \cD}^{\ad }_i \cW =0~,
\ee
and obeys  the Bianchi identity
\bea
\Big({ \cD}^{\a(i} {\cD}_\a^{j)}+4\cS^{ij}\Big)\cW
&=&
\Big(\bar { \cD}_\ad^{(i} \bar{\cD}^{ j) \ad}+ 4{\cS}^{ij}\Big)\bar{\cW}
~.
\label{vectromul}
\eea

Following \cite{KT-M-4D-conf-flat}, a U(1) vector multiplet in AdS$^{4|8}$ is called intrinsic 
if  its field strength is constant, 
\bea
\cW=1~.
\eea
This condition is consistent with the Bianchi identity (\ref{vectromul}). Such a vector multiplet is `frozen'
in the sense that it  has no propagating degrees of freedom.
As will be shown in the next section,
it is completely determined by the geometry of AdS$^{4|8}$. 

\subsection{Intrinsic hypermultiplet}\label{intr_hyper}
The Fayet-Sohnius formulation for the hypermultiplet \cite{Fayet,Sohnius} can be extended
to the case of AdS \cite{BKsigma2,Kuzenko:2011md}. 
A charged off-shell hypermultiplet in AdS is described by a two-component superfield\footnote{Isospinor
indices are raised and lowed using antisymmetric tensors $\ve^{ij}$ and $\ve_{ij}$ normalized by 
$\ve^{\1\2} = \ve_{\2\1} =1$. The rules  are: 
$q^i = \ve^{ij} q_j$ and $q_i = \ve_{ij}q^j$. 
} 
$q_i$ and its conjugate 
${\bar q}{}^i := \overline{q_i}$ (such that $\overline{q^i}=-\bar q_i$) subject to  the constraints
\bea 
{\bm \cD}_\a^{(i} q^{j)} = \bar{\bm \cD}_\ad^{(i} q^{j)} = 0 ~.
\eea
The action of $\cS^{kl}J_{kl} $ on $q_i$ is not assumed to be fixed at our will. 
Instead it is determined by the constraints  to be 
\bea
\cS^{kl}J_{kl} =\D + {\bm J}~, 
\qquad {\bm J} q_i := - \cS_i{}^j q_j ~,
\eea
where  $\D$ takes on the role of a central charge as it commutes with the covariant derivatives, 
\begin{align}\label{eq_SK8}
[\Delta, {\bm \cD}_\alpha^i] = [\Delta, \bar{\bm \cD}_{\ad i}] = 0~.
\end{align} 
Setting $\D q_i=0$  is equivalent to the equation of motion for a massless hypermultiplet.

The covariantly constant torsion tensor of AdS$^{4|8}$, $\cS_{ij}$, 
can always be represented in the form 
\bea
\cS_{ij}= 2\ri \,q_{(i}\bar q_{j)}~, \qquad {\bar q}{}^i := \overline{q_i}~,
\label{quadratic}
\eea
for some isospinor $q_i$ defined modulo arbitrary phase transformations 
$q_i \to \re^{\ri \vf} q_i$, with $\bar \vf = \vf$.
Introducing $|q|^2:= \bar q^i q_i$, we easily obtain 
\bea
\cS_i{}^j q_j =\ri |q|^2 q_i~, \qquad \cS^i{}_j \bar q^j =\ri |q|^2 \bar q^i~,
\label{S-->q}
\eea
as well as 
\bea
s \equiv \sqrt{  \frac{1}{2} { \cS}^{ij} { \cS}_{ij}  }= |q|^2~.
\label{sq}
\eea
This shows that $ |q|$ is constant. The freedom in the definition of $q_i$ can be fixed by requiring it 
to be gauge-covariantly constant, 
\bea
{\bm \cD}_\cA q_i = 0~,\qquad \D q_i=0~, 
\label{2.12cov}
\eea
where the derivatives ${\bm \cD}_\cA$ correspond to the intrinsic vector multiplet. 
In accordance with \eqref{VM-al},  \eqref{S-->q} and \eqref{sq}, 
the integrability condition for this constraint is 
\bea
\hat e q_i = s q_i~.
\eea
This frozen hypermultiplet will be called {\it intrinsic}.

The isometry group of AdS$^{4|8}$,  $\rm OSp(2|4)$,
is generated by the corresponding Killing vector fields.
A real vector field in AdS$^{4|4}$ corresponding to the first-order operator 
\bea
{ \x} := \x^\cA \cD_\cA ={\x}^{a}{\cD}_{ a} + { \x}^\a_i { \cD}_\a^i + \bar { \x}_\ad^i \bar { \cD}^\ad_i 
\label{B.5}
\eea
is said to be a Killing vector field if it obeys the master equation 
\be 
[ { \x} + \hf { \l}^{cd} M_{cd}+ 2 { \ve} \cJ, \cD_\a^i ] = 0~,  \qquad \cJ := \mathcal S^{kl} J_{kl}
\ ,
\ee
 for uniquely determined parameters ${ \l}^{cd}$  and $ \ve$
 generating Lorentz and U(1) transformations respectively. 
 The explicit expressions for these parameters are
 \bea
  \lambda_{ a b} = \cD_{ [a}  \xi_{ b]}~, \qquad
 { \ve} = \frac{1}{8} { \mathcal S}^{ij} { \cD}_{\a i} { \x}^\a_j~,
 \label{B.3}
 \eea
 see \cite{KT-M-4D-conf-flat} for a derivation. 
If $U$ is  a tensor superfield on $\cN=2$ AdS superspace, its infinitesimal transformation associated with 
$ \x$ is  
\bea 
\d_{\text{AdS}} U = - { \x} U - \hf { \l}^{cd} M_{cd} U - 2 { \ve} \cJ U \ .
 \label{killertrans}
\eea
The torsion of $\cN=2$ AdS superspace, $\cS^{ij}$, is an invariant tensor,  $\d_{\text{AdS}} \cS^{ij}=0$.
Combining $\d_{\text{AdS}}$ with a certain U(1) gauge transformation, 
\bea
\hat{\d}_{\text{AdS}}  := \d_{\text{AdS}}  -2\ri \ve \hat{e}~,
\eea
we can see that the intrinsic hypermultiplet, $q_i$,  is invariant, 
\be
\hat{\d}_{\text{AdS}}  q_i =0~.
\ee
The defining property of $\hat{\d}_{\text{AdS}} $ is that the gauge-covariant derivatives do not change, 
\be 
[ { \x}^\cC{\bm \cD}_\cC + \hf { \l}^{cd} M_{cd}+ 2 { \ve} \cJ +2\ri \ve \hat{e} , {\bm \cD}_A ] = 0~,
\ee
where we have used the identity $ \x_\a^i = \cD_\a^i \ve$ 
derived in \cite{KT-M-4D-conf-flat}. 

\subsection{Maximally symmetric solution for $\cN=2$ AdS supergravity} 
The intrinsic vector multiplet and the intrinsic hypermultiplet naturally originate 
in the context of 
a maximally supersymmetric solution for $\cN=2$ supergravity with a cosmological term
if one uses
the off-shell supergravity formulation of \cite{de Wit:1980tn} 
with the following compensators: 
the vector multiplet and the hypermultiplet with an intrinsic central charge
 (see \cite{deWPV} for an early list of off-shell formulations for $\cN=2$ supergravity). 
The supergravity equations of motion can be shown to be 
(see \cite{BK_AdS_supercurrent,KT} for a derivation)
\begin{subequations} \label{2.21}
\bea
\frac{1}{\k^2} \S^{ij} &=&\ri e \bar q^{(i} q^{j)} ~, \label{2.21a}\\
\frac{1}{\k^2} \cW \bar \cW &=& \hf |q|^2~, \label{2.21b}\\
\D q_i &=&0~, \label{2.21c}
\eea
\end{subequations}
where $\k$ is the gravitational constant, $e$ denotes the U(1) charge of $q_i$ related 
to the cosmological constant, and 
\bea
\S^{ij} =\frac{1}{4}\Big({ \cD}^{\a(i} {\cD}_\a^{j)}+4\cS^{ij}\Big)\cW
=\frac{1}{4}
\Big(\bar { \cD}_\ad^{(i} \bar{\cD}^{ j) \ad}+ 4\bar{\cS}^{ij}\Big)\bar{\cW}~.
\label{2.22}
\eea
We assume here  that the Weyl multiplet is described using the superspace formulation 
for $\cN=2$ conformal supergravity given in 
\cite{KLRT-M1} (in this formulation, the torsion $\cS^{ij}$ is complex).
The relations  \eqref{2.21a} and  \eqref{2.21c} are 
 the equations of motion for the vector compensator and the hypermultiplet respectively, 
while eq.  \eqref{2.21b} corresponds to the gravitational superfield (see \cite{KT} for more details). 
The equations \eqref{2.21} can be shown to be super-Weyl invariant (see also subsection 
\ref{geometry_N=2_AdS}).

We are interested in a supergravity solution with vanishing super-Weyl tensor, that is $\cW_{\a\b}=0$. 
The super-Weyl invariance can be used to choose the gauge $\cW=1$, in which
 $\cS^{ij}$ becomes real, as a consequence of \eqref{2.22}.
Then, the relations  \eqref{2.21a},  \eqref{2.21b} and  \eqref{2.22} lead to
\bea
\cS^{ij}=  \frac{2\ri e}{|q|^2} q^{(i}\bar q^{j)}~,
\eea
which implies the  consistency condition $e = s$.
The hypermultiplet  constraints ${\bm \cD}_\a^{(i} q^{j)} = \bar{\bm \cD}_\ad^{(i} q^{j)} = 0$
and the supergravity equation of motion \eqref{2.21b}, $|q|^2 =2/\k^2 =\text{const}$,
imply that $q_i$ is covariantly constant, 
${\bm \cD}_A q_i =0$. As a result, it can be seen that 
the algebra of the supergravity covariant derivatives \cite{KLRT-M1}
 reduces to \eqref{VM-al}, 
and therefore the supergravity solution constructed describes the AdS geometry. 
To completely reproduce the construction of subsection \ref{intr_hyper}, 
it only remains to normalize $|q|^2 = e$. 

Two comments are in order. Firstly, there exists an off-shell formulation for $\cN=2$ supergravity \cite{KLRT-M1}
in which 
the hypermultiplet compensator is described in terms of 
a covariant weight-one arctic multiplet $q^{(1)}(v)$ and its conjugate $\breve{q}{}^{(1)} (v)$, 
which are coupled to the vector compensator in the case of a non-zero cosmological constant.
In this formulation, the hypermultiplet has no central charge, $\D q^{(1)} \equiv 0$ off the mass shell.
The hypermultiplet equation of motion is  $q^{(1)} (v)= q_i v^i $.
The equations  \eqref{2.21a} and  \eqref{2.21b} remain the same. 
Secondly,  one can describe $\cN=2$ AdS supergravity  
using the off-shell
formulation of \cite{deWPV} 
which makes use of two compensators: the vector multiplet  and the tensor multiplet. 
The superspace description of the AdS solution within this supergravity formulation is given 
in \cite{BK_AdS_supercurrent}.

\subsection{Off-shell $\cN=2$ superconformal $\s$-models}

Consider a system of interacting {\it covariant weight-zero arctic} multiplets 
${ \U}^I (v)$   and their smile-conjugates
$ \breve{ \U}^{\bar I} (v)$  described by the Lagrangian
\bea
\cL^{(2)} = \frac{1}{2s} \cS^{(2)}\, {K}({ \U}, \breve{ \U})~.
\label{superconf}
\eea
Here ${K}(\F^I, {\bar \F}^{\bar J}) $ is the K\"ahler potential of a K\"ahler cone; 
it obeys the homogeneity condition
\bea
\F^I \frac{\pa}{\pa \F^I} K(\F, \bar \F) =  K( \F,   \bar \F)~.
\label{Kkahler2}
\eea
With the homogeneity condition imposed, no K\"ahler invariance survives. The K\"ahler cone is 
the target space of an $\cN=1$ superconformal $\s$-model, see e.g. \cite{K-duality}.
We can associate with ${ \U} (v)$ a weight-one arctic multiplet ${\bm \U}^{(1)} (v)$
and its smile-conjugate $\breve{\bm \U}^{(1)} (v)$
defined by 
\bea
{\bm \U}^{(1)} (v) := \frac{1}{\sqrt{s}} q^{(1)} (v) {\U} (v)~, \qquad 
\breve{\bm \U}{}^{(1)} (v) :=  \frac{1}{\sqrt{s}} \breve{q}^{(1)} (v) \breve{\U} (v)~, 
\eea
where we have defined 
$q^{(1)} := q_i v^i $ and $\breve{q}^{(1)} := \bar q_i v^i$, with $q_i$ being the intrinsic 
hypermultiplet. In terms of the weight-one projective superfields ${\bm \U}^{(1)} (v)$
and $\breve{\bm \U}^{(1)} (v)$, the Lagrangian \eqref{superconf}
takes the form 
\bea
\cL^{(2)} = \ri\,  {K}({\bm \U}^{(1)}, \breve{\bm \U}{}^{(1)} )~.
\label{superconf2}
\eea
where we have used the relation  (\ref{quadratic}). 
The $\s$-model obtained is $\cN=2$ superconformal. Its Lagrangian \eqref{superconf2}
has the same form an in the super-Poincar\'e 
case  \cite{K-conf}.

\section{$\sigma$-models from projective superspace: AdS frame}\label{sect_N1AdS}
In the introduction, we briefly reviewed the results of
\cite{BKsigma1, BKsigma2} regarding the general form of $\cN=2$
supersymmetric $\s$-models written in terms of the $\cN=1$ AdS 
superspace AdS$^{4|4}$. In this section, we will demonstrate
explicitly how these models come about from a projective superspace
context.

The starting point is the general projective superspace action
in AdS$^{4|8}$, eq. \eqref{InvarAc1}.
Let us make the choice \eqref{1.3a}. Using the techniques described in \cite{KT-M-4D-conf-flat},
this action can be rewritten in AdS$^{4|4}$ as
\begin{align}\label{3.1sma}
S &= \int \rd^4x\, \rd^2\q\, \rd^2\bar\q\, E\, \cL
\end{align}
where the $\cN=1$ AdS Lagrangian $\cL$ is given by a contour integral
\begin{align}\label{3.2}
\cL &= \oint_C \frac{\rd \zeta}{2\pi \ri \zeta} \, \cL^{[2]}\Big|~, \qquad
\cL^{[2]} (\z):= \frac{1}{\ri (v^\1)^2 \zeta} \cL^{(2)}(v)~,
\end{align}
with the bar-projection defined by \eqref{bar-projection}.
Specializing to the $\s$-model described by \eqref{1.25}, we find
\begin{align}\label{3.3}
\cL &= \frac{1}{2s} \oint_C \frac{\rd \zeta}{2\pi \ri \zeta} \, \cS^{[2]} K(\U, \breve \U)\Big|~, \qquad
\cS^{[2]}(\zeta)  := \frac{1}{\ri (v^\1)^2 \zeta} \cS^{(2)}(v)
= \frac{\ri \mu}{\zeta} + \ri\bar\mu \zeta~.
\end{align}
In this context, $s = |\mu|$. In what follows,  the bar-projection is not indicated explicitly.

We will analyze this $\s$-model action in several stages. First, we will
directly evaluate it without gauging any of the isometries which 
the K\"ahler space with the K\"ahler potential  $K$ might possess. This leads to the class of actions discussed
in \cite{BKsigma1, BKsigma2}. Then we will consider the case where the
K\"ahler space possesses a holomorphic isometry, and demonstrate how this
leads to a tri-holomorphic isometry of the hyperk\"ahler target space,
in terms of the  $\s$-model formulated using $\cN=1$ chiral superfields. 
When one such isometry is gauged by the intrinsic vector
multiplet of AdS a superpotential naturally emerges.\footnote{Due to the properties
of $\cN=1$ AdS, the seemingly more general case involving a gauged isometry
is also contained within the class of actions considered in \cite{BKsigma1, BKsigma2},
even though tri-holomorphic isometries were not considered explicitly in that work.}

\subsection{The ungauged case}
Upon projection to $\cN=1$ AdS superspace, which requires the standard choice \eqref{1.3a},
the weight-zero arctic multiplet $\U^I$ consists of an infinite set of $\cN=1$
superfields, \eqref{1.23}. 
Using the analyticity constraints \eqref{ana},
one can show that the
lowest two components,
\begin{align}
\Phi^I := \U_0^I~, \qquad \S^I := \U_1^I
\end{align}
are constrained $\cN=1$ superfields: 
$\Phi^I$ is covariantly chiral and $\S^I$ is covariantly complex linear,
\begin{align}
\bar \cD_\ad \Phi^I = 0~, \qquad (\bar \cD^2 - 4 \mu) \S^I = 0~.
\end{align}
All the other components of $\U^I$ are unconstrained complex $\cN=1$ superfields.
These superfields appear in the action without derivatives, and therefore they are auxiliary.
The superfields $\Phi^I$ and $\S^I$ are physical.

For our subsequent analysis, it is important to work out the supersymmetry transformation laws 
of the physical superfields $\Phi^I$ and $\S^I$. 
We begin by recalling that the weight-zero arctic multiplet
transforms under the full $\cN=2$ AdS supergroup, $\rm OSp(2|4)$, according to  eq. 
\eqref{harmult1}. Upon $\cN=1$ projection, this $\cN=2$ transformation decomposes into two different 
transformations in $\cN=1$ AdS superspace
\cite{BKsigma2,KT-M-4D-conf-flat} which are:

{\bf (i)}  An $\cN=1$ AdS isometry transformation 
\begin{align}\label{3.66}
\delta \U^I (\z) = 
-\x\U^I (\z)~,
\qquad \x:=\xi^a \cD_a + \xi^\a \cD_\a + \bar\xi_\ad \bar\cD^\ad
\end{align}
generated by an arbitrary $\cN=1$ Killing vector field $\x^A = (\x^a, \x^\a, \bar \x_\ad)$.
Such a vector field obeys the Killing equation 
\bea
[\x +\hf \l^{bc} M_{bc}, \cD_A]\equiv
[\x + \l^{\b\g} M_{\b\g} +\bar \l^{\bd \gd}\bar M_{\bd \gd} , \cD_A]=0~,\qquad \l_{\a\b} = \cD_{(\a} \l_{\b)}~.
\eea 
These Killing vector fields generate the isometry group of AdS$^{4|4}$, 
$\rm OSp(1|4)$.

{\bf (ii)}
An extended supersymmetry transformation
\begin{align}\label{3.8}
\delta \U^I (\z) &= -(\zeta \ve^\a \cD_\a - \frac{1}{\zeta} \bar\ve_\ad \bar \cD^\ad
	- 2\ve \mu \pa_\zeta - 2 \ve \zeta^2 \bar\mu \pa_\zeta) \U^I  (\z) ~, \qquad
\ve_\a := \cD_\a \ve~, 
\end{align}
where $\ve$ is a real  superfield obeying the constraints  \cite{GKS}
\begin{align}
(\bar \cD^2 - 4\mu) \ve = (\cD^2 - 4\bar\mu) \ve = 0~, 
\qquad \bar \cD_\bd \cD_\a \ve = \cD_\a \bar \cD_\bd \ve = 0~.
\end{align}
The first constraint implies  that $\ve$ is real linear.
The superfield parameter $\ve$
contains two components:
(i)  a  bosonic parameter $\r$ which is defined by $ \ve |_{\q=0} = \r  |\m|^{-1} $ 
and describes  the O(2) rotations;  
and (ii) a fermionic parameter $\e_\a := \cD_\a \ve|_{\q=0} $ along with its conjugate, 
which generate the second supersymmetry. 

It follows from \eqref{3.8} that the extended supersymmetry  transformation acts on the lowest
two component fields as
\begin{subequations}\label{eq_dPhi}
\begin{align}\label{eq_dPhi1}
\delta \Phi^I &= \bar\ve_\ad \bar\cD^\ad \S^I + 2 \ve \mu \S^I \\
	&= \frac{1}{2} (\bar \cD^2 - 4 \mu) (\ve \S^I) \label{eq_dPhi2}
\end{align}
\end{subequations}
and
\begin{align}\label{eq_dSigma}
\delta \S^I &= -\ve^\a \cD_\a \Phi^I + \bar\ve_\ad \bar\cD^\ad \U_2^I
	+ 4 \mu \ve \U_2^I 
	= -\ve^\a \cD_\a \Phi^I + \bar\cD_\ad (\bar\ve^\ad \U_2^I)~.
\end{align}
We have written the transformation law of $\Phi^I$ in two ways
by exploiting the complex linearity of $\S^I$; in \eqref{eq_dPhi2}
the transformation is manifestly chiral. Similarly, we have written
\eqref{eq_dSigma} to emphasize that $\delta \S^I$ is complex linear.

The $\cN=1$ Lagrangian $\cL(\U_n, \bar \U_n)$, eq. \eqref{3.3},  is a function of not only $\Phi^I$ and
$\S^I$ but also the infinite set of unconstrained $\cN=1$ superfields
$\U_2^I$, $\U_3^I,  \dots$ 
Because these superfields are unconstrained, they can
be eliminated (at least in principle) using their algebraic equations
of motion,
\begin{subequations}\label{AuxEOMAdS}
\begin{alignat}{2}
0 &= \frac{\pa \cL}{\pa \U_n^I} = \frac{1}{2s} \oint_C \frac{\rd\zeta}{2\pi \ri \zeta} \cS^{[2]} 
\frac{\pa K}{\pa \U^I} \zeta^n~,
	&\qquad &n\geq 2~; \\
0 &= \frac{\pa \cL}{\pa \bar\U_n^{\bar J}} = \frac{1}{2s} \oint_C \frac{\rd\zeta}{2\pi \ri \zeta} 
\cS^{[2]} \frac{\pa K}{\pa \breve\U^{\bar J}} {(-\zeta)}^{-n}~,
	&\qquad &n\geq 2~.
\end{alignat}
\end{subequations}
This step is the nontrivial technical challenge of the projective superspace approach:
we must solve an infinite set of algebraic equations. This issue was
resolved in Minkowski superspace for a large class of $\s$-models 
on cotangent bundles of Hermitian symmetric spaces
 \cite{GK1,GK2,AN,AKL1,AKL2,KN} (see \cite{K-Srni}
for a review). For these cases, the equations which must be solved in
a flat background correspond to the choice $\mathcal S^{[2]} = 2s$ --
in other words, the choice \eqref{1.3b} discussed in the introduction.
For the case of interest to us, $\cS^{[2]} = \ri \mu / \z + \ri \bar\mu \z$,
the problem remains unsolved except for the special case described in section \ref{section9}.
Nevertheless, we may proceed formally by generalizing the flat-superspace
analysis of  \cite{K-duality,K-comments}. 

The $\s$-model action is manifestly invariant under the $\cN=1$ AdS transformation \eqref{3.66}, 
both before and after elimination of the auxiliary superfields. 
If we assume that the auxiliaries have been eliminated, 
with $\cL(\F, \S, \bar \F , \bar \S)$ being the resulting Lagrangian, 
the action must be invariant under
the transformations \eqref{eq_dPhi1} and \eqref{eq_dSigma} where $\U_2^I$
is now understood as a function of $\Phi^I$, $\S^I$, and their complex conjugates. 
This follows from
the contour integral definition of $\cL$. 
It will be important later
that the proof of invariance of the action does \emph{not} require that
$\S^I$ be complex linear, provided one uses the form \eqref{eq_dPhi1} 
for the transformation $\delta \Phi^I$.

Alternatively, one can look for a Lagrangian $\cL(\F, \S, \bar \F , \bar \S)$ such that the corresponding action 
is invariant under extended supersymmetry transformations  \eqref{eq_dPhi} and \eqref{eq_dSigma}, 
with $\U^I_2$ some unknown function of $\Phi^I$, $\S^I$, and their complex conjugates. 
One can then 
check that invariance of the action implies the existence of
a complex function $\Xi$ obeying the equations,\footnote{It was argued for the first
time in \cite{K-duality,K-comments} that such a function $\Xi$ must exist in flat backgrounds.
We find the same condition in AdS.}
\begin{subequations}\label{eq_XiEqnsAdS}
\begin{align}
\frac{\partial \cL}{\partial \Phi^I} +
	\frac{\partial \cL}{\partial \Sigma^J} \frac{\partial \Upsilon_2^J}{\partial \Sigma^I}
	&= \frac{\partial \Xi}{\partial \Sigma^I}~, \\
-\frac{\partial \cL}{\partial \bar\Sigma^{\bar I}} +
	\frac{\partial \cL}{\partial \Sigma^J} \frac{\partial \Upsilon_2^J}{\partial \bar \Phi^{\bar I}}
	&= \frac{\partial \Xi}{\partial \bar\Phi^{\bar I}}~, \\
\frac{\partial \cL}{\partial \Sigma^J} \frac{\partial \Upsilon_2^J}{\partial \bar \Sigma^{\bar I}}
	&= \frac{\partial \Xi}{\partial \bar\Sigma^{\bar I}}~.
\end{align}
\end{subequations}
From the contour integral definition of $\cL$, one can show that $\Xi$
is given by
\begin{align}
\Xi &:= \frac{1}{2s} \oint_C\frac{\rd\zeta}{2\pi \ri \zeta} \frac{\cS^{[2]}}{\zeta} K~.
\end{align}

In
the case of Minkowski space \cite{K-comments}, 
the equations \eqref{eq_XiEqnsAdS} guarantee invariance of the action 
under the second supersymmetry transformation.\footnote{The contour integral representation 
for $\X$ is different in flat space, see eq. \eqref{6.20}.} In the case of AdS space, however, 
one finds an additional requirement
\begin{align}
 \mu \Xi - \hf \mu \S^I \frac{\pa\cL}{\pa \Phi^I} -  \mu \U_2^I \frac{\pa \cL}{\pa \S^I}
	+ \textrm{c.c.} = \L(\Phi) + \bar \L(\bar\Phi)
\end{align}
for some chiral superfield $\L(\Phi)$. Making use of the contour integral
definitions of $\Xi$ and $\cL$, we actually find that a stronger version of this
condition is satisfied,
\begin{align}\label{3.16AdS}
 \mu \Xi - \hf \mu \S^I \frac{\pa\cL}{\pa \Phi^I} -  \mu \U_2^I \frac{\pa \cL}{\pa \S^I}
+ \textrm{c.c.}=0
~.
\end{align}
This condition (in either form) should be equivalent to the existence of a Killing
vector $V^\m$ in the dual geometry which acts as a rotation on the complex structures.

The next step is to perform a duality, exchanging the complex linear superfield
$\S^I$ for a chiral superfield $\Psi_I$. One introduces the so-called
``first-order'' action
\begin{align}\label{3.17FO}
S_{\rm F.O.} &= \int \rd^4x\, \rd^2\q\, \rd^2\bar\q\, E\,
	\Big(\cL + \S^I \Psi_I + \bar \S^{\bar J} \bar \Psi_{\bar J}\Big)~,
\end{align}
where complex linearity of $\S^I$ is enforced by the $\Psi_I$ equation of motion.
Instead if we apply the equation of motion for the unconstrained $\S^I$, we find
\begin{align}\label{eq_dualActionAdS}
S_{\rm dual} &= \int \rd^4x\, \rd^2\q\, \rd^2\bar\q\, E\, \cK(\F^I, \J_I , \bar \F^{\bar J}, \bar \J_{\bar J} )~,
\end{align}
where $\cK$ is defined as
\begin{align}
\cK := \cL + \S^I \Psi_I + \bar \S^{\bar J} \bar \Psi_{\bar J}
\end{align}
and $\S^I$ is chosen to obey
\begin{align}
\frac{\pa \cL}{\pa \S^I} = - \Psi_I~.
\end{align}

We must construct the transformation law of $\Psi_I$ for the dual action.
This is easiest to do by first constructing it for the first-order action.
We begin by postulating the transformation law of $\Phi^I$. To
maintain its chirality, we must choose
\begin{align}
\delta \Phi^I &:= \frac{1}{2} (\bar \cD^2 - 4 \mu) (\ve \S^I)
	= \bar\ve_\ad \bar\cD^\ad \S^I + 2 \ve \mu \S^I 
	+ \frac{1}{2} \ve (\bar \cD^2 - 4 \mu) \S^I~,
\end{align}
where the last term no longer vanishes. Keeping the same transformation
law \eqref{eq_dSigma} for $\S^I$, we find that
\begin{align}
\int \rd^4x\, \rd^2\q\, \rd^2\bar\q\, E\, \delta \cL 
	&= \frac{1}{2} \int \rd^4x\, \rd^2\q\, \rd^2\bar\q\, E\, \S^I 
	(\bar \cD^2 - 4 \mu)\left(\ve \frac{\pa \cL}{\pa \Phi^I}\right) + \textrm{c.c.}
\end{align}
It follows that we must choose
\begin{align}
\delta \Psi_I &= -\frac{1}{2} (\bar \cD^2 - 4 \mu) \left(\ve \frac{\pa \cL}{\pa \Phi^I}\right)~.
\end{align}
The dual action \eqref{eq_dualActionAdS} is then invariant under
\begin{align}\label{3.24Dar}
\delta \Phi^I = \frac{1}{2} (\bar \cD^2 - 4 \mu) \left(\ve \frac{\pa \cK}{\pa \Psi_I}\right)~,\quad
\delta \Psi_I = -\frac{1}{2} (\bar \cD^2 - 4 \mu) \left(\ve \frac{\pa \cK}{\pa \Phi^I}\right)~.
\end{align}

Now, we have to recall the structure of the extended supersymmetry transformation 
in the model \eqref{4DN1action}. In accordance with  \cite{BKsigma1, BKsigma2},
let $\vf^a$ denote the general chiral coordinate of the K\"ahler potential $\cK$,
\begin{align}
\delta \vf^a = \frac{1}{2} (\bar \cD^2 - 4 \mu) (\ve \omega^{ab} \cK_b)~,
\label{3.25SUSY}
\end{align}
where 
$(-\o^{ab} )$ is the inverse of the 
covariantly constant $(2,0)$ holomorphic form $\o_{ab} $
in terms of which 
two other complex structures $J_1$ and $J_2$ of the hyperk\"ahler target space
are constructed, eq. \eqref{complex_structure2}. 
In our case, the set of chiral superfields is $\vf^a = (\F^I, \J_I)$. 
It follows from \eqref{3.24Dar} that 
\begin{align}
\omega^{ab} =
\begin{pmatrix}
0 & \delta^I{}_J \\
-\delta_I{}^J & 0
\end{pmatrix}~, \qquad
\omega_{ab} =
\begin{pmatrix}
0 & \delta_I{}^J \\
-\delta^I{}_J & 0
\end{pmatrix}~.
\end{align}
In other words, the coordinates $\Phi^I$ and $\Psi_I$ we have found from projective
superspace are holomorphic Darboux coordinates for the target space. 

\subsection{Superpotentials and tri-holomorphic isometries}\label{sect_AdSSuper}
An obvious question to ask is whether, given an $\cN=2$ supersymmetric $\s$-model 
\eqref{4DN1action} invariant under the second supersymmetry transformation  \eqref{3.25SUSY}, 
it is possible to deform the
action to include a superpotential,
\begin{align}\label{eq_AdS4pot}
\int \rd^4x\, \rd^2 \q\,\rd^2\bar\q\, E\,\cK (\vf, \bar \vf)+
	\left(\int \rd^4x\, \rd^2 \q\, \cE\, W (\vf) + \textrm{c.c.}\right)~, 
\end{align}
with $\cE$ the chiral density. 
However, there is no way to distort the supersymmetry transformation rule \eqref{3.25SUSY}
modulo a trivial symmetry transformation \cite{BKsigma1, BKsigma2}. 
It follows that 
the superpotential terms in \eqref{eq_AdS4pot} must be separately invariant. In order
for this to occur, the integral
\begin{align}
\int \rd^4x\, \rd^2 \q\,\rd^2\bar\q\, E\, \ve\, (W_a \omega^{ab} \cK_b + \textrm{c.c.})
\end{align}
must vanish. Since $\ve$ is real linear, we find
\begin{align}
X^a \cK_a + \bar X^{\bar a} \cK_{\bar a} = \L(\phi) + \bar \L(\bar\phi)
\end{align}
where
\begin{align}
X^a := - \omega^{ab} W_b
\end{align}
is a holomorphic Killing vector. Note that $X^a W_a = 0$ by construction.
Moreover, from its construction, one can further show that $X^a$ must be tri-holomorphic,
\begin{align}
\cL_X \omega_{ab} = 0 \quad \implies\quad \cL_X J_A = 0~.
\end{align}

If we impose the additional requirement that the action \eqref{4DN1action} itself be invariant under
the holomorphic isometry $X^a$, we find the stricter condition
\begin{align}
\cL_X \cK =X^a \cK_a + \bar X^{\bar a} \cK_{\bar a} = 0~.
\end{align}
Note that the superpotential is automatically invariant, $X^a W_a = 0$.
It is straightforward to check that
\begin{align}
\cL_X \cK = 0 \quad \Longleftrightarrow \quad \cL_V \left(\frac{W}{\mu} + \frac{\bar W}{\bar \mu}\right) = 0
	\quad \implies \quad [X,V] = 0~.
\end{align}

There is an interesting geometric interpretation of this construction.
We observe that within $\cN=1$ AdS, a superpotential is not distinct from a purely
holomorphic contribution to $\cK$,
\begin{align}
\int \rd^4x\, \rd^2 \q\, \cE\, W + \textrm{c.c.} =
	\int \rd^4x\, \rd^2 \q\,\rd^2\bar\q\, E\, \left(\frac{W}{\mu} + \textrm{c.c.}\right)~.
\end{align}
So the addition of a superpotential corresponds to a modification of the
Lagrangian $\cK$ by
\begin{align}
\cK \rightarrow \cK' = \cK + \frac{W}{\mu} + \frac{\bar W}{\bar\mu}~.
\end{align}
Recall that the target space geometry possesses a U(1) Killing vector
$V^a$ which rotates the complex structures. It is given by eq. \eqref{eq_VAds}.
If we modify the original action to include a superpotential, it is natural
to absorb this superpotential back into $\cK$. This induces the transformation
\begin{align}
V^a \rightarrow V'^a = V^a - \frac{1}{2|\mu|} X^a~.
\end{align}
Since both $V^a$ and $V'^a$ should rotate the complex structures in the
same way, it follows that $X^a$ must leave them invariant. In other words,
$X^a$ must be tri-holomorphic.

This structure emerges naturally in projective superspace if we consider
the gauging of holomorphic symmetries of the original 
$\s$-model action in projective superspace.

\subsubsection{Tri-holomorphic isometries}
We now return to 
the off-shell $\cN=2$ supersymmetric $\s$-model \eqref{3.3}.
The reader should keep in mind that this theory is associated with some real analytic  K\"ahler manifold $\cX$, 
and $K(\F, \bar \F)$ is its K\"ahler potential in local complex coordinates $\F^I$.  Suppose $\cX$ has a 
U(1) isometry group generated by a holomorphic Killing vector field
\begin{align}
X=  X^I(\F) \pa_I +  \bar X^{\bar I}(\bar \F) \pa_{\bar I}~.
\end{align}
Under an infinitesimal isometry transformation 
\begin{align} \label{3.38}
\ri \hat e 
 \F^I  \equiv X \F^I = X^I(\F)~, 
\end{align}
the K\"ahler potential changes as 
\begin{align}\label{3.39}
\ri \hat e 
 K(\F, \bar \F) = X^I \pa_I K + \bar X^{\bar I} \pa_{\bar I} K= F(\F) + \bar F(\bar \F)~,
\end{align}
for some holomorphic function $F(\F)$.
Here we have formally introduced the U(1) generator, $\hat e$, 
in order to make contact with the description in terms of the gauge-covariant derivatives 
\eqref{g-cd}.

The isometry transformation can be extended to the arctic variables of the $\s$-model  \eqref{3.3}
by analytic continuation:
\begin{align}
\ri \hat e \U^I = X^I(\U)~,
\label{3.40}
\end{align}
which implies the following transformation of  the antarctic variables
\begin{align}
\ri \hat e \breve{\U}^{\bar I} = \bar X^{\bar I} (\breve{\U})~.
\end{align}
It immediately follows from \eqref{3.40} that the tangent bundle variables $(\F^I, \S^I)$ 
transform as 
\begin{align}
\ri \hat e \Phi^I = X^I(\Phi)~, \qquad
\ri \hat e \S^I = \S^J \pa_J X^I(\Phi)~.
\end{align}
Due to \eqref{3.39}, it holds that 
\begin{align}
\ri \hat e K(\U, \breve \U) = X^I \pa_I K + \breve X^{\bar J} \pa_{\bar J} K
	= F(\U) + \bar F(\breve \U)~.
\end{align}
Because $K(\U, \breve \U)$ is not invariant in the case $F \neq0 $,
it turns out that the tangent bundle
Lagrangian $\cL$ is not invariant either. Observe that
\begin{align}
\ri \hat e \cL = \frac{1}{2s} \oint_C\frac{\rd\zeta}{2\pi \ri \zeta} \cS^{[2]} \Big(F (\U) + \bar F(\breve \U)\Big)
	= \frac{1}{2s} \oint_C\frac{\rd\zeta}{2\pi \ri \zeta} \Big(\frac{\ri \mu}{\z} F(\U) + \ri \bar\mu \z
	\bar F(\breve \U)\Big)~,
\end{align}
where we have dropped terms
which do not contribute under the contour integral. 
Assuming that the contour is around the origin in the $\z$-plane, this gives
\begin{align}
\ri \hat e \cL = \frac{\ri \mu}{2|\mu|} \S^I F_I(\Phi) - \frac{\ri \bar\mu}{2|\mu|} \S^{\bar J} \bar F_{\bar J}(\bar\Phi)~,
\end{align}
where $F_I = \pa_I F$.
So long as $\S^I$ is complex linear, this remains a symmetry of the action  \eqref{3.1sma}. 
However, when we go to the first order form, eq. \eqref{3.17FO}, 
we must modify the transformation
law of $\Psi_I$:
\begin{align}\label{3.46}
\ri \hat e \Psi_I = -\pa_I X^J \Psi_J - \frac{\ri \mu}{2|\mu|} F_I~.
\end{align}
The presence of the $\m$-dependent term implies that $\J_I$ does not transform
as a holomorphic one-form at the point $(\F, \bar \F)$ of $\cX$, 
unlike in the super-Poincar\'e case \cite{Kuzenko:Superpotentials}.

We end up with a holomorphic vector field 
\bea
X= X^a (\vf) \pa_a + \bar X^{\bar a} (\bar \vf) \pa_{\bar a}
\eea
on the 
hyperk\"ahler target space
parametrized by local complex coordinates $\vf^a = (\F^I, \J_I)$.
This vector field acts on the complex coordinates 
as follows:
\begin{align}\label{3.48}
\ri \hat e \Phi^I \equiv X\F^I = X^I (\F)~,\qquad
\ri \hat e \Psi_I \equiv X \J_I
= -\pa_I X^J(\Phi) \Psi_J - \frac{\ri \mu}{2|\mu|} F_I(\Phi)~.
\end{align}
The K\"ahler potential $\cK$, which serves as the $\cN=1$ AdS Lagrangian
for the cotangent bundle \eqref{eq_dualActionAdS}, remains invariant, 
\bea
\ri\hat e \cK \equiv X \cK  = 0~. 
\eea
This is 
 completely natural 
 since in AdS the K\"ahler potential \emph{must} be invariant -- there are no K\"ahler
transformations. 
Therefore the vector field $X$ constructed is a Killing vector on the hyperk\"ahler target space. 
Moreover, one can check that
this vector field is actually tri-holomorphic. This follows from the fact that $X$ is 
Hamiltonian
with respect
to the canonical holomorphic symplectic two-form $\omega = \rd \Phi^I \wedge \rd \Psi_I$.

In the super-Poincar\'e case, the complex variables $\vf^a = (\F^I, \J_I)$ parametrize 
(an open domain of the zero section of) the cotangent bundle of $\cX$. The curious feature 
of the AdS case is that $\J_I$ does not transform as a (1,0) form at the point $(\F, \bar \F)$ of $\cX$.
Indeed, let us consider two coordinate charts $U$ and $U'$ for $\cX$ parametrized by complex coordinates 
$\F^I$ and $\F^{I'}$ respectively, which are related to each other by a holomorphic transformation 
$\F^{I'} = f^{I'} (\F)$, on the intersection of the charts, $U \bigcap U' \subset \cX$.
Let $K(\F, \bar \F)$ and  $K'(\F', \bar \F')$ be the K\"ahler potentials defined in the charts $U$ and $U'$ 
respectively. On the intersection $U \bigcap U' \subset \cX$, we have 
\bea
K'(\F', \bar \F') = K(\F, \bar \F) +\l(\F) + \bar \l (\bar \F) ~. 
\eea
It can be seen that the  variables $\J_I$ and $\J_{I'}$ corresponding to  the charts $U$ and $U'$
should be related to each other by the rule
\bea
\J_{I' } = \frac{\pa \F^J }{ \pa \F^{ I' }  } \Big( \J_J - \frac{\ri \mu}{2 |\mu| } \pa_J \l (\F ) \Big) ~,
\label{3.51}
\eea
in order to be consistent with the isometry transformation law \eqref{3.46}.
The transformation $(\F^I, \J_I) \to (\F^{I'}, \J_{I'}) $ proves to be symplectic with respect to $\o$, 
\bea
 \rd \Phi^I \wedge \rd \Psi_I =  \rd \Phi^{I'} \wedge \rd \Psi_{I'}~.
 \eea
One can think of  \eqref{3.51} as the transformation law of a deformed  holomorphic  (1,0) form. 

We can add some more flavor to the above discussion. First of all, 
let us consider the K\"ahler one-form on $\cX$, 
\bea
\r = \frac{\ri}{2} K_I \rd \F^I -\frac{\ri}{2} K_{\bar I} \rd \bar \F^{\bar I}~.
\eea 
On the intersection of two charts, we have 
\bea
\r' = \r +\frac{\ri}{2} \rd(\l - \bar \l)~,  
\eea
so $\r$ is not globally defined. Secondly, defining $\J:= \J_I \rd \F^I $ we observe that 
\bea
\J' = \J  - \frac{\ri \mu}{2 |\mu| } \rd \l ~. 
\eea
This result means that 
\bea
{\bm \r} := \r + \frac{\bar \m}{|\m|} \J +  \frac{ \m}{|\m|} \bar \J
\eea
is a well-defined one-form on the hyperk\"ahler target space. 

\subsubsection{A convenient fictitious coordinate}
We would like to gauge the tri-holomorphic isometry introduced above
using the intrinsic vector multiplet. This can be done using old results on gauged
$\cN=1$ supersymmetric $\s$-models \cite{BWitten,CL,S,HKLR,BWess}.
However, in order to gauge a Lagrangian, it is easiest to deal with
\emph{gauge invariant} Lagrangians. The standard way to deal with this,
developed in \cite{HKLR},
is to introduce a fictitious field which is a pure gauge degree of
freedom that counters the gauge transformation of the original Lagrangian.
That is, if $K$ is the original polar multiplet K\"ahler potential
transforming as
\begin{align}
\ri \hat e K = F(\U) + \breve F(\breve \U)~,
\end{align}
we introduce a new arctic multiplet $\U^0$ 
with the U(1) transformation law 
$\ri\hat e \U^0 = F(\U)$
and consider the modified K\"ahler potential
\begin{align}
K' = K - \U^0 - \breve \U^0~.
\end{align}
This new K\"ahler potential is U(1) invariant.
The theory with Lagrangian $K'$ is invariant under gauge transformations 
\begin{align}
\U^0 \rightarrow \U^0 + \L~,
\end{align}
with $\L$ an arbitrary arctic superfield. 
This gauge symmetry allows us to gauge away $\U^0$. In the gauge $\U^0 =0$ 
we return to the original symmetry. 
In other words, the theory with Lagrangian $K'$ is completely equivalent to the original $\s$-model.

Let us group all the arctic multiplets together as
$\U^{I'} = (\U^0, \U^I)$. Their isometry transformation is generated by the holomorphic vector field
 $X^{I'} = (F, X^I)$ where both $F$ and $X^I$
depend only on $\U^I$. On the tangent bundle coordinates, we find
\begin{subequations}
\begin{alignat}{4}
\ri \hat e \Phi^{I'} &= X^{I'} &\quad& \implies \quad \ri\hat e \Phi^0 = F(\Phi)~, &\qquad \ri\hat e \Phi^I &= X^I(\Phi)~, \\
\ri \hat e \S^{I'} &= \S^{J'} \pa_{J'} X^{I'} &\quad &\implies \quad \ri\hat e\S^0 = \S^J \pa_J F(\Phi)~,
	&\qquad \ri\hat e \S^I &= \S^J \pa_J X^I(\Phi)~,
\end{alignat}
\end{subequations}
where $F(\Phi)$ and $X^I(\Phi)$ depend only on $\Phi^I$.

Making use 
of $K'$ leads to a modified Lagrangian $\cL'$:
\begin{align}
\cL' = \frac{1}{2s} \oint_C\frac{\rd\zeta}{2\pi \ri \zeta} \cS^{[2]} K' = 
	\frac{1}{2s} \oint_C \frac{\rd\zeta}{2\pi \ri \zeta} \cS^{[2]} (K - \U^0 - \breve \U^0)
	= \cL - \frac{\ri \mu}{2|\mu|} \S^0 + \frac{\ri \bar \mu}{2|\mu|} \bar\S^0~.
\end{align}
By construction $\cL'$ is U(1) invariant, but we can see this explicitly by noting
that the transformation of $\cL$ is cancelled by the transformation of the
additional terms.

Now we construct
$\cK' = \cL' + \S^{I'} \Psi_{I'} + \bar \S^{\bar J'} \bar\Psi_{\bar J'}$.
Because $\cL'$ is invariant, we choose
\begin{align}
\ri \hat e \Psi_{I'} = -\pa_{I'} X^{J'} \Psi_{J'}~.
\end{align}
This implies
\begin{align}
\ri \hat e \Psi_{0} = 0~,\qquad \ri \hat e \Psi_{I} = -\pa_I F \Psi_0 - \pa_I \S^J \psi_J~.
\end{align}
Now let us  make an observation:
\begin{align}
\cK' = \cL + \S^I \Psi_I + \bar \S^{\bar I} \bar \Psi_{\bar I}
	+ \S^0\left(\J_0 - \frac{\ri \mu}{2|\mu|}\right) + \bar\S^0 \left(\bar\J_0 + \frac{\ri \bar\mu}{2|\mu|}\right).
\end{align}
Because $\cL$ is independent of $\S^0$, the equation of motion
of $\S^0$ merely enforces $\J_0 = \ri \mu / 2|\mu|$. This is consistent with
$\ri\hat e \J_0 = 0$. For the other $\S^I$ coordinates, we dualize
as usual and end up with $\cK' = \cK$.
The new K\"ahler potential is the same as the old! By construction it
is gauge invariant. As before, 
we stay with the dynamical variables $(\F^I, \J_I)$ with the U(1) transformation law
\eqref{3.48}. 
There are no other dynamical fields.

\subsection{Gauged isometries}
Now let us gauge the Lagrangian $K'(\U, \breve \U)$ by covariantizing
the arctic and antarctic superfields by the introduction of
new covariant derivatives $\bm\cD_\cA$, eq. \eqref{g-cd},  
where the vector multiplet associated
with the generator $\hat e$ is the intrinsic vector multiplet of
AdS, as discussed in section \ref{IntrinsicVH}.
Instead of ordinary arctic multiplets $\U^{I'} $ obeying the analyticity conditions 
\eqref{ana}, we have to consider gauge covariantly arctic multiplets constrained by 
\bea
{\bm \cD}^{(1)}_{\a} \U^{I'}   = {\bar {\bm \cD}}^{(1)}_{\ad} \U^{I'}   =0~, \qquad 
{\bm \cD}^{(1)}_\a := v_i {\bm \cD}^{i}_\a ~, \quad 
{\bar {\bm \cD}}^{(1)}_\ad := v_i {\bar {\bm \cD}}^{i}_\ad ~,
\eea
and similarly for their smile-conjugates $\breve{\U}^{\bar I'} $. 
The gauge covariant superfields $\U^{I'}$ and $\breve{\U}^{\bar I'} $ 
are assumed to have the  functional form \eqref{1.23} and \eqref{1.24} respectively. 

By the usual argument, one can show that $\Phi^{I'} = \U^{I'}_0$ is
covariantly chiral. However, $\S^{I'} = \U^{I'}_1$ is no longer complex linear,
but is instead \emph{modified} complex linear,
\begin{align}
-\frac{1}{4} (\bar {\bm\cD}^2 - 4 \mu) \S^{I'} = X^{I'}~.
\end{align}
The transformation law of the arctic multiplet
\begin{align}
\delta \U^{I'} = -(\xi^A \bm\cD_A + \frac{1}{2} \l^{cd} M_{cd} + 2 \ve \mathcal S^{ij} J_{ij} + 2 \ri \ve \hat e) \U^{I'}
\end{align}
leads to the transformation laws of $\F^{I'}$ and $\S^{I'} $:
\begin{subequations}
\begin{align}
\delta \Phi^{I'} &= \bar\ve_\ad \bar{\bm\cD}^\ad \S^{I'} + 2 \ve \mu \S^{I'} - 2 \ri \ve \hat e \Phi^{I'} \\
	&= \frac{1}{2} (\bar {\bm\cD}^2 - 4 \mu) (\ve \S^{I'})
\end{align}
\end{subequations}
and
\begin{align}
\delta \S^{I'} &= -\ve^\a \bm\cD_\a \Phi^{I'} + \bar\ve_\ad \bar{\bm\cD}^\ad \U_2^{I'}
	+ 4 \mu \ve \U_2^{I'} - 2 \ri \ve \hat e \S^{I'} \eol
	&= -\ve^\a \bm\cD_\a \Phi^{I'} + \bar{\bm\cD}_\ad (\bar\ve^\ad \U_2^{I'}) - 2 \ri \ve \hat e \S^{I'}~.
\end{align}
As before, we have
$\ri \hat e \Phi^{I'} = X^{I'}$
and $\ri \hat e \S^{I'} = \S^{J'} \pa_{J'} X^{I'}$.

The  tangent bundle Lagrangian  is
\begin{align}
\cL' = \frac{1}{2s} \oint_C \frac{\rd\zeta}{2\pi \ri \zeta} \cS^{[2]} K'~.
\end{align}
We introduce the first order action
\begin{align}
S_{\rm F.O.} &= \int \rd^4x\, \rd^2\q\, \rd^2\bar\q\, E\, \cK' -
	\Big(\int \rd^4x\, \rd^2\q\,\cE\, \Psi_{I'} X^{I'} + \textrm{c.c.}\Big) ~,
\end{align}
where 	
\begin{align}
\cK' &= \cL' + \S^{I'} \Psi_{I'} + \bar \S^{\bar J'} \bar \Psi_{\bar J'}~.
\end{align}
The additional holomorphic terms we have added are necessary so
that the equation of motion  for  $\Psi_{I'}$
imposes the modified
complex linear condition on $\S^{I'}$.
We must take $\ri\hat e \Psi_{I'} = - \pa_{I'} \S^{J'}(\Phi)\, \Psi_{J'}$ so that
each of the terms above is separately gauge invariant.

Next, we must determine the transformation rules.
As before, $\delta\Phi^{I'}$ must be manifestly covariantly chiral,
\begin{align}
\delta \Phi^{I'} &:= \frac{1}{2} (\bar {\bm\cD}^2 - 4 \mu) (\ve \S^{I'})~.
\end{align}
This can be rewritten
\begin{align}
\delta \Phi^{I'} &:= \bar\ve_\ad \bar{\bm\cD}^\ad \S^{I'} + 2 \ve \mu \S^{I'} 
	+ \frac{1}{2} \ve (\bar {\bm\cD}^2 - 4 \mu) \S^{I'}~.
\end{align}
We keep the same rule for $\delta\S^{I'}$. This leads to
\begin{align}
&\int \rd^4x\, \rd^2\q\, \rd^2\bar\q\, E\, \delta \cL' \non \\
&{}\qquad =
	\int \rd^4x\, \rd^2\q\, \rd^2\bar\q\, E\,
	\left(\frac{1}{2} \S^{I'} (\bar {\bm\cD}^2 - 4 \mu) \Big(\ve \frac{\pa \cL'}{\pa \Phi^{I'}}\Big)
	- 2 \ve \ri \hat e \S^{I'} \frac{\pa \cL'}{\pa \S^{I'}} + \textrm{c.c.}\right)~.
\end{align}
We postulate that
\begin{align}
\delta \Psi_{I'} &= -\frac{1}{2} (\bar {\bm\cD}^2 - 4 \mu) \left(\ve \frac{\pa \cL'}{\pa \Phi^{I'}}\right)~,
\end{align}
which leads to
\begin{align}
&\int \rd^4x\, \rd^2\q\, \rd^2\bar\q\, E\, \delta \left(\cL' + \S^{I'} \Psi_{I'} + \bar \S^{\bar J'} \bar \Psi_{\bar J'}\right)
	\eol & \quad
= \int \rd^4x\, \rd^2\q\, \rd^2\bar\q\, E\, \left(
	-2 \ve \ri\hat e \S^{I'} \frac{\pa \cL'}{\pa \S^{I'}} - 2 \ve \ri \hat e \S^{I'} \Psi_{I'} + \textrm{c.c.}\right)~.
\end{align}
For the superpotential piece, we have
\begin{align}
- \int \rd^4x\, \rd^2\q\, \cE\,\delta (\Psi_{I'} X^{I'}) &=
	\int \rd^4x\, \rd^2\q\, \rd^2\bar\q\, E\, \left(
	- 2 \ve \frac{\pa \cL'}{\pa \Phi^{I'}} X^{I'}
	+ 2 \ve \Psi_{I'} \pa_{J'} X^{I'} \S^{J'} \right)~.
\end{align}
Adding everything together gives
\begin{align}
\delta S_{\rm F.O.}
	= -2 \int \rd^4x\, \rd^2\q\, \rd^2\bar\q\, E\, \ve \,\ri\hat e \cL' = 0~.
\end{align}

Now we want to finish the duality by eliminating $\S^{I'}$. Recall that
\begin{align}
\cL' &= \frac{1}{2s} \oint_C \frac{\rd\zeta}{2\pi \ri \zeta} \cS^{[2]} (K - \U^0 - \breve \U^0)
	= \cL - \frac{\ri \mu}{2|\mu|} \S^0 + \frac{\ri \bar \mu}{2|\mu|} \bar \S^0~.
\end{align}
Here $\cL$ is independent of $\S^0$ and $\Phi^0$. Moreover, $\Phi^0$
is completely absent from the action. The dual kinetic Lagrangian is
\begin{align}
\cK' &= \cK + \S^0 \left(\Psi_0 - \frac{\ri \mu}{2|\mu|} \right)
	+ \bar\S^0 \left(\bar\Psi_0 + \frac{\ri \bar\mu}{2|\mu|} \right)~.
\end{align}
The equation of motion for $\S^0$ again fixes $\Psi_0 = \ri \mu$,
so we find $\cK' = \cK$. The elimination of the physical $\S^I$ proceeds
as usual. However, now we have a superpotential,
\begin{align}
W = -\Psi_{I'} X^{I'} = - \frac{\ri \mu}{2|\mu|} F - \Psi_I X^I~.
\end{align}
So the full dual action can be written in the form 
\begin{align}
S_{\rm dual} = \int \rd^4x\, \rd^2\q\, \rd^2\bar\q\, E\, 
	\left( \cK + \frac{W}{\mu} + \frac{\bar W}{\bar \mu}\right)~.
\end{align}
It is invariant under the extended supersymmetry transformation 
\begin{align}
\delta \Phi^I = +\frac{1}{2} (\bar {\bm\cD}^2 - 4 \mu) \left(\ve \frac{\pa \cK}{\pa \Psi_I}\right)~,\qquad
\delta \Psi_I = -\frac{1}{2} (\bar {\bm\cD}^2 - 4 \mu) \left(\ve \frac{\pa \cK}{\pa \Phi^I}\right)~.
\end{align}

Some comments are in order. First, we have written the transformation
laws in terms of the  gauge-covariant $\cN=1$ derivatives $\bm\cD_A$. However,
because we chose the intrinsic vector multiplet to gauge the U(1)
group, in this AdS frame the  $\cN=1$ AdS derivatives possess no
U(1) curvature. In other words, the U(1) connection is \emph{pure gauge}
and we can adopt a gauge where it vanishes, $\bm \cD_A \rightarrow \cD_A$.
This removes all trace of the gauging from the $\cN=1$ superspace geometry.

Second, the tri-holomorphic isometry $X^a = (X^I, X_I)$ with
\begin{align}
X^I &= \ri \hat e \Phi^I = X^I(\Phi) = -\frac{\pa W}{\pa \Psi_I}~, \eol
X_I &= \ri \hat e \Psi_I =  -\frac{\ri \mu}{2|\mu|} F_I - \pa_I X^J \Psi_J = +\frac{\pa W}{\pa \Phi_I}~,
\end{align}
indeed obeys $X^a = -\omega^{ab} W_b$, as required.

\section{Poincar\'e coordinates for AdS$^{4|4}$ and AdS$^{4|8}$}
\setcounter{equation}{0}
\label{geometry}

The aim of this section is to describe a new conformally flat realization for 
four-dimensional (4D) $\cN=2$ AdS superspace with the property that ${\rm AdS^{4|8} }$ is foliated
into a union of 3D $\cN=4$ flat superspaces with a real central charge corresponding to 
a derivative in the fourth dimension. 
This realization will be used in the next sections. As a warm-up exercise, 
we first consider the case of 4D $\cN=1$ AdS superspace.

\subsection{AdS$^{4|4}$}
The conformal flatness of the superspace AdS$^{4|4}$ was established by Ivanov and Sorin  \cite{IS} and later on
reviewed, in the modern form, in  textbooks \cite{GGRS,BK}. 
The approaches pursued in \cite{IS,GGRS,BK} made use of stereographic coordinates in AdS${}_4$ 
in which  the space-time metric is
\bea
{\rm d}s^2 
= \frac{{\rm d}x^a  {\rm d}\,x_a }{\big(1-\frac{1}{4}|\m|^2x^2\big)^2}~.
\label{metric}
\eea  
This metric is manifestly invariant under the group of four-dimensional Lorentz 
transformations, O(3,1). Here we would like to derive an alternative conformally flat realization 
of AdS$^{4|4}$ which is characterized by the space-time metric (\ref{1.4}).
The latter metric is invariant under the group of three-dimensional Poincar\'e 
transformations, IO(2,1).

Let us start by recalling the structure of the super-Weyl transformations 
in 4D $\cN=1$ old minimal supergravity \cite{GWZ}. 
The superspace geometry of  supergravity
is described by covariant derivatives 
\be
\cD_{A}=(\cD_a,\cD_\a,{\bar \cD}^\ad)
= E_A{}^M\pa_M+\hf \O_A{}^{bc}M_{bc}
\ee 
and a set of constrained superfields, $\cR,\,\cG_{\a\ad}$ and $\cW_{\a\b\g}$
in terms of which the torsion and curvature tensors are constructed \cite{GWZ}. 
We refer the reader to \cite{BK} for a detailed 
description of the geometry of old minimal supergravity.
Let  $D_{A}=(D_a,D_\a,\DB^\ad)$ be another set of superspace covariant derivatives  
characterized by the torsion superfields $R,\,G_{\a\ad}$ and $W_{\a\b\g}$. 
The two supergeometries, which are associated with $\cD_A$ and $D_A$, 
are said to be conformally related  if their covariant derivatives  are related by a 
super-Weyl transformation of the form\footnote{Here
$M_{\a\b}=\hf(\s^{ab})_{\a\b}M_{ab}$,  $\bar{M}_{\ad\bd}=-\hf(\tilde{\s}^{ab})_{\ad\bd}M_{ab}$
and  $M_{ab}$ are the Lorentz generators with spinor  and vector indices, respectively, see  \cite{BK}
for more details.} \cite{HT}
\bsubeq
\bea
\cD_\a&=&\re^{\hf\s-\bar{\s}}\Big(D_\a-(D^\b\s)M_{\a\b}\Big)
~,
\label{4D-N=1-1}
\\
\cDB_\ad&=&\re^{\hf\sba-\s}\Big(\DB_\ad
-(\DB^\bd\sba)\bar{M}_{\ad\bd}\Big)
~,
\label{4D-N=1-2}
\\
\cD_{\a\bd}&=&\frac{\ri}{2}\{\cD_\a,\cDB_\bd\}~,
\label{4D-N=1-3}
\eea
\esubeq
where the parameter $\s$ is covariantly chiral, $\DB_\ad\s=0$. 
The  components of the torsion transform 
as
\bsubeq
\bea
\cR&=&-\frac{1}{4}\re^{-2\s}(\DB^2-4R)\re^{\sba}~,
\\
\cG_{\a\ad}&=&\re^{-(\s-\sba)/2}\Big(G_{\a\ad}+\hf(D_\a\s)(\DB_\ad\sba)+\ri D_{\a\ad}(\sba-\s)\Big)
~, \\
\cW_{\a\b\g}&=&\re^{-3\s/2}W_{\a\b\g}~.
\eea
\esubeq

If the covariant derivatives $D_A$ are flat, and hence $R=G_{\a\ad}=W_{\a\b\g}=0$, then 
the geometry described by $\cD_A$ is said to be conformally flat.
A well-known  example of conformally flat supergeometry is AdS$^{4|4}$.
The geometry of this superspace is characterized by $\cG_{\a\ad}=\cW_{\a\b\g}=0$ and 
$\cR=\mu ={\rm const}$, $\m \neq0$. 
The covariant derivatives obey the (anti-) commutation relations \eqref{1.5}.
The requirement  that  AdS$^{4|4}$ is conformally flat
means that there exists a `flat' chiral superfield $\s$, $\bar D_\ad \s =0$,   such that
\begin{subequations}
\bea
\mu&=&-\frac{1}{4}\re^{-2\s}\DB^2\re^{\sba}~, \label{4D_N=1_conf-flat-a}\\
0&=&\hf(D_\a\s)(\DB_\ad\sba)+\ri\pa_{\a\ad}(\sba-\s)
~,
\label{4D_N=1_conf-flat-b}
\eea
\end{subequations}
 where
  $D_A=(\pa_a,D_\a,\DB^\ad)$ are the  covariant derivatives 
 of  4D $\cN=1$ Minkowski superspace parametrized by Cartesian coordinates $(x^a,\q^\a,\qb_\ad)$,
  \bea
D_\a=\frac{\pa}{\pa\q^\a}+\ri\qb^\bd\pa_{\a\bd}~,\qquad
\DB_\ad=-\frac{\pa}{\pa\qb^\ad}-\ri\q^\b\pa_{\b\ad}~. 
\eea
Note that  the equation  (\ref{4D_N=1_conf-flat-b}) can be equivalently rewritten as
\bea
0={[}D_\a,\DB_\bd{]}\re^{-\hf(\s+\sba)}~.
\label{4D_N=1_conf-flat-bis}
\eea

At this stage, it is convenient  to introduce 
a $3+1$ splitting 
of the 4D vector indices 
that is suitable for a 3D foliation of AdS${}_4$.
We adopt the 3D spinor notation introduced in \cite{KPT-MvU,KLT-M}.
The 4D sigma-matrices are
\bea
(\s_{a } )_{\a \dot \b}:= ({\mathbbm 1}, \vec{\s} ) ~, \qquad
(\tilde{\s}_{a} )^{{\dot \a}  \b}:= 
\ve^{\b \g}\ve^{\ad\dd}(\s_{a} )_{\g \dot \d}=
({\mathbbm 1}, - \vec{\s} ) ~, \qquad {m}=0,1,2,3
~,
\label{4DsigmaM}
\eea
where $\vec{\s}=(\s_1,\s_2,\s_3)$ are the Pauli matrices.
By deleting the matrices with space index $a =2$, we obtain the 
real and symmetric
3D   gamma-matrices
\begin{subequations}
\bea
(\s_{a} )_{\a \dot \b}\quad & \longrightarrow & \quad (\g_{\hat a} )_{\a  \b} = (\g_{\hat a})_{\b\a} ~
=({\mathbbm 1}, \s_1, \s_3) ~,\\
(\tilde{\s}_{a} )^{\dot \a  \b}\quad & \longrightarrow & \quad (\g_{\hat a}  )^{\a  \b} = (\g_{\hat a})^{\b\a}
=\ve^{\a \g} \ve^{\b \d} (\g_{\hat a})_{\g \d} ~,
\eea
\end{subequations}
where the spinor indices are  raised and lowered using
the SL(2,${\mathbb R}$) invariant tensors
\bea
\ve_{\a\b}=\left(\begin{array}{cc}0~&-1\\1~&0\end{array}\right)~,\qquad
\ve^{\a\b}=\left(\begin{array}{cc}0~&1\\-1~&0\end{array}\right)~,\qquad
\ve^{\a\g}\ve_{\g\b}=\d^\a_\b
\eea
by the rule:
\bea
\psi^{\a}=\ve^{\a\b}\psi_\b~, \qquad \psi_{\a}=\ve_{\a\b}\psi^\b~.
\eea
Note that the 3D gamma-matrices satisfy\footnote{The 3D Minkowski
metric is $\eta_{mn}=\eta^{mn}={\rm diag}(-1,1,1)$
and the Levi-Civita tensor is normalised as $\ve_{012}=-\ve^{012}=-1$.}
\bsubeq
\bea
&\g_{\hat{a}}:=(\g_{\hat{a}})_\a{}^{\b}=\ve^{\b\g}(\g_{\hat{a}})_{\a\g}
~,
\\
&\{\g_{\hat{a}},\g_{\hat{b}}\}=2\eta_{\hat{a}\hat{b}} {\mathbbm 1}
~,~~~~
\g_{\hat{a}}\g_{\hat{b}}=\eta_{\hat{a}\hat{b}} {\mathbbm 1} +\ve_{\hat{a}\hat{b}\hat
{c}}\g^{\hat{c}}
~.
\eea
\esubeq

There is no difference between dotted and undotted spinor indices in three dimensions.
Given a four-vector $V_a$, 
it decomposes as follows
\begin{subequations}
\bea
V_{\a\bd}=(\s^{{a}})_{\a\bd}V_{{a}} ~&\to&~ V_{\a\b}+\ri \, \ve_{\a\b}V_z~,\qquad
V_{\a\b}:=(\g^{\hat a} )_{\a\b}V_{\hat a}~,
\\
V^{\ad \b}=(\tilde{\s}^{{a}})^{\ad\b}V_{{a}} ~&\to& ~V^{\a\b} + \ri \, \ve^{\a\b}V_z~,\qquad
V^{\a\b}:=(\g^{\hat a} )^{\a\b}V_{\hat a}~,
\eea
\end{subequations}
where $\hat a =0,1,3$.
In particular, the 4D vector coordinates $x^a$  split as $x^a=(x^{\hat a},z)$, where
$z:=x^2$.
Then, the 4D $\cN=1$ flat covariant derivatives 
take the form
\bea
D_\a=\frac{\pa}{\pa\q^\a}+\ri\qb^\b\pa_{\a\b}
-\qb_\a\pa_z
~,~~~
\DB_\a=-\frac{\pa}{\pa\qb^\a}
-\ri\q^\b\pa_{\a\b}
-\q_\a\pa_z~.~~~
\label{4D-N=1-3D}
\eea
They obey the anti-commutation relations
\bea
\{D_\a,D_\b\} = \{\DB_\a,\DB_\b\}=0~, \qquad
\{D_\a,\DB_\b\}
=
-2\ri\pa_{\a\b}
+2\ve_{\a\b}\pa_z 
\eea
which correspond to the 3D $\cN=2$ Poincar\'e supersymmetry with central charge.
The central charge is identified with $\pa_z$.

We are now prepared to look for a solution of the equations (\ref{4D_N=1_conf-flat-a})
and (\ref{4D_N=1_conf-flat-bis}).
Let us  define the bosonic coordinates
\bea
z_L=z-\q^\a\qb_\a~,~~~~~~
z_R=z+\q^\a\qb_\a~,
\eea
which are respectively chiral, $\DB_\a z_L=0$, and antichiral, $D_\a z_R=0$,
with respect to the derivatives (\ref{4D-N=1-3D}).
Then, it is a short computation to prove that the superfields
\bea
&\re^{-\s}=|\mu|z_L-\mub\q^2
~,\qquad
\re^{-\sba}=|\mu|z_R-\mu\qb^2
\eea
satisfy the equations (\ref{4D_N=1_conf-flat-a}) and (\ref{4D_N=1_conf-flat-bis}).
As a result, the relations (\ref{4D-N=1-1})--(\ref{4D-N=1-3}) with  $\re^{-\s}$  given above 
define a
conformally flat 
realization of AdS$^{4|4}$. 

Given a superfield $U(x^{\hat a}, z, \q^\a, {\bar \q}_\a)$, we introduce  the projection 
$U|:= U(x^{\hat a}, z, 0, 0)$.
For the covariant derivatives the projection is defined similarly
\bea
\cD_A|:=E_A{}^M|\pa_M+\hf\F_A{}^{bc}|M_{bc}
~.
\eea
It follows that 
\bea
\de_a:= \cD_a| = 
\re^{-\hf(\s +\sba)}|\pa_{a}+\cdots
=|\mu| z\,\pa_{a}+\cdots
\eea
where the ellipses  denote the  Lorentz connection. 
For the vierbein we get
\bea
e_a{}^m= |\m| z\, \d_a^m
~.
\eea
Therefore the space-time metric has the form \eqref{1.4} with $s=|\m|$.

%%%%%%%%%%%%%%%%%%%%%%%%%%%%%%%%%%%%%%%%%%%%%%%%
%%%%%%%%%%%%%%%%%%%%%%%%%%%%%%%%%%%%%%%%%%%%%%%%
%%%%%%%%%%%%%%%%%%%%%%%%%%%%%%%%%%%%%%%%%%%%%%%%

\subsection{AdS$^{4|8}$}
\label{geometry_N=2_AdS}

The $\cN=2$ AdS superspace was briefly introduced in subsection \ref{N=2AdSgeneral}. 
We recall that AdS$^{4|8}$ is a maximally symmetric geometry that originates within 
the superspace formulation of $\cN=2$ conformal supergravity developed in  \cite{KLRT-M1}.
This formulation is based on the curved superspace geometry 
given by Grimm \cite{Grimm}. What makes this geometry suitable to describe 
conformal supergravity is the invariance of the corresponding constraints under certain super-Weyl 
transformations discovered in \cite{KLRT-M1}.

Let us summarize  
the main ingredients of  the formulation of 4D $\cN=2$ conformal supergravity 
given in \cite{KLRT-M1}.
The  superspace geometry is described
 by covariant derivatives of the form
 \bea
 \cD_\cA =(\cD_a,\cD_\a^i,{\bar \cD}^\ad_i)
 =E_\cA{}^\cM\pa_\cM+\hf \O_\cA{}^{bc}M_{bc}+\F_\cA{}^{kl}J_{kl}
 ~,
 \eea
 where $E_\cA{}^\cM$ is the supervielbein, $\O_\cA{}^{bc}$ the Lorentz connection 
 and $\F_\cA{}^{kl}$ the SU(2) connection 
 (with $J_{kl}$ being the corresponding generators).
The covariant derivatives are subject to certain conventional constraints \cite{Grimm}
which are solved in terms of 
several dimension-1 constrained superfields,
 $\cS_{ij} = \cS_{ji}, \,\cG_{\a\ad},\,\cY_{\a\b} = \cY_{\b\a}$ and $\cW_{\a\b} = \cW_{\b\a}$, 
 and their covariant derivatives.
 The superfield $\cG_{\a\ad}$ is real,  $\overline{\cG_{\a\ad}}=\cG_{\a\ad}$, 
 while the other torsion components are in general complex 
 (the torsion $\cS_{ij}$ can be made real in a special super-Weyl gauge). 
 The constraints turn out to be invariant under super-Weyl transformations generated 
 by a covariantly chiral parameter $\s$. The super-Weyl transformations were given 
 originally in \cite{KLRT-M1} in the infinitesimal form, 
 and then in \cite{KT-M-4D-conf-flat}  in the finite form.

Two superspace geometries described by covariant derivatives $\cD_\cA =(\cD_a,\cD_\a^i,{\bar \cD}^\ad_i)$
and $D_{{\cA}}=(D_a,D_\a^i,\DB^\ad_i)$ are conformally related if 
the covariant derivatives $\cD_\cA$ are obtained from  $D_\cA$ 
by applying a finite 
super-Weyl transformation
\begin{subequations}\label{2.23}
\bea
\cD_\a^i&=&\re^{\hf\sba}\Big(D_\a^i+(D^{\g i}\s)M_{\g\a}-(D_{\a k}\s)J^{ki}\Big)~,
\label{Finite-D}\\
\cDB_{\ad i}&=&\re^{\hf\s}\Big(\DB_{\ad i}+(\DB^{\gd}_{i}\sba)\bar{M}_{\gd\ad}
+(\DB_{\ad}^{k}\sba)J_{ki}\Big)~,
\label{Finite-Db} 
\\
\cD_a
&=&\re^{\hf(\s+\sba)}\Big(
D_a
+{\ri\over 4}(\s_a)^{\a}{}_\bd(\DB^{\bd}_{k}\sba)D_\a^k
+{\ri\over 4}(\s_a)^{\a}{}_\bd(D_\a^k\s)\DB^\bd_{k}
-{1\over 2}\big(D^b(\s+\sba)\big)M_{ab}
\non\\
&&~~~~~~~~~~
+{\ri\over 8}(\ts_a)^{\ad\a}(D^{\b k}\s)(\DB_{\ad k}\sba)M_{\a\b}
+{\ri\over 8}(\ts_a)^{\ad\a}(\DB^{\bd}_{k}\sba)(D_\a^k\s)\bar{M}_{\ad\bd}
\non\\
&&~~~~~~~~~~
-{\ri\over 4}(\ts_a)^{\ad\a}(D_{\a}^k\s)(\DB_\ad^{l}\sba)J_{kl}
\Big)
~,~~~~~~~~
\label{Finite-D_c}
\eea
\end{subequations}
where the parameter  $\s$ is covariantly chiral $\DB^\ad_i\s=0$.
The  dimension-1 components of the torsion in the two geometries are related to each other as follows:
\begin{subequations}
\bea
\cS_{ij}&=&\re^{\sba}\Big(S_{ij}
-{1\over 4}(D^\g_{(i}D_{\g j)}\s)
+{1\over 4}(D^\g_{(i}\s)(D_{\g j)}\s)\Big)
\label{Finite-S}~,
\\
\cG_\a{}^\bd&=&
\re^{\hf(\s+\sba)}\Big(G_\a{}^\bd
-{\ri\over 4}(\s^c)_\a{}^\bd D_c(\s-\sba)
-{1\over 8}(D_\a^k\s)(\DB^{\bd}_k\sba)
\Big)
~,
\label{Finite-G}
\\
\cY_{\a\b}&=&\re^{\sba}\Big(Y_{\a\b}
-{1\over 4}(D^k_{(\a}D_{\b)k}\s)
-{1\over 4}(D^k_{(\a}\s)(D_{\b)k}\s)\Big)
\label{Finite-Y}~,
\\
\cW_{\a\b}&=&\re^{\s}{W}_{\a \b}~.
\label{Finite-W}
\eea
\end{subequations}
The geometry described by $\cD_{{\cA}}$ is said to be {\it conformally flat} if 
the covariant derivatives $D_{{\cA}}$ correspond to a flat superspace
characterized by  $\cS_{ij}=G_a=Y_{\a\b}=W_{\a\b}=0$.
An example of a conformally flat superspace is AdS$^{4|8}$ \cite{KT-M-4D-conf-flat}.\footnote{In the framework 
of nonlinear realizations, the conformal flatness of  AdS$^{4|8}$  was shown in \cite{BILS}.} 
Its geometry is completely determined by a nonzero, real, covariantly constant isotriplet
\bea
\cS^{ij}=\cS^{ji}~,~~~~~~
\overline{\cS^{ij}}=\cS_{ij}~,~~~~~~
\cD_\cA \cS^{jk}
=0
~,
\eea
while the other component of the torsion vanish, 
\be
\cW_{\a\b}=\cY_{\a\b}=\cG_{\a \bd}=0~.
\label{AdS-geometry1}
\ee
The covariant derivatives of AdS$^{4|8}$ obey the (anti-)commutation relations (\ref{1.2}). 
The proof of the conformal flatness of AdS$^{4|8}$ given in  \cite{KT-M-4D-conf-flat} was based on the
use of the stereographic coordinates for AdS${}_4$. We now
derive a new conformally flat realization for  AdS$^{4|8}$  that 
makes
use of the Poincar\'e coordinates. 

The condition of conformal flatness of AdS$^{4|8}$ means that there  exists a chiral  superfield $\s$ 
such that $\cY_{\a\b}=\cG_{\a\bd}=0$ and $\cS^{ij}=\bar{\cS}^{ij}$ is covariantly constant.
Let $D_\cA =(\pa_a,D_\a^i,\DB^\ad_i)$  be
the covariant derivatives of the $\cN=2$ Minkowski superspace, 
\bea
&&D_\a^i={\pa\over\pa\q_i^\a}-\ri(\s^b)_{\a}{}^{\bd}\qb_\bd^{i}\pa_b~, \qquad
\DB^{\ad}_{i}={\pa\over\pa\qb_{\ad}^{i}}-\ri(\s^b)_\b{}^{\ad}\q^\b_{i}\pa_b~, 
\label{2.30}
\eea
with the standard anti-commutation relations
\bea
&&\{D_\a^i,D_\b^j\}=\{\DB^\ad_i,\DB^\bd_j\}=0~,~~~~~~
\{D_\a^i,\DB^\bd_j\}=-2\d^i_j(\s^a)_\a{}^\bd\pa_{a} ~.
\label{2.333}
\eea
It follows from (\ref{Finite-G})  that 
the equation $\cG_{\a\bd}=0$ is equivalent to
\bea
{[}D_\a^k,\DB^\ad_k{]}\re^{\s+\sba}=0~.
\label{G=0}
\eea
In accordance with (\ref{Finite-Y}), 
the condition $\cY_{\a\b}=0$ 
is equivalent to
\bea
D^k_{(\a}D_{\b)k}\,\re^{\s}=0~.~~~~~~
\label{Y=0}
\eea
The equation 
(\ref{Finite-S}) 
leads to 
\bea
\cS^{ij}&=&\frac{1}{ 4}\re^{\s+\sba}(D^{ij}\re^{-\s})=\frac{1}{ 4}\re^{\s+\sba}(\bar D^{ij}\re^{-\s})
~,
\label{Finite-S-2}
\eea
where $D^{ij}:=D^{\a i}D_\a^j=D^{ji}$ and $\bar D^{ij}:=\bar D_\ad^i \bar D^{\ad j}=\bar D^{ji}$.
Due to \eqref{Finite-W}, the equation $\cW_{\a\b}=0$  is satisfied automatically. 
The condition \eqref{Finite-S-2}
tells us that the chiral superfield 
\bea
W:=\re^{-\s}~, \qquad \bar D^\ad_i W =0
\label{W2.32}
\eea
obeys the reality condition
\bea
D^{ij}W=\DB^{ij}\bar{W}~,
\label{2.33}
\eea
which is the Bianchi identity for the chiral field strength, $W$, 
of an Abelian vector multiplet in flat superspace \cite{GSW}.
Associated with  this vector multiplet in flat superspace is 
the {\it intrinsic  vector multiplet} 
in AdS$^{4|8}$ (introduced in section \ref{IntrinsicVH})
such that its covariantly chiral field strength, $\cW$, is constant, 
\bea
\cW = \re^{\s} \, W =1~,
\eea
where we have used the super-Weyl transformation law of the vector multiplet \cite{KLRT-M1}.
It should also be mentioned that the Bianchi identity \eqref{2.33} implies that the real isotriplet
\bea
\S^{ij}:=\frac{1}{ 4}D^{ij}W
~, \qquad 
\overline{\S^{ij}}:= \frac{1}{4}\DB_{ij}\bar{W} = \ve_{ik} \ve_{jl}\, \S^{kl}
\label{Sigma_0-0-0}
\eea
 satisfies the constraints
\bea
D_\a^{(i}\S^{jk)}=\DB_\ad^{(i}\S^{jk)}=0
\label{ana-S_0}
\eea
which are  characteristic of the $\cN=2$ linear multiplet.
The above relations are completely general in the sense that they hold for any super-Weyl 
parameter $\s$ which conformally relates  AdS$^{4|8}$ to flat superspace.

Now, we turn to deriving 
an IO(2,1)-invariant solution of the equations  (\ref{Y=0}) and (\ref{Finite-S-2}) 
which leads to the Poincar\'e coordinates for AdS$_4$.
In complete analogy with the  $\cN=1$ case described in the previous subsection, 
we introduce a 3D foliation of the space-time coordinates. 
Then, the  4D $\cN=2$ flat covariant derivatives take the form
\bea
&&
D_\a^i={\pa\over\pa\q_i^\a}
+\ri(\g^{\hat m})_{\a\b}\qb^{\b i}\pa_{\hat m}
-\qb_\a^{i}\pa_z
~,~~~
\DB_{\a i}=-{\pa\over\pa\qb^{\a i}}
-\ri(\g^{\hat m})_{\a\b}\q^\b_i\pa_{\hat m}
-\q_{\a i}\pa_z~.~~~
\label{3D-N=2-1}
\eea
The anti-commutation relations for the covariant derivatives turn into
\bea
&&\{D_\a^i,D_\b^{j}\}=
\{\DB_{\a i},\DB_{\b j}\}=0
~,\quad
\{D_\a^i,\DB_{\b j}\}=
-2\ri\d^i_j(\g^{\hat m})_{\a\b}\pa_{\hat m}
+2\d^i_j\ve_{\a\b}\pa_z~.
\label{3D-N=2-2}
\eea
These relations correspond to the 3D $\cN=4$ super-Poincar\'e algebra with a real central charge. 

We make the most general IO(2,1)-invariant  ansatz for $\s$ in (\ref{2.23})
\bea
\re^{\s}:=A( z_L)+\q_{ij}B^{ij} (z_L) + \q^{ij} \q_{ij} C(z_L)
~,\qquad \q_{ij}:=\q^\a_i\q_{\a j}
~,
\label{ansatz}
\eea
where $z_L :=z-\q^{\a}_k\qb_{\a}^k$ is the chiral completion of the space coordinate $z$,
$\DB_{\a i}z_L=0$. The unknown functions $A(z_L)$, $B^{ij} (z_L) $ and $C(z_L)$ 
can  be determined by requiring the equations \eqref{G=0}, \eqref{Y=0} 
and \eqref{Finite-S-2} to hold. The solution to these equations (compare with the five-dimensional case
\cite{KT-M_5D_conf-flat}) is
\bea
\re^{\s} = s\,z_L+ s^{ij}\q_{ij}~,
\label{2.45}
\eea
where $s^{ij}$ is a constant real isotriplet,
\bea
s^{ij}=\bar{s}^{ij}~, \qquad 
\qquad s^2:=\hf s^{ij}s_{ij}~.
\eea

Evaluating the torsion superfield $\cS^{ij}$  gives
\bea
\cS^{ij}=
{1\over 4}\re^{\s+\sba}(D^{ij}\re^{-\s})=
{1\over 4}\re^{\s+\sba}(\DB^{ij}\re^{-\sba})
= s^{ij} + O(\q) ~.
\label{bmS-1}
\eea
It can be checked that $\cS^{ij}$ is real, $\cS^{ij}=\bar{\cS}^{ij}$.
It is covariantly constant by construction. 
It follows from (\ref{bmS-1}) that 
\be 
\cS^2 := \hf\cS^{ij}\cS_{ij} = \hf s^{ij}s_{ij}= s^2~.
\ee
 This completes the derivation of the  conformally flat representation for AdS$^{4|8}$.
In what follows we do not distinguish between $\cS$ and $s$.

It should be remarked that ${\cS}^{ij}$ in (\ref{bmS-1}) is  covariantly 
constant, $\cD_\cA {\cS}^{kl} = 0$,  but not constant. 
The point is that the  conformally flat representation 
for the covariant derivatives, eqs. (\ref{Finite-D})  and (\ref{Finite-Db}), 
is given  in terms of a linear combination of all the generators 
of the group SU(2)$_R$. 
As discussed in subsection \ref{N=2AdSgeneral},  the SU(2) gauge freedom 
can be used to bring the SU(2) connection of  AdS$^{4|8}$
to the form $ \F_{\cA}{}^{ i j } =  \F_{\cA}\cS^{ij}$, for some one-form $\F_\cA$. 
In such a gauge, $\cS^{ij}$ becomes constant, $\cS^{ij} =s^{ij}={\rm const}$.

For our analysis in the next sections, it is important to know explicit expressions for the
components of $\S^{ij}$, eq. \eqref{Sigma_0-0-0}. 
To derive them, it proves useful to replace $\S^{ij}$ with an index-free object $\S^{(2)}(v)$
obtained by contracting the SU(2) indices of $\S^{ij}$ with an auxiliary  bosonic isotwistor 
$v^i \in {\mathbb C}^2 \setminus \{0\}$: 
\bea
\S^{(2)}(v):=v_iv_j\S^{ij}~.
\label{def-Sigma_0}
\eea
The significance of this definition will become clear later in the paper when the projective superspace
techniques will play a central role. For now, let us focus on presenting some technical  
results concerning $\S^{(2)}$.

In terms of $\S^{(2)}$, the equations (\ref{ana-S_0}) take the form of analyticity constraints
\bea
D_\a^{(1)}\S^{(2)}=\DB_\a^{(1)}\S^{(2)}=0~,
\eea
where 
we have introduced
\bea
D_\a^{(1)}:=v_iD_\a^i~,~~~
\DB_\a^{(1)}:=v_i\DB_\a^i
~.
\label{Sigma_0-0}
\eea
It follows from the anti-commutation relations (\ref{2.333}) that the fermionic operators
$D_\a^{(1)}$ and $\DB_\a^{(1)}$ strictly anti-commute with each other, 
\bea
\{ D_\a^{(1)}, D_\b^{(1)} \} = \{ D_\a^{(1)}, \DB_\b^{(1)} \} = \{ \DB_\a^{(1)}, \DB_\b^{(1)}\}=0~.
\label{2.48}
\eea
These properties allow us to define flat projective supermultiplets 
(compare with the definition given in subsection \ref{subsection1.3}). 

We introduce a new basis for the superspace variables 
$z$, $\q^\a_i $ and $\bar \q_i^\a$ 
(which is analogous to the analytic basis
in harmonic superspace \cite{GIKOS}) defined as follows:
\bsubeq
\bea
&&\q_\a^{(1)}:=v_i\q_\a^i
~,~~~
\q_\a^{(-1)}:=\frac{1}{(v,u)}u_i\q_\a^i~,\qquad
\qb_\a^{(1)}:=v_i\qb_\a^i~,~~~\qb_\a^{(-1)}:=\frac{1}{(v,u)}u_i\qb_\a^i~,~~~~~~~
\\
&&
z_A:=z+\q^{\a (1)}\qb_{\a}^{(-1)}+\q^{\a (-1)}\qb_\a^{(1)}~.
\eea
\esubeq
Here we have introduced a second isotwistor  
$u_i \in {\mathbb C}^2 \setminus \{0\}$ which is subject to the condition that
$(v,u):=v^iu_i=\ve_{ij}v^i u^j\ne0 $,
but  otherwise is completely arbitrary.
The coordinates $Z_A :=(z_A,\,\q_\a^{(1)},\,\qb_\a^{(1)})$ are annihilated by the derivatives
$D_\a^{(1)}$ and $\DB_\a^{(1)}$,  
\bea
D_\a^{(1)}Z_A=\DB_\a^{(1)}Z_A=0
~.
\label{an-basis}
\eea
In terms of  the variables $Z_A$, the superfield
 $\S^{(2)}$ can be shown to have the form:
\bea
\S^{(2)}&=&
s^{-2}(z_A)^{-2}s^{(2)}
-2s^{-1}(z_A)^{-3}\(\q^{(2)}+\qb^{(2)}\)
+4s^{-2}(z_A)^{-3}s^{(0)}\q^{\a (1)}\qb_{\a}^{(1)}
\non
\\
&&
+6 s^{-2}(z_A)^{-4} s^{(-2)}\q^{(2)}\qb^{(2)}
~,
\label{Sigma_0^{++}-analytic}
\eea
where $\q^{(2)}=\q^{\a(1)}\q_\a^{(1)}$, $\qb^{(2)}=\qb_\a^{(1)}\qb^{\a(1)}$ and
\bea
{ s}^{(2)}:= v_i v_j{ s}^{ij} 
~,\qquad
{ s}^{(0)}:= \frac{v_i u_j}{(v,u)}{ s}^{ij}
~,\qquad
{ s}^{(-2)}:= \frac{u_i u_j}{(v,u)^2}{ s}^{ij}  ~.
\eea
The expression  (\ref{Sigma_0^{++}-analytic}) makes manifest the fact 
that $\S^{(2)}$ satisfies the constraints $D_\a^{(1)}\S^{(2)}=\DB_\ad^{(1)}\S^{(2)}=0$.
Moreover,
despite the fact that the separate contributions
in the right-hand side of (\ref{Sigma_0^{++}-analytic}) explicitly depend
on $u_i$, it is easy to prove that $\S^{(2)}$ is independent of $u_i$,
\be
\frac{\pa }{\pa u_i} \S^{(2)}=0~,
\ee
as it should be in accordance with (\ref{def-Sigma_0}).

\section{Off-shell supersymmetric theories in AdS$^{\bf 4|8}$ using the 3D foliation}\label{Section_Offshell}
\setcounter{equation}{0}

As discussed in subsection \ref{subsection1.3}, general $\cN=2$ supersymmetric theories 
in $\rm AdS_4$ can be formulated in terms of covariant projective supermultiplets living in
the projective superspace AdS$^{4|8} \times {\mathbb C}P^1$. Here we would like to reformulate, 
following the five-dimensional analysis of \cite{KT-M_5D_conf-flat},  
the dynamics of these theories in terms of flat projective supermultiplets by using the conformally flat 
realization of AdS$^{ 4|8}  $ introduced in the previous section.

\subsection{Supersymmetric action}

The manifestly supersymmetric action principle in AdS$^{4|8}  \times {\mathbb C}P^1$ 
is  given by \eqref{InvarAc1}. Our goal is to bring this action to a 
flat superspace form which is based on the 
use of the conformally flat realization of AdS$^{4|8}  $ 
 introduced in the previous section.  In order to achieve this, the key technical result 
is the super-Weyl transformation law of covariant projective supermultiplets \cite{KLRT-M1}. 
Under the super-Weyl transformation given by eqs. (\ref{2.23}) and \eqref{2.45}, 
which relates the AdS covariant derivatives to the flat ones, 
the AdS projective supermultiplet  can be represented as follows:
\bea
\cQ^{(n)} (v) = \re^{\hf n(\s +\bar \s)} Q^{(n)}(v)~, 
\label{3.6}
\eea
where $Q^{(n)} (v)$ is a projective multiplet in flat superspace, 
\bea
D^{(1)}_{\a} Q^{(n)}  = {\bar D}^{(1)}_{\a} Q^{(n)}  =0~,
\label{3.7}
\eea
with the derivatives $D^{(1)}_{\a} $ and $ {\bar D}^{(1)}_{\a}$ defined in \eqref{Sigma_0-0}.
Using eq. \eqref{3.6}, the action  \eqref{InvarAc1}
can be transformed to a form that involves integration over four Grassmann directions
\cite{KT-M-4D-conf-flat}, instead of the eight-dimensional Grassmann variables in  \eqref{InvarAc1}.

It follows from eqs. (\ref{2.23}),  \eqref{Finite-S-2}, \eqref{Sigma_0-0-0}
 and \eqref{3.6} that
\bea
\cL^{(2)}&=& \re^{\s+\sba}L^{(2)}~,
\qquad
\cS^{(2)}=\re^{\s+\sba}\S^{(2)}~, \qquad {\mathbb E}=1~.
\eea
Then the  action (\ref{InvarAc1}) can be rewritten as
\bea
S&=&
\frac{1}{2\pi} \oint_C (v, \rd v)
\int \rd^4 x \, {\rm d}^4\q \, {\rm d}^4{\bar \q}\, \frac{\re^{-\s-\sba}L^{(2)}}{(\S^{(2)})^2} \non \\
&=&
\frac{1}{2\pi} \oint_C (v, \rd v)
\int \rd^4 x \,D^{(-4)}D^{(4)}\,\frac{\re^{-\s-\sba}L^{(2)}}{(\S^{(2)})^2}\Big|_{\q=0}~,
\label{InvarAc2}
\eea
where we have defined
\begin{subequations}
\bea
D^{(4)}&:=&\frac{1}{1 6} D^{(2)}\DB^{(2)}~, \qquad D^{(2)}:=v_iv_j D^{ij}~, 
\quad \bar D^{(2)}:=v_iv_j \bar D^{ij}~;\\
D^{(-4)}&:=&
\frac{1}{16} \frac{u_iu_ju_ku_l}{(v,u)^4}D^{ij}\DB^{kl}~.
\eea
\end{subequations}
Here again $u_i$ is  an auxiliary  isotwistor which is only subject 
to the condition $(v,u) := v^{i}u_i \neq 0$, and is otherwise completely arbitrary.
Since $L^{(2)}$ and $\S^{(2)}$ are flat projective supermultiplets, 
it remains to use the identity
\bea
D^{(4)}\re^{-\s-\sba}=\Big( \frac{1}{ 4}D^{(2)}W\Big)\Big({1\over 4}\DB^{(2)}\bar{W}\Big)
=(\S^{(2)})^2
\eea
to end up with
\bea
S&=&
\frac{1}{2\pi} \oint_C (v, \rd v)
\int \rd^4 x \,D^{(-4)}\,L^{(2)}\Big|_{\q=0}~.
\label{InvarAc3}
\eea

The new form of the supersymmetric action obtained, eq. \eqref {InvarAc3}, 
is much simpler than the original one, eq.  \eqref{InvarAc1}.
However, this  is not our final representation for the action to work with. 
There are reasons to look for further simplifications. 
The point is that we are mostly interested in the off-shell $\s$-model \eqref{1.25}. 
The flat-superspace version of the corresponding Lagrangian is 
\bea
L^{(2)} = \frac{1}{2s} \S^{(2)}\,
{ K}({ \U}, \breve{ \U})~,
\label{model}
\eea
with $\S^{(2)} $ given by \eqref{Sigma_0^{++}-analytic}. Upon projection of this Lagrangian 
to the $\cN=2$ subspace of the 3D $\cN=4$ superspace, $L^{(2)}|_{\q_{\2} = \bar \q^{2} =0}$, 
it turns out that $\S^{(2)}$ has a nontrivial dependence on the isotwistor $v^i$, and this 
complicates the evaluation of the contour integral  \eqref {InvarAc3}. 
This problem can be avoided by choosing an alternative $\cN=2$ subspace 
of the 3D $\cN=4$ superspace.

\subsection{New Grassmann coordinates for 3D $\cN=4$ central charge superspace}\label{section5.2}

In performing reduction from four to three dimensions, 
the manifest 4D Lorentz symmetry gets broken down to the 3D one, and  
the 4D $\cN=2$ flat spinor derivatives 
(\ref{2.30}) turn into 
the 3D $\cN=4$ covariant derivatives 
defined in  (\ref{3D-N=2-1}).
Since the difference between dotted and undotted indices disappears in three dimensions, 
we are in a position to introduce a new basis for the spinor covariant derivatives defined as
\begin{subequations}\label{3.17bold}
\bea
{\bm D}_\a^i := \frac{1}{\sqrt{2}} (D^i_\a +\l \bar D^i_\a)~, \qquad \l \in \mathbb C~.
\eea
By complex conjugation we find
\bea
\bar{\bm D}_{\a i} = \frac{1}{\sqrt{2}} (\bar D_{\a i}  - \bar \l  D_{\a i} )
~, 
\qquad 
\overline{\big( {\bm  D}_\a^i \J \big)} = (-1)^{\ve (\J)} \bar {\bm D}_{\a i} \bar \J
~,
\eea
\end{subequations}
where $\ve (\J)$ denotes the Grassmann parity of a complex superfield $\J$.
Using (\ref{3D-N=2-2})
 we derive the anti-commutation relations obeyed by the operators introduced:
\begin{subequations}
\bea
\big\{ {\bm D}_\a^i , {\bm D}_\b^j \big\} &=& - 2\l \ve^{ij}\ve_{\a\b} \pa_z ~, 
\qquad
\big\{ \bar{\bm D}_{\a i} , \bar{\bm D}_{\b j} \big\} = - 2\bar{\l} \ve_{ij}\ve_{\a\b} \pa_z 
\label{new_deriv-1}\\
\big\{ {\bm D}_\a^i , \bar{\bm D}_{\b j} \big\} &=& -\ri (1+|\l|^2 ) \d^i_j \pa_{\a\b} 
+(1-|\l|^2) \d^i_j \ve_{\a\b} \pa_z
\label{new_deriv-2}
~.
\eea
\end{subequations}
Making the choice
\be
\l \bar \l =1
\ee
simplifies the above algebra 
\begin{subequations}
\bea
&\big\{ {\bm D}_\a^i , {\bm D}_\b^j \big\} = - 2\l \ve^{ij}\ve_{\a\b} \pa_z ~,\qquad
\big\{ \bar{\bm D}_{\a i} , \bar{\bm D}_{\b j} \big\} = - 2\bar{\l} \ve_{ij}\ve_{\a\b} \pa_z 
~,
 \\
&\big\{ {\bm D}_\a^i , \bar{\bm D}_{\b j} \big\} = - 2 \ri  \d^i_j \pa_{\a\b} 
~.
\eea
\end{subequations}
An explicit realization of the covariant derivatives introduced is as follows:
\begin{subequations}
\bea
{\bm D}_\a^i&=&\phantom{-}
\frac{\pa}{\pa{\bm \q}^\a_i}
+\ri(\g^{\hat m})_{\a\b}{\bm\qb}^{\b i}\pa_{\hat m}
-\l{\bm\q}_\a^i\pa_z
~,~~~ \\
{\bm \DB}_{\a i}&=&
-\frac{\pa}{\pa{\bm \qb}^{\a i}}
-\ri(\g^{\hat m})_{\a\b}{\bm\q}^{\b}_{i}\pa_{\hat m}
+\bar\l{\bm\qb}_{\a i}\pa_z
~,
\eea
\end{subequations}
where we have defined
\bea
{\bm\q}^\a_i&:=&\frac{1}{\sqrt{2}}\(\q^\a_i+\bar\l\qb^\a_i\)~,\qquad
{\bm\qb}_\a^i:=\frac{1}{\sqrt{2}}\(\qb_\a^i-\l\q_\a^i\) ~.
\label{3.23}
\eea

Let $Q^{(n)} (v)$ be a weight-$n$ projective multiplet, 
\bea
D^{(1)}_{\a} Q^{(n)}  = {\bar D}^{(1)}_{\a} Q^{(n)}  =0~,
\eea 
with respect to the covariant derivatives 
(\ref{Sigma_0-0})  obeying the anti-commutation relations \eqref{2.48}.
We can extend the definition (\ref{Sigma_0-0}) to the case of the covariant derivatives
\eqref{3.17bold} by defining 
\bea
{\bm D}^{(1)}_\a := {\bm D}_{\a}^i v_i ~, \qquad \bar {\bm D} ^{(1)}_\a := \bar {\bm D}_\a^i v_i  ~.
\eea
It follows from the definition of ${\bm D}_{\a}^i $ and $\bar {\bm D}_{\a}^i$ that 
\bea
{\bm D}^{(1)}_\a  = \frac{1}{\sqrt{2} } ( D^{(1)}_\a  + \l \bar D^{(1)}_\a ) ~, \qquad 
\bar{\bm D}^{(1)}_{\a } = \frac{1}{\sqrt{2}} (\bar D^{(1)}_{\a }  - \bar \l  D^{(1)}_{\a } )~.
\eea
In other words, the operators ${\bm D}^{(1)}_\a $ and $\bar{\bm D}^{(1)}_{\a } $
are related to ${ D}^{(1)}_\a $ and $\bar{ D}^{(1)}_{\a } $ by a linear unimodular  transformation.
Therefore, if $Q^{(n)}$ is a projective supermultiplet 
with respect to the covariant derivatives 
(\ref{Sigma_0-0}),
it is also a projective supermultiplet 
with respect to  the spinor derivatives ${\bm D}^{(1)}_\a $ and $ \bar {\bm D} ^{(1)}_\a$,  
\bea
{\bm D}^{(1)}_\a Q^{(n)} = \bar {\bm D}^{(1)}_\a Q^{(n)} =0~,
\eea
and vice versa. 
We conclude that the two sets of operators, $({ D}^{(1)}_\a , \bar { D} ^{(1)}_\a)$ and
$({\bm D}^{(1)}_\a , \bar {\bm D} ^{(1)}_\a)$,  are completely equivalent to use 
when dealing with the projective supermultiplets.

There are two technical reasons why the use of covariant derivatives \eqref{3.17bold},
and the associated Grassmann variables  (\ref{3.23}),
are advantageous  in the context of supersymmetric $\s$-models in projective superspace.  
First of all, it is  this realization of 3D $\cN=4$ central charge superspace which
provides the simplest embedding of 3D $\cN=2$ superspace {\it without central charge}. 
Given an arbitrary $\cN=4$ superfield $U({\bm\q}^\a_i, {\bm\q}^i_\a)$, we define its $\cN=2$ projection 
as follows
\bea
U|:= U\Big|_{{\bm\q}_\2={\bm\q}^\2=0}~.
\eea
Defining also the Grassmann variables of $\cN=2$ superspace,
\bea
{\bm\q}^\a:={\bm\q}^\a_\1 = \frac{1}{\sqrt{2}}(\q^\a_\1-\bar{\l}\qb^{\a\2})~, \qquad
\bar{\bm\q }_\a:=\bar{\bm\q}^\1_\a=\frac{1}{\sqrt{2}}(\qb_\a^\1-\l\q_{\a\2})
~,
\eea 
  it is easy to see that 
\begin{subequations}
\bea
{\bm D}_{\a}&:=&{\bm D}_{\a}^{\1}\Big|_{{\bm\q}_\2=0} 
= \phantom{-}
\frac{\pa}{\pa{\bm \q}^\a}
+\ri(\g^{\hat m})_{\a\b}{\bm\qb}^{\b }\pa_{\hat m}
~,~~~ \\
{\bm \DB}_{\a}&:=&{\bm \DB}_{\a \1}\Big|_{{\bm\q}_\2=0}
=-\frac{\pa}{\pa{\bm \qb}^{\a }}
 -\ri(\g^{\hat m})_{\a\b}{\bm\q}^{\b}_{}\pa_{\hat m}
~.
\eea
\end{subequations}
These operators have no dependence on $z$, and 
obey  the anti-commutation relations
\bea
&
\{{\bm D}_\a,{\bm D}_\b\}=\{{\bm \DB}_\a,{\bm \DB}_\b\}=0~,\qquad
\{{\bm D}_\a,{\bm \DB}_\b\}=-2\ri(\g^{\hat m})_{\a\b}\pa_{\hat m}
\eea
corresponding to  3D $\cN=2$ Poincar\'e supersymmetry  without central charge.
Note that the similar $\q^\2=\qb_\2=0$ reduction of the 
derivatives $(D_\a^\1,\DB_{\a \1})$,  eq. (\ref{3D-N=2-1}), 
does not decouple the $z$-dependence, leading to the 3D $\cN=2$
superspace with central charge.

The second reason to use the covariant derivatives \eqref{3.17bold}
is that this realization turns out to provide, for a special value of $\l$, 
the simplest expression for the $\cN=2$ projection of $\S^{[2]}$.
Let us look again at $\S^{(2)}(v)$. 
If we restrict our attention to the open domain of ${\mathbb C}P^1$ where $v^\1\ne 0$, 
which is the north chart of $\mathbb{C}P^{1}$,
we can write 
\bea
\S^{(2)}(v) =\ri(v^{\1})^2\z \S^{[2]}(\z)~, \qquad
 \S^{[2]}(\z)= - \frac{\ri}{ \z}\,\S^{ij}\z_i\z_j
~,
\eea
where 
\bea
v^{i}=v^{\1}(1,\z)=v^{\1}\z^i~,~~~\z^i=(1,\z)~,~~~\z_i=(-\z,1)
~.
\eea
Now, we consider the case in which only one of the three independent
components of $\bms^{ij}$ is nonzero
\bea\label{eq_sijvalues}
s^{\1\1}= s^{\2\2}=0~,\quad s^{\1\2}=\a s~,\qquad \a=\pm \ri
~,
\eea
which is exactly the choice \eqref{1.3a}.
Note that  $s$ is  real and positive.
A straightforward computation of the 3D $\cN=2$ reduction of $\S^{[2]}(\z)$ gives
\bea
\S^{[2]}(\z)\big{|}&=&
2\ri \a s^{-1}z^{-2}
\non\\
&&
-\ri s^{-1}z^{-3}\Big(
2(\l+\bar{\l})
\bmq^\a\bmqb_\a
+\frac{1}{\z}(\l^2-1+2\a \l)\bmq^2
+\z (\bar{\l}^2-1-2\a \bar{\l})\bmqb^2
\Big)
\non\\
&&
-3\ri s^{-1}z^{-4} (\l-\bar{\l}+2\a )\bmq^2\bmqb^2~. \label{5.27}
\eea
So far, the parameter $\l$ has been restricted by $|\l|=1$. 
We now observe that choosing
\bea
\l=-\a=\mp\,\ri~,
\label{good-choice}
\eea
allows us to  eliminate all the $\bmq$ and $\z$ dependence of $\S_0^
{(2)}\big{|}$
(in complete analogy to  the 5D case \cite{KT-M_5D_conf-flat}).
The condition (\ref{good-choice}) completely fixes the 3D $\cN=2$
reduction to be
used below.

Note that the choice of sign in (\ref{good-choice}) is conventional.
In fact one can flip the sign by changing everywhere the $\bmq$
coordinates to $\bmqb$ and vice versa.
Throughout the main body of the paper we will use $\l=-\a=\ri$,
which simplifies \eqref{5.27} to
\bea
\S^{[2]}\big{|}=
2s^{-1}z^{-2}
~.
\label{S_0-reduced}
\eea

%%%%%%%%%%%%%%%%%%%%%%%%%%%%%%%%%%%%%%%%%%%%%%%%

\subsection{The supersymmetric action revisited}

We now return to the supersymmetric action  \eqref {InvarAc3}.
It is given in terms of the flat covariant derivatives 
$(D^i_\a,\DB_\a^i)$. Above we have introduced  
the new basis for the spinor covariant derivatives, 
$({\bm D}^i_\a,{\bm \DB}_\a^i)$.
We have also shown 
that any projective supermultiplet 
with respect to  $(D^i_\a,\DB_\a^i)$ 
is also projective with respect to $({\bm D}^i_\a,{\bm \DB}_\a^i)$.
In terms of the new spinor derivatives, the action can be seen to be
\bea
S&=&
\frac{1}{2\pi} \oint_C (v, \rd v)
\int \rd^3 x\,\rd z \,{\bm D}^{(-4)}\,L^{(2)}\Big|_{\q=0}~,
\eea
where
\be
{\bm D}^{(-4)}:={1\over 16(v,u)^4}u_iu_ju_ku_l{\bm D}^{ij}{\bm \DB}^{kl}
~.
\ee

Without loss of generality, we can assume the north pole of ${\mathbb C}P^1$, 
$v^i \propto (0,1)$,  
to lie  outside of the integration contour in   \eqref {InvarAc3}, 
and hence  we introduce the complex inhomogeneous coordinate $\z$ for ${\mathbb C}P^1$ 
defined by $v^{i}=v^{\1}(1,\z)$.
Since the action   \eqref {InvarAc3} is independent of $u_i$, we can also choose
 $u_i=(1,0)$. In addition, it is standard to represent $L^{(2)}$ in the form
\be
L^{(2)}(v) =  {\rm i}\, v^{\1} v^{\2}\,
L(\z) =  {\rm i} \big( v^{\1} \big)^2 \z\, L(\z)~.
\ee
By using the fact that $L^{(2)}$ enjoys the constraints
$\z_i {\bm D}^i_\a L=\z_i {\bar {\bm D}}^i_\a L=0$, we can finally rewrite  
$S$ as an integral over the 3D $\cN=2$  superspace
followed by an integral over ${\mathbb R}_+$
\bea
S=
\int \rd z 
 \oint_C \frac{\rd \z}{2\p \ri \z}
\int \rd^3 x \,{\rm d}^2\bmq\,{\rm d}^2{\bmqb}\,  L(\z) \Big|_{{\bm\q}_\2={\bm\qb}^\2=0}~.
\label{InvarAc5}
\eea
This action is manifestly invariant under the 3D $\cN=2$ super-Poicar\'e group without central charge. 
By construction, the action  is in fact invariant under the larger 4D $\cN=2$ AdS supergroup $\rm OSp(2|4)$.

Using our earlier result \eqref{S_0-reduced}, 
for the supersymmetric $\s$-model  \eqref{model} we end up with the following action:
\bea
S=
\int \frac {\rd z}{(sz)^2}
\oint_C \frac{\rd \z}{ 2\p \ri \z}
\int \rd^3 x\, {\rm d}^2\bmq\,{\rm d}^2\bmqb\,  
{ K}({ \U}^I, \breve{ \U}^{\bar I} )\Big|_{{\bm\q}_\2={\bm\qb}^\2=0}~.
\label{3DProjAction}
\eea
The arctic $\U^I (\z)$ and  antarctic $\breve{\U}^{\bar I} (\z)$ dynamical variables  
are generated by an infinite set of ordinary 3D $\cN=2$ superfields parametrically depending on $z$:
\begin{subequations}
\bea
 \U^I (\z) &=& \sum_{n=0}^{\infty}  \z^n  \, \U^I_n= 
\F^I +  \z \, \S^I + O(\z^2) ~, \\
\breve{\U}^{\bar I} (\z) &=& \sum_{n=0}^{\infty}  (-\z)^{-n} \, {\bar
\U}^{\bar I}_n
=\bar \F^{\bar I} -\frac{1}{\z} \bar \S^{\bar I} +O(\z^{-2}) ~.
\eea
\end{subequations} 
Here the physical superfields
$\F^I:=\U^I_0 $  and  $\S^I:=\U^I_1 $  are 3D $\cN=2$ chiral and complex linear respectively, 
 \bea
 \bar{\bm D}_\a \F^I =0~, \qquad 
 \bar{\bm D}^2 \S^I =0~,
 \eea 
while the remaining components, $\U^I_2, \U^I_3, \dots, $ are complex unconstrained  
3D $\cN=2$ superfields.  
The crucial point is that, 
except for the overall integral $ \int {\rd z}\,(sz)^{-2}$,
the action \eqref{3DProjAction} looks exactly like 
a flat 3D $\cN=4$ $\sigma$-model, see e.g. \cite{KPT-MvU}.

\newcommand{\newomega}{{\hat\omega}}

\section{$\sigma$-models from projective superspace: 3D foliated frame}\label{section6}
Our next task is to reformulate the AdS supersymmetric $\s$-model
\eqref{3DProjAction} in terms of 3D $\cN=2$ chiral superfields by formally eliminating 
the auxiliary superfields $\U^I_2, \U^I_3, \dots, $ and performing an appropriate duality transformation.  
In this and subsequent sections, we will
be dealing only with 3D $\cN=2$ superfields (parametrically depending on 
the forth space variable $z$), so we 
will always drop
the explicit bar-projection.

The action \eqref{3DProjAction}  possesses off-shell 4D $\cN=2$ AdS supersymmetry, $\rm OSp(2|4)$,  
by construction.
Part of this supersymmetry is manifest as 3D $\cN=2$ Poincar\'e supersymmetry.
To exhibit explicitly the 
remaining symmetries, we need to know
the structure of the Killing vector fields in this foliated superspace. Since
the generic form of the Killing vector fields in $\rm AdS^{4|8}$ is known \cite{KT-M-4D-conf-flat}, 
it is a straightforward task to specialize them to the specific frame
we have chosen. We leave the details to appendix \ref{app_KV} and give only
the result here when projected to $\bm\q_\2 = \bar{\bm\q}^\2 = 0$.

\subsection{Extended supersymmetry and Killing vector fields}
Within the 3D foliated frame of AdS, a projective supermultiplet $\cQ (\z) $ of weight zero transforms as
\begin{align}\label{deltaQ}
\delta \cQ  = - \left(\xi^\ha \pa_\ha 
	+ \xi^z \pa_z
	+ \xi^\a \bm D_\a 
	+ \bar{\xi}_\a \bar{\bm D}^\a 
	+ \zeta \rho^\a {\bm D}_\a 
	- \frac{1}{\zeta} \bar\rho_\a \bar{\bm D}^\a 
	- 2\ri \Lambda \zeta \partial_\zeta \right)\cQ~.
\end{align}
The parameters $\xi^\ha$, $\xi^z$, etc., can be decomposed into those
associated with the manifest 3D $\cN=2$ Poincar\'e supersymmetry
\begin{subequations}\label{3DKilling}
\begin{align} \label{KV1}
\xi^\ha &= p^\ha + \omega^\ha{}_\hb x^\hb
	- \frac{\ri}{2} \ve^{\ha \hb \hc} \omega_{\hb \hc} \bm\q^\a \bar{\bm\q}_\a
	- 2\ri\, \bar{\e}_\a (\g^\ha)^{\a\b} \bm\theta_\b
	- 2\ri\, {\e}^\a (\g^\ha)_{\a\b} \bar{\bm\theta}^\b~, \\
\xi^\a &= \e^\a
	- \frac{1}{4} \omega^{\hb \hc} \ve_{\hb\hc\hd} \bm\q^\b (\g^\hd)_\b{}^\a~, \\
\xi^z &= \rho^\a = \bar\rho_\a = \Lambda = 0~,
\end{align}
\end{subequations}
and those associated with the remaining isometries 
\begin{subequations}\label{3DKillingSecond}
\begin{align} \label{KV2}
\xi^\ha &=
        2 r x^\ha
        + 2\ri \Lambda_\1{}^\1 {\bm\theta}^\a (\g^\ha)_{\a\b} \bar{\bm\theta} 
^\b
        - 2 x^\hb k_\hb x^\ha + x^2 k^\ha - 2\ri \ve^{\ha \hb \hc} k_\hb x_ 
\hc \bm\theta^\a \bar{\bm\theta}_\a
        + z^2 k^\ha
        \eol & \quad
        -\hf \bm\q^2 \bar{\bm\theta}^2 k^\ha
        + 2 x^\ha \eta^\a \bm\theta_\a
        + 2 x^\ha \bar{\eta}_\a \bar{\bm\theta}^\a
        - 2 x^\hb \ve^\ha{}_{\hb \hc} \eta^\a (\g^\hc)_\a{}^\b \bm\theta_\b
        + 2 x^\hb \ve^\ha{}_{\hb \hc} \bar{\eta}_\a (\g^\hc)^\a{}_\b \bar{\bm 
\theta}^\b
        \eol & \quad
        -\ri \,\eta^\a  (\g^\ha)_{\a\b} \bar{\bm\q}^\b\bm\q^2
        -\ri \,\bar{\eta}^\a   (\g^\ha)_{\a\b}\bm\q^\b \bar{\bm\q}^2 ~,\\
\xi^\a &=
        - \Lambda_\1{}^\1 \bm\q^\a
        - k_\hb x^\hb \bm\q^\a
        - k^\hb x^\hc \ve_{\hb \hc \hd} \bm\q^\b (\g^\hd)_\b{}^\a
        -\frac{\ri}{2}k_\hb \bm\q^2 \bar {\bm \q}^\b (\g^\hb)_{\b}{}^\a
        - \ri x^\hb \bar{\bm \eta}_\b (\g_\hb)^{\b \a}
        \eol & \quad
        + \bm\q^\b \bar {\bm \q}_\b \bar{\eta}^\a
        +  \eta^\a \bm\q^2
        + 2 \bm\q^\b \bar{\eta}_\b \bar{\bm\q}^\a~, \\
\xi^z &= 2 r z - 2 z x^\hb k_\hb + 2 z \eta^\a \bm\theta_\a + 2z \bar 
{\eta}_\a \bar{\bm\theta}^\a~, \\
\rho^\a &= \ri z \eta^\a + z k_\hb \bar{\bm\q}^\b (\g^\hb)_{\b}{}^ 
\a~, \\
\Lambda &= -\ri \Lambda_\1{}^\1 - 2 \ri \eta^\a\bm\q_\a + 2 \ri \bar 
{\eta}_\a \bar{\bm\q}^\a
        - 2 k_\ha \bm\q^\a (\g^\ha)_{\a\b} \bar{\bm\q}^\b ~.
\end{align}
\end{subequations}
A general isometry is the combination of the two given. From now on, we will
always consider the combination.

Several observations should be made. The isometry transformations  \eqref{3DKilling}
can be easily identified with those of 3D $\cN=2$ Minkowski superspace: the
constant parameters $p^\ha$, $\omega_{\ha\hb}$, and $\e^\a$ correspond
exactly to the 3D translations, Lorentz rotations, and supersymmetry.
In contrast, the isometries \eqref{3DKillingSecond} are more
complicated. The parameter $r$ can be identified with translations in
the $z$ direction, $\eta^\a$ parametrizes the additional supersymmetry,
and $\Lambda_{\1}{}^\1$ is plainly the SO(2) $R$-symmetry.
The parameter $k_\ha$ may be 
interpreted as 
a remnant of the four-dimensional Lorentz symmetry.

An alternative interpretation of these isometries is suggested by the well-known fact that 
the 4D $\cN=2$ AdS supergroup and the 3D
$\cN=2$ superconformal group
are isomorphic to the supergroup $\rm OSp(2|4)$. From the 3D superconformal point of view,  
the constant parameters $r$, $\eta^\a$, $\L_\1{}^\1$, and $k_\ha$ are associated
with the 3D dilatations, $S$-supersymmetry, $R$-symmetry, and
special conformal transformations, respectively. The above set of
transformations is a certain realization of the 3D $\cN=2$ superconformal
group. Observe that
\begin{gather}
\bar{\bm D}^\b \xi^\a = 0~, \qquad
\bar{\bm D}^\b \xi^\ha = 2\ri \xi_\g (\g^\ha)^{\g \b}~, \qquad
\bar{\bm D}^{(\g} \xi^{\b \a)} = 0~,
\end{gather}
which imply that 
 $\xi^\ha$ and $\xi^\a$ are respectively  the vector and spinor components of 
a 3D $\cN=2$ \emph{superconformal} Killing vector field \cite{KPT-MvU}.
In fact, except for the $z^2 k^\ha$ term in $\xi^\ha$, the general
expression for $\xi^\ha$ is the most general solution of these
constraints in 3D $\cN=2$ superspace. From the 3D point of view, the $z^2 k^\ha$ term is just a constant 
three-vector which can be combined with $ p^\ha$ into a constant parameter of translations. 

The SO(2) superfield parameter $\Lambda$ in \eqref{deltaQ} obeys
\begin{align}
\Lambda = \frac{\ri}{4} (\bm D_\a \xi^\a - \bar{\bm D}^\a \bar\xi_\a)
\end{align}
which identifies it as the SO(2) superfield parameter of the 3D superconformal
group. The 3D superconformal scale parameter $\sigma$
may be identified as
\begin{align}
\sigma = \frac{1}{2} (\bm D_\a \xi^\a + \bar{\bm D}^\a \bar\xi_\a)
	= \frac{1}{3} \pa_\ha \xi^\ha = 2 r - 2 x^\ha k_\ha
	+ 2 \eta^\a\bm\q_\a + 2 \bar\eta_\a\bar{\bm\q}^\a~,
\end{align}
and a certain combination of $\s$ and $\Lambda$ must be chiral,
\begin{align}
\bm \s := \s + \ri \Lambda = 2 r + \Lambda_\1{}^\1 + 4 \eta^\a \bm\q_\a
		- 2 k_\hb (x^\hb + \ri \bm\q^\a (\g^\hb)_{\a\b} \bar{\bm\q}^\b)~, \qquad
\bar{\bm D}^\a \bm\s = 0~.
\end{align}
The remaining parameters
 in \eqref{deltaQ}
obey
\begin{gather}
\xi^z = z \s~,\qquad
\rho^\a = \frac{\ri z}{2} \bm D^\a \s~,\qquad
\bm D_\b \rho^\a = \bar{\bm D}^\b \bar\rho_\a = 0 ~.
\end{gather}
All of the parameters discussed above can be derived from $\xi^\ha$
using the equations given. There is one difference between the parameter $\xi^\ha$ and the
usual 3D $\cN=2$ superconformal parameter $\xi^\ha$: the explicit $z$
dependence in the $z^2 k^\ha$ term. This leads to the additional identity
\begin{align}
\pa_z \xi_\ha = \pa_\ha \xi^z = z \pa_\ha \s~.
\end{align}
Note that $\xi^\a$ and $\s$ possess no $z$-dependence.

It is straightforward to work out the algebra of the isometry transformations.
Defining $\delta_{21} = [\delta_2,\delta_1]$, one finds that
\begin{align}
\xi_{21}^\ha = 2 \xi_{[2}^B \bm D_B \xi_{1]}^\ha
	+ 4 \ri \,\xi_{[2}^\a\bar\xi_{1]}^\b (\g^\ha)_{\a\b}
	+ 2 \xi_{[2}^z \pa_z \xi_{1]}^\ha
	+ 4 \ri \,\rho_{[2}^\a \bar\rho_{1]}^\b (\g^\ha)_{\a\b} ~.
\end{align}
From this parameter, one can derive the rest. Note that this
differs from the usual 3D $\cN=2$ superconformal algebra in
the presence of the third and fourth terms, both of which are
proportional to $z^2$. These are consistent with the presence
of the $z^2 k^\ha$ term in $\xi^\ha$. Although the specific
transformations of the Killing vectors have been modified,
the transformation induced on the constant parameters
$p^\ha$, $\omega^\ha{}_\hb$, $\epsilon$, $r$, $\L_\1{}^\1$, $k_\ha$, $\eta^\a$
remains that of the supergroup $\rm OSp(2|4)$.

The above identities allow us to easily check that the
action \eqref{3DProjAction} is invariant. Since the Lagrangian $K$ is
a weight-zero projective multiplet, the variation of the action is
\begin{align}
\delta S = -\int \frac{\rd z}{(sz)^2} \oint_C \frac{\rd \z}{2\pi \rm i\, \z  }
\int \rd^3 x\, {\rm d}^2\bmq \, {\rm d}^2\bmqb\,  
&\Big(\xi^\ha \pa_\ha K
	+ (\xi^\a + \z \rho^\a) \bm D_\a K
	+ (\bar{\xi}_\a - \frac{1}{\zeta} \bar\rho_\a ) \bar{\bm D}^\a K
	\eol & \quad
	+ \xi^z \pa_z K
	- 2\ri \Lambda \zeta \partial_\zeta K\Big)~.
\end{align}
The last term is a total contour derivative and vanishes. The first
three terms can be integrated by parts to give
\begin{align}
\delta S &=  \oint_C \frac{\rd \z}{2\pi \rm i\z}
\int \frac{\rd z}{(s z)^2} \int \rd^3 x\, {\rm d}^2\bmq \, {\rm d}^2\bmqb\,  
	\Big(
	(\pa_\ha \xi^\ha - \bm D_\a \xi^\a \bm - \bar{\bm D}^\a \bar{\xi}_\a) K
	- \xi^z \pa_z K\Big) \eol
	&=  \oint_C \frac{\rd \z}{2\pi \rm i\z}
\int \frac{\rd z}{(s z)^2} \int \rd^3 x\, {\rm d}^2\bmq \, {\rm d}^2\bmqb\,  
\Big(
	\s K - z\s \pa_z K\Big) \eol
	&= -\oint_C \frac{\rd \z}{2\pi \rm i\z}
\int \rd z \int \rd^3 x\, {\rm d}^2\bmq \, {\rm d}^2\bmqb\,  
	\pa_z \Big(\frac{\s K}{s^2 z}\Big)~,
\end{align}
which vanishes provided we can dispense with total derivatives in $z$.\footnote{When
analyzing invariance of the action under infinitesimal isometry transformations,
it suffices to restrict the fields to be supported inside the Poincar\'e patch.}

\subsection{Elimination of auxiliaries}
We emphasize that the $\s$-model action \eqref{3DProjAction} is a \emph{four-dimensional} action
involving hypermultiplets with an infinite number of auxiliary fields.
Upon elimination of the auxiliaries and a duality transformation, this action
is naturally associated with a four-dimensional $\cN=2$ 
supersymmetric $\s$-model in AdS.
Our peculiar parametrization of AdS$^{4|8}$ has been chosen to render this
elimination of auxiliaries as simple as possible. Let us now turn to this task.

The elimination of auxiliaries  begins with the
following observation. Performing the contour integral yields
\begin{align}
\oint_C \frac{\rd\zeta}{2\pi \ri \zeta} K(\U^I, \breve \U^{\bar J})
	= \cL(\U^I_n, \bar\U^{\bar J}_n)
\end{align}
where $\cL$ is some function depending on all of the components $\U^I_n$
of the arctic multiplets and their conjugates. For $n \geq 2$, these components are unconstrained
from the point of view of 3D $\cN=2$ superspace. Putting them on-shell
naturally leads to an infinite set of nonlinear algebraic equations
\begin{subequations}\label{AuxEOM}
\begin{alignat}{2}
0 &= \frac{\pa \cL}{\pa \U_n^I} = \oint_C \frac{\rd\zeta}{2\pi \ri \zeta} \frac{\pa K}{\pa \U^I} \zeta^n~,
	&\qquad &n\geq 2 \\
0 &= \frac{\pa \cL}{\pa \bar\U_n^{\bar J}} 
= \oint_C \frac{\rd\zeta}{2\pi \ri \zeta} \frac{\pa K}{\pa \breve\U^{\bar J}} {(-\zeta)}^{-n}~,
	&\qquad &n\geq 2
\end{alignat}
\end{subequations}
which are difficult to solve in general. 

However (and this is the point) the equations \eqref{AuxEOM} are \emph{exactly} the same
as those originating in the   $\cN=2$ supersymmetric $\s$-model 
in four-dimensional Minkowski space 
\bea
S=
\oint_C \frac{\rd \z}{ 2\p \ri \z}
\int \rd^4 x\, {\rm d}^2 \q\,{\rm d}^2\bar \q\,  
{ K}({ \U}^I, \breve{ \U}^{\bar J} )~,
\label{6.15}
\eea
which was  first studied in  \cite{GK1,GK2}.\footnote{The most general $\cN=2$ supersymmetric
$\s$-model in four-dimensional projective superspace \cite{LR88} is obtained 
from \eqref{6.15} by allowing the superfield Lagrangian to possess an arbitrary $\z$-dependence, 
$ { K}({ \U}, \breve{ \U} ) \to {\bm K}({ \U} ,  \breve{ \U},\z )$.}
This naturally allows us to appropriate with little modification 
the formal technique of reformulating \eqref{6.15} in terms of $\cN=1$ chiral superfields 
\cite{K-duality} (see also \cite{K-comments}) and devote it
to solving the problem in AdS. 
The price we pay is the loss of manifest
four-dimensional Lorentz invariance.

Assuming that the equations \eqref{AuxEOM} have been satisfied for some choice of
the auxiliaries, the action is reduced to
\begin{align}
S &= \int \frac{\rd z}{(s z)^2} \int \rd^3 x\,{\rm d}^2\bmq \,{\rm d}^2\bmqb\,  
\cL(\Phi, \Sigma, \bar\Phi, \bar\Sigma) ~,
\label{6.16}
\end{align}
where
\begin{align}
\cL &:= \oint_C \frac{\rd \zeta}{2\pi \ri \zeta}\, K(\U, \breve \U)~.
\label{6.17}
\end{align}
Here $\Phi^I = \U_0^I$ and $\Sigma^I = \U_1^I$ are the lowest two components of the arctic
multiplet $\U^I$. These are constrained as
\begin{align}\label{PhiSigmaConstraints}
\bar{\bm D}^\a \Phi^I = 0~, \qquad
-\frac{1}{4} \bar {\bm D}^2 \Sigma^I= \ri \,\partial_z \Phi^I~.
\end{align}
In other words, $\Phi^I$ is chiral. The condition $\Sigma^I$ obeys is
a modified version of the complex linear condition, so we will
refer to $\Sigma^I$ as a modified complex linear
superfield.\footnote{Similar modifications to the complex linearity condition
are standard in 5D \cite{KL_5D} and 6D \cite{GPT-M_6D}.}

The transformation law \eqref{deltaQ} for $\U^I$ implies transformations for $\Phi^I$
and $\Sigma^I$:
\begin{subequations}\label{eq_TangentBundleSusy}
\begin{align}
\delta \Phi^I
	&= - \xi^\ha \partial_\ha \Phi^I - \xi^\a {\bm D}_\a \Phi^I
	+ \bar\rho_\a \bar {\bm D}^\a \Sigma^I - \xi^z \partial_z \Phi^I~, \\
\delta \Sigma^I &= -\xi^\ha \partial_\ha \Sigma^I - \xi^\a {\bm D}_\a \Sigma^I - \bar \xi_\a \bar {\bm D}^\a \Sigma^I
	- \rho^\a {\bm D}_\a \Phi^I
	\eol & \quad
	+ \bar\rho_\a \bar {\bm D}^\a \Upsilon_2^I
	- \xi^z \partial_z \Sigma^I + 2 \ri \Lambda \Sigma^I~,
\end{align}
\end{subequations}
where $\U_2^I = \U_2^I(\Phi,\bar\Phi,\S,\bar\S)$.
It is straightforward to check that these transformation laws respect
the constraints \eqref{PhiSigmaConstraints}. In order for $\cL$
to be invariant, it must obey a number of constraints which are
derivable from its contour definition \eqref{6.17}.
Defining
\begin{align} \label{6.20}
\Xi &:= \oint_C \frac{\rd\zeta}{2\pi \ri \zeta} \frac{K}{\zeta}~,
\end{align}
it can be shown that the following conditions hold 
identically when the auxiliaries have been eliminated:
\begin{subequations}\label{eq_XiEqns}
\begin{align}
\frac{\partial \cL}{\partial \Phi^I} +
	\frac{\partial \cL}{\partial \Sigma^J} \frac{\partial \Upsilon_2^J}{\partial \Sigma^I}
	&= \frac{\partial \Xi}{\partial \Sigma^I}~, \\
-\frac{\partial \cL}{\partial \bar\Sigma^{\bar I}} +
	\frac{\partial \cL}{\partial \Sigma^J} \frac{\partial \Upsilon_2^J}{\partial \bar \Phi^{\bar I}}
	&= \frac{\partial \Xi}{\partial \bar\Phi^{\bar I}}~, \\
\frac{\partial \cL}{\partial \Sigma^J} \frac{\partial \Upsilon_2^J}{\partial \bar \Sigma^{\bar I}}
	&= \frac{\partial \Xi}{\partial \bar\Sigma^{\bar I}}~.
\end{align}
\end{subequations}
They may be proven using the contour definitions of $\cL$
and $\Xi$. In addition, since $K$ lacks any explicit $\zeta$ dependence,
one can show \cite{GK1,GK2} that
\begin{align}\label{6.22}
\S^I \frac{\pa \cL}{\pa \S^I} &= \bar\S^{\bar J} \frac{\pa \cL}{\pa \bar\S^{\bar J}}~.
\end{align}

Alternatively, requiring that a  Lagrangian $\cL (\Phi, \Sigma, \bar\Phi, \bar\Sigma)$ be invariant under 
\eqref{eq_TangentBundleSusy}, for some unknown function  $\U_2^I = \U_2^I(\Phi,\bar\Phi,\S,\bar\S)$, 
leads to a set of equations which imply
the existence of some function $\Xi$ obeying the equations
\eqref{eq_XiEqns} \cite{K-duality,K-comments}.
Making use of the equations \eqref{eq_XiEqns} and \eqref{6.22} 
leads to the following relation \cite{K-duality}:
\bea
\X = \S^I \frac{\pa \cL}{\pa \F^I} + 2 \U^I_2  \frac{\pa \cL}{\pa \S^I}~.
\label{6.23}
\eea
This result also follows from the contour integral representation \eqref{6.20}. 
Eq. \eqref{6.23} is the 3D foliated version of 
the condition \eqref{3.16AdS} which originates in the AdS frame.

Our next task is to perform a duality transformation converting the complex linear variables
to a set of purely chiral variables. 
For this, 
one relaxes $\Sigma^I$ to an
unconstrained superfield and introduces a Lagrange multiplier
chiral superfield $\Psi_I$, $\bar{\bm D}^\a \J_I = 0$, 
and the first-order action
\begin{align}
S_{\text{F.O.}} = \int \frac{\rd z}{(s z)^2} &\Bigg\{\int \rd^3 x\, {\rm d}^2\bmq \,{\rm d}^2\bmqb\,
	\left(\cL(\Phi, \Sigma, \bar\Phi, \bar\Sigma)
	+ \Sigma^I \Psi_I + \bar\Sigma^{\bar I} \bar\Psi_{\bar I}\right)
	\eol & \quad
	- \ri \int \rd^3 x\, {\rm d}^2\bmq \,
	\Psi_I \partial_z \Phi^I
	+ \ri \int \rd^3 x\, {\rm d}^2\bmqb \,
	\bar\Psi_{\bar I} \partial_z \bar\Phi^{\bar I} \Bigg\}~.
\end{align}
Varying $S_{\text{F.O.}} $ with respect to $\J_I$ 
leads to the 
constraint on $\Sigma^I$ as in \eqref{PhiSigmaConstraints}, 
and then $S_{\text{F.O.}}$ reduces to the original action, eq. \eqref{6.16}.   
Instead, making use of the equation of motion for $\Sigma^I$,
\begin{align}\label{SigmaEOM}
\frac{\pa \cL}{\pa \Sigma^I} = -\Psi_I~,
\end{align}
leads to the dual action
\begin{align}\label{SDarboux}
S_{\rm dual} = \int \frac{\rd z}{(s z)^2} &\Bigg\{
	\int \rd^3 x\, {\rm d}^2\bmq \,{\rm d}^2\bmqb\,
	\mathbb K(\Phi, \Psi, \bar\Phi,\bar\Psi)
	\eol & \quad
	-\ri \int \rd^3 x\, {\rm d}^2\bmq \,
	\Psi_I \partial_z \Phi^I
	+ \ri \int \rd^3 x\, {\rm d}^2\bmqb \,
	\bar\Psi_{\bar I} \partial_z \bar\Phi^{\bar I} \Bigg\}~,
\end{align}
where
\begin{align} \label{6.27}
\mathbb K(\Phi, \Psi, \bar\Phi,\bar\Psi) := \cL(\Phi, \Sigma, \bar\Phi, \bar\Sigma)
	+ \Sigma^I \Psi_I + \bar\Sigma^{\bar I} \bar\Psi_{\bar I}
\end{align}
with $\S$ understood to obey its equation of motion \eqref{SigmaEOM}.
The dual action is invariant under the AdS transformations
\begin{subequations}
\begin{align}
\delta \Phi^I &= -\xi^\ha \partial_\ha \Phi^I - {\xi}^\a {\bm D}_\a \Phi^I
	- \frac{\ri}{4} \bar {\bm D}^2 \left(z \s \frac{\partial \mathbb K}{\partial \Psi_I}\right)~, \\
\delta \Psi_I &= -\xi^\ha \partial_\ha \Psi_I - {\xi}^\a {\bm D}_\a \Psi_I
	- 2 \bm\s \Psi_I
	+ \frac{\ri}{4} \bar {\bm D}^2 \left(z\s \frac{\partial \mathbb K}{\partial \Phi^I} \right)~.
\end{align}
\end{subequations}
A new feature of the 3D foliation of AdS is the
appearance of chiral superspace integrals involving an explicit $z$-derivative.
In this dual formulation, the full 4D Lorentz symmetry is no longer
manifest. Because the steps we took to construct $\mathbb K$ are formally
identical to what occurs in four dimensions, we
conclude that the target space must be hyperk\"ahler. We will demonstrate
this more explicitly in the next section.

\subsection{$\cN=2$ supersymmetric $\s$-models on the cotangent bundles of Hermitian symmetric spaces}
\label{subsection6.3}
The procedure of converting the off-shell $\s$-model \eqref{3DProjAction} to the chiral form \eqref{SDarboux}, 
which we  employed in the previous subsection, was purely formal, since we assumed 
the auxiliary field equations \eqref{AuxEOM}  to be solved. 
But the actual solution of this problem is the most difficult part of the construction!
In 4D Minkowski space, this problem was solved
in a series of papers
\cite{GK1,GK2,AN,AKL1,AKL2,KN} for a large class of $\cN=2$ supersymmetric  $\s$-models \eqref{6.15}
in which $K(\F,  \bar \F )$ is the K\"ahler potential 
of a Hermitian symmetric space,  and therefore
 the corresponding  curvature tensor is covariantly constant,
\bea
\nabla_L  R_{I_1 {\bar  J}_1 I_2 {\bar J}_2}
= {\bar \nabla}_{\bar L} R_{I_1 {\bar  J}_1 I_2 {\bar J}_2} =0~.
\label{covar-const}
\eea
Here we can immediately apply the results obtained in \cite{GK1,GK2,AN,AKL1,AKL2,KN}
to the case of $\s$-models in AdS.

If the Riemann tensor associated with $K(\F, \bar \F)$ is covariantly constant, 
eq. \eqref{covar-const}, then 
the auxiliary field equations \eqref{AuxEOM}   are equivalent 
to the geodesic equation with complex evolution parameter \cite{GK1,GK2}
\bea
\frac{ {\rm d}^2 \U^I (\z) }{ {\rm d} \z^2 } + 
\G^I_{~JK} \Big( \U (\z), \bar{\F} \Big)\,
\frac{ {\rm d} \U^J (\z) }{ {\rm d} \z } \,
\frac{ {\rm d} \U^K (\z) }{ {\rm d} \z }  =0~.
\label{OldGeodesic}
\eea
This equation has a unique solution under the initial conditions
\bea
\U^I (0) =\F^I~, \qquad 
\dot{\U}^I (0) = \S^I~.
\eea
In particular, from \eqref{OldGeodesic} we derive 
\bea
\U^I_2 (\F,  \bar \F, \S , \bar \S)&=& -\hf \G^I_{JK} \big( \F, \bar{\F} \big) \, \S^J\S^K~,
\eea
with $\G^I_{JK} 
( \F , \bar{\F} )$  the Christoffel symbols for the  
K\"ahler metric $g_{I \bar J} ( \F , \bar{\F} )$. 
The function $\U^I_2$ determines the supersymmetry transformation law \eqref{eq_TangentBundleSusy}.

Upon elimination of the auxiliary superfields, 
the Lagrangian $\cL(\Phi, \Sigma, \bar\Phi, \bar\Sigma)$ appearing in \eqref{6.16}
can be shown to take the form  \cite{KN}:
\bea
\cL(\Phi, \Sigma, \bar\Phi, \bar\Sigma)
&=&    
 K \big( \F, \bar \F  \big) 
- \hf {\bm \S}^{\rm T} {\bm g} \,
\frac{ \ln \big( {\mathbbm 1} + {\bm R}_{\S,\bar \S}\big)}{\bm R_{\S,\bar \S}}
\, {\bm \S}
~, \quad 
{\bm \S} :=\left(
\begin{array}{c}
\S^I\\
{\bar \S}^{\bar I} 
\end{array}
\right) ~, ~~~~~
\label{act-HS}
\eea
where 
\bea
{\bm R}_{\S,\bar \S}
:=\left(
\begin{array}{cc}
0 & (R_\S)^I{}_{\bar J}\\
(R_{\bar \S})^{\bar I}{}_J &0 
\end{array}
\right)~, 
\quad (R_\S)^I{}_{\bar J}:=\hf R_K{}^I{}_{L \bar J}\, \S^K \S^L~, 
\quad (R_{\bar \S})^{\bar I}{}_J := \overline{(R_\S)^I{}_{\bar J}}~,~~~~
\label{R-Sigma}
\eea
and
\bea
{\bm g}
:=\left(
\begin{array}{cc}
0 & g_{I \bar J}\\
g_{{\bar I}J} &0 
\end{array}
\right)~.
\eea
A different universal representation for $\cL(\Phi, \Sigma, \bar\Phi, \bar\Sigma)$  can be found in 
\cite{AKL2}.

The hyperk\"ahler potential \eqref{6.27} can be shown  \cite{KN} to be 
\bea
{\mathbb K} (\F, \bar \F, \J, \bar \J ) =
K(\F, \bar \F) +  \hf {\bm \J}^{\rm T}{\bm g}^{-1}  \cF \big( - {\bm R}_{\J,\bar \J} \big)\, 
{\bm \J} ~,
\eea
where
\bea
\cF(x) = \frac{1}{x} \,\Bigg\{ \sqrt{1+4x} -1 -\ln \frac{1+ \sqrt{1+4x} }{2} \Bigg\}~, 
\qquad \cF(0)=1~
\eea
and the operator $ {\bm R}_{\J,\bar \J} $ is defined as
\bea
{\bm R}_{\J,\bar \J}
&:=&\left(
\begin{array}{cc}
0 & (R_\J)_I{}^{\bar J}\\
(R_{\bar \J})_{\bar I}{}^J &0 
\end{array}
\right)~, \non \\ 
(R_\J)_I{}^{\bar J}&=& (R_\J)_{IK} \,g^{K \bar J}~, \qquad 
(R_\J )_{K L}:= \hf R_{K}{}^I{}_{L}{}^J \,\J_I \J_J~. 
\eea

\section{The most general $\cN=2$ supersymmetric $\sigma$-model using the 3D foliation}\label{section7}
In this section, we attempt to generalize the models discussed in the previous
section. 

\subsection{The most general $\cN=2$ supersymmetric $\sigma$-model}
\label{subsection7.1}
We take the action
\begin{align}\label{3DSigma}
S &= \int \frac{\rd z}{(sz)^2} \Bigg\{\int \rd^3 x\,{\rm d}^2\bmq \,{\rm d}^2\bmqb\,
	\mathbb K(\phi, \bar\phi)
	+ \Big(\int \rd^3 x\,{\rm d}^2\bmq \,
	\ri H_\ra(\phi) \partial_z \phi^\ra + \textrm{c.c.} \Big)\Bigg\}
\end{align}
where $\mathbb K$ is the K\"ahler potential and $H_{\ra}$ is a holomorphic (1,0) form.
This action is manifestly 3D $\cN=2$ supersymmetric. We 
make an ansatz for the
transformation
law under the full 4D AdS supersymmetry of the form
\begin{align}
\delta \phi^{\ra} = -\xi^\ha \pa_\ha \phi^\ra - \xi^\a \bm D_\a \phi^\ra
	+ 2 \bm \sigma \psi^\ra
	- \frac{\ri}{4} \bar {\bm D}^2 (z \s \Omega^\ra)~,
\end{align}
where $\psi^\ra = \psi^\ra(\phi)$ and $\Omega^\ra = \Omega^\ra(\phi,\bar\phi)$
are for the moment arbitrary. Requiring invariance of the action
under this transformation implies conditions on both the functions
$\psi^\ra$ and $\Omega^\ra$. Similarly, closure of the algebra 
imposes additional restrictions. Ultimately, one discovers that the
transformation law must take the form
\begin{align}
\delta \phi^{\ra} = -\xi^\ha \pa_\ha \phi^\ra - \xi^\a \bm D_\a \phi^\ra
	- 2 \bm \sigma \newomega^{\ra \rb} H_\rb
	- \frac{\ri}{4} \bar {\bm D}^2 (z \s \newomega^{\ra \rb} \mathbb K_\rb)~,
\end{align}
where $\newomega^{\ra \rb}$ is an antisymmetric chiral quantity obeying a number
of conditions.  First, it must be covariantly constant,
\begin{align}
\nabla_\rc \newomega^{\ra \rb} = 0~, \qquad \nabla_{\bar \rc} \newomega^{\ra \rb} = \pa_{\bar \rc} \newomega^{\ra \rb} = 0~.
\end{align}
Secondly, it must obey
\begin{align}
\newomega^{\ra \rb} \newomega_{\rb \rc} = -\delta^\ra_\rc~, \qquad
\newomega_{\ra \rb} = g_{\ra \bar \rc} g_{\rb \bar\rd} \newomega^{\bar\rc \bar\rd}~.
\end{align}
These two conditions imply that the target space is hyperk\"ahler.

A third requirement is that the holomorphic $(2,0)$ form
$\newomega_{\ra\rb}$ must be exact, with its one-form potential given by $H_\ra$,
\begin{align}
\newomega_{\ra\rb} = \pa_\ra H_\rb - \pa_\rb H_\ra~.
\end{align}
In addition, $H_\ra$ must be related to a holomorphic U(1) Killing vector
\begin{align}\label{VHeqn}
V^\ra := \ri \,\newomega^{\ra \rb} H_\rb~, \qquad
H_\rb = \ri \,\newomega_{\rb \rc} V^\rc
\end{align}
which from its definition can be shown to rotate the complex structure,
\begin{align}\label{Vomega}
\cL_V \newomega_{\ra \rb} = \nabla_\ra V^\rc \newomega_{\rc \rb} + \nabla_\rb V^\rc \newomega_{\ra \rc}
	= \ri \,\nabla_\ra H_\rb - \ri \,\nabla_\rb H_\ra
	= \ri \, \newomega_{\ra \rb}~,
\end{align}
as well as the holomorphic one-form $H_\ra$,
\begin{align}
\cL_V H_\ra = \ri \,\cL_V \newomega_{\ra \rb} V^\rb = \ri\, H_\ra~.
\end{align}

Remarkably, when all of these conditions are imposed, the 
supersymmetry algebra closes \emph{off-shell}. The same behavior
was observed in the conventional foliation of AdS \cite{BKsigma1, BKsigma2}
discussed in section \ref{sect_N1AdS}.

The action can be written in terms of the Killing vector $V^\ra$,
\begin{align}\label{3DSigma2}
S &= \int \frac{\rd z}{(sz)^2} \Bigg\{
	\int \rd^3 x\, {\rm d}^2\bmq \,{\rm d}^2\bmqb\,
	\mathbb K(\phi, \bar\phi)
	+ \left(\int \rd^3 x\,{\rm d}^2\bmq \,
	V^\ra \newomega_{\ra \rb} \partial_z \phi^\rb + \textrm{c.c.} \right)\Bigg\}~.
\end{align}
The transformation law for $\phi^\ra$ similarly can be rewritten
\begin{align}
\delta \phi^{\ra} &= -\xi^\ha \pa_\ha \phi^\ra - \xi^\a \bm D_\a \phi^\ra
	+ 2 \ri \bm \sigma V^\ra
	- \frac{\ri}{4} \bar {\bm D}^2 (z \s \newomega^{\ra \rb} \mathbb K_\rb)~. \label{dphi2}
\end{align}

The action derived in the previous section from projective superspace
corresponds to a choice of Darboux coordinates where
\begin{subequations}
\bea
\phi^\ra = (\Phi^I, \Psi_I)~,\qquad H_\ra &=& (-\Psi_I, 0)~, \qquad
\newomega^{\ra \rb} = \newomega_{\ra \rb} =\left(
\begin{array}{rr}
0 & \phantom{+}{\mathbbm 1} \\
-{\mathbbm 1} & 0
\end{array}
\right)~,  \label{Dar1}\\
V &=& \ri \Psi_I \frac{\pa}{\pa \Psi_I} - \ri \bar\Psi_{\bar J} \frac{\pa}{\pa \bar\Psi_{\bar J}}~.
\label{Dar2}
\eea
\end{subequations}

We know from the earlier works \cite{GK1,GK2,K-duality} that 
\bea
{\mathbb K} (\F ,  \J, \bar \F,  \bar \J) = K \big( \F, \bar{\F} \big)+    
\cH \big(\F,  \J ,\bar \F, \bar \J \big)~, 
\label{h-master}
\eea
where
\bea
\cH \big(\F,  \J , \bar \F,  \bar \J \big)&=& 
\sum_{n=1}^{\infty} \cH^{I_1 \cdots I_n {\bar J}_1 \cdots {\bar 
J}_n }  \big( \F, \bar{\F} \big) \J_{I_1} \dots \J_{I_n} 
{\bar \J}_{ {\bar J}_1 } \dots {\bar \J}_{ {\bar J}_n } ~,\non \\
\cH^{I {\bar J}} \big( \F, \bar{\F} \big) 
&=& g^{I {\bar J}} \big( \F, \bar{\F} \big)  ~.
\label{h}
\eea
Here the coefficients $\cH^{I_1 \cdots I_n {\bar J}_1 \cdots {\bar 
J}_n }$, for  $n>1$, 
are tensor functions of the K\"ahler metric
$g_{I \bar{J}} \big( \F, \bar{\F}  \big) 
= \pa_I 
\pa_ {\bar J}K ( \F , \bar{\F} )$,  the Riemann curvature $R_{I {\bar 
J} K {\bar L}} \big( \F, \bar{\F} \big) $ and its covariant 
derivatives. 
Using this result, we can compute a Killing potential, 
${\cK} = \bar \cK $,
 corresponding to  the Killing vector field \eqref{Dar2}. In accordance with  \cite{BWitten}, 
 it is defined by $V^a (\f) \pa_\ra {\mathbb K} (\f , \bar \f) = (\ri /2) \cK(\f, \bar \f)   + \l(\f) $, 
 for some holomorphic function $\l$. In our case, it is immediately seen that  $\l=0$ and 
 \bea
 \cK \big(\F,  \J , \bar \F,  \bar \J \big)&=& 2
\sum_{n=1}^{\infty} n\,  \cH^{I_1 \cdots I_n {\bar J}_1 \cdots {\bar 
J}_n }  \big( \F, \bar{\F} \big) \J_{I_1} \dots \J_{I_n} 
{\bar \J}_{ {\bar J}_1 } \dots {\bar \J}_{ {\bar J}_n } ~.
 \label{KP7.15}
\eea
It follows that $\cK (\f, \bar \f)$ is a globally defined function on the hyperk\"ahler target space. 

As an example, we can consider $\cX ={\mathbb C}P^n  $. 
In standard inhomogeneous  coordinates for ${\mathbb C}P^n  $,
the K\"ahler potential is
\bea 
K (\F, {\bar \F}) = r^2 \ln \left(1 + \frac{1}{r^2}  
\F^L \overline{\F^L} \right)
~.
\eea
The hyperk\"ahler potential on $T^*{\mathbb C}P^n$  is known to be 
(see e.g. \cite{AKL1} for a derivation)
 \bea
{\mathbb K}(\F,  \J , \bar \F, \bar \J) &=& K(\Phi, {\bar \Phi})
+ r^2\Big\{   \sqrt{  1 + 4 | \J  |^2/r^2 } 
-  \ln \Big( 1+  \sqrt{ 1 + 4 | \J  |^2/r^2 }  \Big) \Big\}~,~~~~~~~
\eea
with 
$
|\J|^2 := g^{I\bar J} (\F, \bar \F) \J_I \bar \J_{\bar J}$. 
The Killing potential is 
\bea
 \cK \big(\F,  \J , \bar \F,  \bar \J \big)&=&  r^2\Big(  \sqrt{  1 + 4 | \J  |^2/r^2 } -1\Big)~.
 \label{7.18}
 \eea

We can now demonstrate that the most general supersymmetric $\s$-model described above, eq. \eqref{3DSigma}, 
can be derived from a model in projective superspace of the form \eqref{3DProjAction}.
Our starting point is the holomorphic symplectic (2,0) form $\hat \o  = \rd H$.   
According to Darboux's theorem (see, e.g., \cite{Sternberg}),  locally we can choose  new complex 
coordinates $\phi^\ra = (\Phi^I, \Psi_I)$, centred around the origin of ${\mathbb C}^{2n}$,   
in which  the holomorphic (1,0) form $H$ looks like a Liouville form
\bea
H = -\J_I \rd \F^I~.
\eea
Then $\hat \o$ coincides with the canonical symplectic form, 
\bea
\hat \o = \rd \F^I \wedge \rd \J_I~,
\eea
which is equivalent to \eqref{Dar1}.
As a consequence of eq. \eqref{VHeqn}, we also observe that the Killing vector field $V$ takes the form 
(\ref{Dar2}). Since the vector field $V$ is Killing, from (\ref{Dar2}) 
we conclude that 
\bea 
{\mathbb K} (\F , \re^{\ri \a } \J, \bar \F,  \re^{-\ri \a } \bar \J) 
= {\mathbb K} (\F ,  \J, \bar \F,  \bar \J) +  \Big\{ \L (\F, \J) +{\rm c.c.} \Big\}~, \qquad \a \in \mathbb R
\eea
for some holomorphic function $\L (\F, \J)$. 
We assume the K\"ahler potential $ {\mathbb K} (\F ,  \J, \bar \F,  \bar \J) $ to be a real analytic function 
on the coordinate chart chosen. Then, the previous result tells us that modulo a K\"ahler transformation 
we can choose $ {\mathbb K} (\F ,  \J, \bar \F,  \bar \J) $ to be invariant under the U(1) isometry group 
generated by $V$, 
\bea 
{\mathbb K} (\F , \re^{\ri \a } \J, \bar \F,  \re^{-\ri \a } \bar \J) 
= {\mathbb K} (\F ,  \J, \bar \F,  \bar \J) ~.
\eea
We now consider a submanifold defined by $\J_I= \bar \J_{\bar J}=0$. 
On this submanifold,  we introduce a K\"ahler potential $K(\F, \bar \F)$ as follows:
\bea
K(\F, \bar \F) :=  {\mathbb K} (\F ,  \J, \bar \F,  \bar \J) \Big|_{\J = \bar \J =0} ~.
\eea
Associated with $K(\F, \bar \F) $ is the desired supersymmetric $\s$-model  \eqref{3DProjAction}.

\subsection{Superpotentials and tri-holomorphic isometries}
It is natural to ask whether it is possible to modify the action
\eqref{3DSigma} to include a holomorphic superpotential. From experience
with the Minkowski limit, we expect the superpotential to be associated
with a tri-holomorphic isometry $X^\ra$, which obeys
\begin{gather}\label{eq_XTriHolo}
\cL_X \newomega_{\ra \rb} = 0~, \qquad \partial_{\bar\rb} X^{\ra} = 0~,\qquad
\nabla_{\ra} X_{\bar \rb} + \nabla_{\bar \rb} X_{\ra} = 0~.
\end{gather}
The tri-holomorphy requirement implies that $X^\ra$ can be locally written as
\begin{align}\label{eq_XfromW}
X^\ra = -\newomega^{\ra \rb} \pa_\rb W~,
\end{align}
for a holomorphic function $W$. (The choice of phase on the right-hand side of
this expression is conventional at this point, but we will soon see that this
is the correct choice.)

Let us make the additional requirement that the action \eqref{3DSigma2}
be invariant under the tri-holomorphic isometry $\delta \phi^\ra = X^\ra$.
This leads to
\begin{align}
\delta S &= \int \frac{\rd z}{(sz)^2} \Big\{\int \rd^3 x\,{\rm d}^2\bmq \,{\rm d}^2\bmqb\,
	\cL_X \mathbb K
	\eol & \qquad \qquad
	+ \int \rd^3 x\,{\rm d}^2\bmq \,(\cL_X V)^\ra \newomega_{\ra \rb} \partial_z \phi^\rb
	+ \int \rd^3 x\,\rd z\, {\rm d}^2\bmqb \,
		(\cL_X V)^{\bar \ra} \newomega_{\bar\ra \bar\rb} \partial_z \bar\phi^{\bar\rb} \Big\}~.
\end{align}
The K\"ahler term is already invariant up to the real part of a holomorphic
superfield, so it vanishes. This leaves the superpotential term, and we find
the additional requirement
\begin{align}\label{LxV}
\cL_X V = [X,V] = 0~.
\end{align}
Making use of this condition along with \eqref{Vomega}, one can show that
there exists a globally defined choice for the holomorphic function $W$,
\begin{align}\label{eq_WfromX}
W = -\ri\,V^\ra \newomega_{\ra\rb} X^\rb~, \qquad
\cL_V W = \ri\, W~.
\end{align}
Let us add to the action \eqref{3DSigma2} the term
\begin{align}\label{eq_superpotential}
\int \rd^3x\, \rd z \, \rd^2\bm\q\, \frac{1}{(sz)^3} W + \textrm{c.c.}~.
\end{align}
From our experience with the situation in section \ref{sect_AdSSuper},
we expect the addition of the superpotential to correspond to the
modification of the U(1) Killing vector, so we postulate the modified
transformation law
\begin{align}
\delta \phi^{\ra} &= -\xi^\ha \pa_\ha \phi^\ra - \xi^\a \bm D_\a \phi^\ra
	+ 2 \ri \bm \sigma \left(V^\ra - \frac{1}{2s} X^\ra\right)
	- \frac{\ri}{4} \bar {\bm D}^2 (z \s \newomega^{\ra \rb} \mathbb K_\rb)~.
\end{align}

At this point, it is easy to show that the new action and transformation law
is equivalent to the old action and transformation law for the choice
\begin{align}
V'^\ra = V^\ra - \frac{1}{2s} X^\ra~.
\end{align}
To prove this, observe that
\begin{align}
&\int \rd^3 x\,\rd z\, {\rm d}^2\bmq \,
	\frac{1}{(sz)^2} V'^\ra \newomega_{\ra \rb} \pa_z \phi^\rb \eol
&= \int \rd^3 x\,\rd z\, {\rm d}^2\bmq \,
	\frac{1}{(sz)^2} \left(V^\ra \newomega_{\ra \rb} \pa_z \phi^\rb
	- \frac{1}{2s} X^\ra \newomega_{\ra\rb} \pa_z\phi^\rb\right)~.
\end{align}
One can show that $\pa_\ra W = \newomega_{\ra\rb} X^\rb$, and so we have
\begin{align}
&\int \rd^3 x\,\rd z\, {\rm d}^2\bmq \,
	\frac{1}{(sz)^2} \left(V^\ra \newomega_{\ra \rb} \pa_z \phi^\rb
	+ \frac{1}{2s} \pa_\rb W \pa_z\phi^\rb\right) \eol
&= \int \rd^3 x\,\rd z\, {\rm d}^2\bmq \,
	\frac{1}{(sz)^2} \left(V^\ra \newomega_{\ra \rb} \pa_z \phi^\rb
	+ \frac{1}{sz} W \right)~.
\end{align}
Note that since $X^\ra$ is tri-holomorphic, we are allowed to add it to
$V^\ra$ without modifying any of the conditions that $V^\ra$ obeys.

So the general form of the action with a superpotential
\begin{align}\label{eq_generalSigmaPot}
S &= \int \frac{\rd z}{(sz)^2} \, \Bigg\{\int \rd^3 x\,{\rm d}^2\bmq \,{\rm d}^2\bmqb\,
	\mathbb K(\phi, \bar\phi)
	\eol & \qquad\qquad
	+ \Big[\int \rd^3 x\,{\rm d}^2\bmq \,
	V^\ra \newomega_{\ra \rb} \Big(
	\pa_z \phi^\rb - \frac{\ri}{sz} X^\rb\Big)
	+ \textrm{c.c.} \Big]\Bigg\}~,
\end{align}
with the AdS isometry
\begin{align}
\delta \phi^{\ra} &= -\xi^\ha \pa_\ha \phi^\ra - \xi^\a \bm D_\a \phi^\ra
	+ 2 \ri \bm \sigma \left(V^\ra - \frac{1}{2s} X^\ra \right)
	- \frac{\ri}{4} \bar {\bm D}^2 (z \s \newomega^{\ra \rb} \mathbb K_\rb) \label{eq_generalSigmaTrans}
\end{align}
is completely equivalent to the original form \eqref{3DSigma2}
with $V'^\ra = V^\ra - X^\ra / 2s$. This equivalence is completely
analogous to the situation in the AdS frame, which we will elaborate
upon in section \ref{HK_geometry}.

For completeness, we should mention that there is one other possibility for
adding a superpotential term; however, its geometric significance is quite different.
Let us take the original action \eqref{3DSigma2} and add a superpotential term
\eqref{eq_superpotential}. However, in contrast to the choice of phase made in
\eqref{eq_XfromW} and \eqref{eq_WfromX}, let us take
\begin{align}\label{eq_gaugeW}
W = f\, V^\ra \newomega_{\ra \rb} X^\rb~, \qquad f \in \mathbb R~.
\end{align}
Note that this still obeys
\begin{align}
V^\ra W_\ra = \ri\, W~,\qquad W_\ra =  \ri \, f\,\newomega_{\ra \rb} X^\rb~.
\end{align}
It turns out that for this choice, the superpotential term and the
original action \eqref{3DSigma2} are separately invariant under the original
isometry \eqref{dphi2}. What then is the physical significance of this choice?

The answer lies in the following observation: there is no barrier to choosing
the real proportionality factor $f$ in \eqref{eq_gaugeW} to possess \emph{arbitrary}
$z$ dependence! So we arrive at a chiral integral of the form
\begin{align}\label{eq_dphiX}
\int \rd^3x\, \rd z \, \rd^2\bm\q\, \frac{1}{(sz)^2} V^\ra \newomega_{\ra\rb}
	\left(\pa_z \phi^\rb + \frac{f(z)}{sz} X^\rb\right)+ \textrm{c.c.}
\end{align}
There is an obvious interpretation of this additional term: it is
a $z$-dependent gauge connection which gauges the tri-holomorphic isometry.
That is, we may identify
\begin{align}
\cD_z \phi^\rb := \pa_z \phi^\rb + \frac{f(z)}{sz} X^\rb~.
\end{align}
Note that this connection is actually pure gauge, and so we can remove
it if we perform a gauge transformation
\begin{align}
\delta_{g} \phi^\ra = -\Lambda(z) X^\ra~,\qquad
\Lambda'(z) = \frac{f(z)}{sz}~.
\end{align}

In fact, we should always choose $f(z)=0$ in this way.
Eq. \eqref{eq_dphiX} for $f(z) \neq 0$ is clearly problematic from the point of
view of 4D Lorentz invariance of the component action. Any 4D Lorentz transformation
for $f(z)\neq 0$ must be accompanied by a gauge transformation to restore $f(z)$ to this
form. While this in a very technical sense respects 4D Lorentz
invariance, it clearly violates its spirit. So we will always restrict
to the case $f(z) = 0$. We emphasize that this is the natural gauge choice.

The actions considered above emerge automatically when one considers projective
superspace Lagrangians with holomorphic isometries on the original K\"ahler
manifold. These become tri-holomorphic isometries on the cotangent bundle. When
gauged with the intrinsic vector multiplet prepotential of AdS, the above
structure can be shown to emerge, including the bizarre factor $f(z)$.
It is to this construction which we now turn.

\subsection{Gauged $\s$-models from projective superspace}
We return to the original action
\begin{align}
S= \frac{1}{2\pi \rm i}  \oint_C \frac{\rd \z}{\z}
\int \rd^3 x\,\rd z\, {\rm d}^2\bmq \, {\rm d}^2\bmqb\,  
\frac{1}{(s z)^2}
{ K}({ \U}, \breve{ \U})~,
\end{align}
written in projective superspace. Let us suppose $K$
possesses a holomorphic isometry $X^I$ under which
\begin{align}\label{eq_Xisometry}
X^I K_I + \breve X^{\bar J} K_{\bar J} = F(\U) + \breve F(\breve \U)
\end{align}
where $F$ is a holomorphic function of the arctic superfields $\U^I$.
Such transformations are symmetries of the action.
The transformation $\delta \U^I = X^I$ on the arctic superfield leads to
\begin{align}
\delta \Phi^I = X(\Phi)^I~,\qquad \delta \Psi_I = -\pa_I X(\Phi)^J \Psi_J
\end{align}
for the cotangent bundle coordinates. Denoting this holomorphic isometry as
$X^\ra$ when acting on the complex coordinates
$\phi^\ra = (\Phi^I, \Psi_I)$ of the hyperk\"ahler manifold,
it is easy to see that it is tri-holomorphic,
$\cL_X \newomega_{\ra \rb} = 0$
where $\newomega_{\ra\rb}$ is the canonical symplectic form.

Now let us consider gauging the isometry. In section \ref{sect_N1AdS},
we did this in a manifestly covariant way to avoid dealing specifically
with the vector prepotential. This required the addition of a fictitious
target space coordinate when the function $F$ discussed above was nonzero.
We could follow that same procedure here, but for the sake of clarity we
will take an alternative approach where the modifications induced by the
gauging are more transparent.

We follow the procedure described in \cite{Kuzenko:Superpotentials}, which
was based on \cite{HKLR}. From \eqref{eq_Xisometry}, it follows that \cite{BWitten}
\begin{align}
X^I K_I = \ri D + F
\end{align}
where $D$ is a real function, the Killing potential.
We consider a complexified transformation
$\delta \U^I = -\Lambda X^I$
where $\L$ is an arctic superfield and introduce a
tropical abelian prepotential $V$ which transforms as
$V \rightarrow V + \ri \L - \ri \breve \L$. The original
Lagrangian $K$ can then be modified to
\begin{align}
K + V \frac{e^{\ri \,V L_{\bar X} - 1}}{\ri \,V L_{\bar X}} D
	= K + V D + \frac{\ri}{2} V^2 L_{\bar X} D + \cdots
\end{align}
where $L_{\bar X} D = \bar X^{\bar J} \pa_{\bar J} D$.
This new Lagrangian is the gauged $\s$-model.

The AdS geometry comes equipped with an intrinsic vector multiplet
whose prepotential $V_{\rm AdS}$ is defined up to a gauge transformation.
The construction of $V_{\rm AdS}$ is given in appendix \ref{app_IntrinsicVector}.
The relevant details here are that
$V_{\rm AdS}$, when written in terms of the rotated $\bm\q_i$ and $\bar{\bm\q}^i$
and projected to $\bm\q_\2 = \bar{\bm\q}^\2 = 0$, is given by
\begin{align}
V_{\rm AdS}\vert = 
	\frac{1}{\zeta} \bm\q^2 \left(\frac{\ri}{sz} - \L'(z) \right)
	+ \zeta \bar {\bm\q}^2 \left(\frac{\ri}{sz} + \L'(z) \right)~.
\end{align}
The first term in both sets of parentheses is dictated by the requirement
that $V_{\rm AdS}$ possess the correct frozen vector multiplet field strength
$W_0 = 1 / sz$. The second term turns out to be pure gauge and has no effect
on $W_0$. It is possible to show that the gauge connection associated with $V_{\rm AdS}$,
when projected to $\bm \q_\2 = \bar{\bm\q}^\2 = 0$ obeys
$\bm A_\a^\1 \vert= \bar {\bm A}_{\a \1} \vert= A_{\a\b} \vert= 0$, while $A_z \vert= \L'(z)$.
In other words, from the 3D $\cN=2$ superspace point of view, the connection
is pure gauge arising from a $z$-dependent gauge transformation.
We present a proof of this in appendix \ref{app_IntrinsicVector} for the
curious reader.

We can separate $V_{\rm AdS}\vert$ into two pieces, $V_{(+)}$ and $V_{(-)}$,
which represent the arctic and antarctic components,
\begin{align}
V_{\rm AdS}\vert = V_{(+)} + V_{(-)}~,\quad
V_{(+)} = \zeta \bar {\bm\q}^2 \left(\frac{\ri}{sz} + \L'(z) \right)~,\quad
V_{(-)} = \frac{1}{\zeta} \bm\q^2 \left(\frac{\ri}{sz} - \L'(z) \right)~.
\end{align}
Note that $V_{(+)}$ and $V_{(-)}$ are nilpotent, $(V_{(+)})^2 = (V_{(-)})^2 = 0$.
From this observation it is possible to show that the gauged Lagrangian can
be written
\begin{align}
S= \frac{1}{2\pi \rm i}  \oint_C \frac{\rd \z}{\z}
\int \rd^3 x\,\rd z\, {\rm d}^2\bmq \, {\rm d}^2\bmqb\,  
\frac{1}{(s z)^2}
K(\hat\U, \breve{\hat \U})~,\qquad
\hat \U^I = \U^I - \ri V_{(+)} X^I(\U)~.
\end{align}
The superfield $\hat\U^I$ is the covariant arctic superfield which we
discussed in section \ref{sect_N1AdS}; here we have constructed it
explicitly and denoted it with a circumflex.
One can easily see its lowest component $\hat\Phi^I = \U_0^I$
remains chiral, but the next component $\hat \S^I = \U_1^I$ obeys a new modified
complex linearity constraint,
\begin{align}
-\frac{1}{4} \bar{\bm D}^2 \hat\S^I = \ri \,\pa_z \hat \Phi^I 
	+ \left(\frac{1}{sz} + \ri \L'(z) \right) X^I(\hat\Phi)~.
\end{align}
Because $\L'(z)$ may be interpreted as a $z$-dependent gauge connection $A_z$,
the above expression can be interpreted as
\begin{align}
-\frac{1}{4} \bar{\bm D}^2 \hat\S^I = \ri \,\cD_z \hat \Phi^I 
	+ \frac{1}{sz} X^I(\hat\Phi)~,\qquad
\cD_z \hat \Phi^I = \pa_z \hat\Phi^I + \L'(z) X^I(\hat\Phi)~.
\end{align}

The elimination of auxiliaries proceeds exactly as before and the form of the
Lagrangian on the tangent bundle is unchanged, except for the modification to
the complex linearity constraint. However, upon dualizing to the cotangent
bundle, we find
\begin{align}\label{eq_gaugedSigma}
S_{\rm dual} &= \int \rd^3 x\,\rd z\, {\rm d}^2\bmq \,{\rm d}^2\bmqb\,
	\frac{1}{(s z)^2} \mathbb K(\hat\Phi, \hat\Psi, \hat{\bar\Phi},\hat{\bar\Psi})
	\eol & \quad
	- \Bigg[\int \rd^3 x\,\rd z\, {\rm d}^2\bmq \,
	\frac{1}{(sz)^2} \hat\Psi_I \left(\ri \,\cD_z \hat\Phi^I + \frac{1}{sz} X^I(\hat\Phi)\right)
	+ \textrm{c.c.}\Bigg]
\end{align}
where as before
\begin{align}
\mathbb K(\hat\Phi, \hat\Psi, \hat{\bar\Phi},\hat{\bar\Psi}) := \cL(\hat\Phi, \hat\Sigma, \hat{\bar\Phi}, \hat{\bar\Sigma})
	+ \hat\Sigma^I \hat\Psi_I + \hat{\bar\Sigma}^{\bar J} \hat{\bar\Psi}_{\bar J}
\end{align}
When recast into general chiral coordinates, this Lagrangian matches
that postulated in \eqref{eq_generalSigmaPot}.

To complete the equivalence, we must determine the modified transformation laws
for $\hat \Phi^I$ and $\hat\Psi_I$.
Letting $\hat e$ correspond to the U(1) generator, we define a covariant weight-zero
arctic multiplet $\hat\U$ by
\begin{align}
\hat \U = \exp\left(V_{(+)} \hat e \right) \U~.
\end{align}
(For the arctic multiplets we have been discussing, $\ri \hat e \U^I = X^I$.)
The corresponding covariant derivatives are $\bm\cD_A$.
The covariant arctic multiplet $\hat \U$ is defined to transform covariantly,
\begin{align}
\delta \hat\U = -\Big(\xi^A \bm \cD_A  + \lambda^{ij} J_{ij} + \ri\, \Gamma \hat e\Big) \hat \U~,
\end{align}
where $\G$ is real, while the original arctic multiplet transforms as
\begin{align}
\delta \U = -\Big(\xi^A \bm D_A  + \lambda^{ij} J_{ij} + \ri \,\Lambda \hat e\Big) \hat \U~,
\end{align}
where $\L$ is arctic.
Together, these imply that $V_{(+)}$ should transform as
\begin{align}
\delta V_{(+)} = -(\xi^A \bm D_A + \lambda^{ij} J_{ij}) V_{(+)} - \ri \, (\xi^B \bm A_B + \Gamma - \Lambda)~.
\end{align}
Note that this implies that $V$ transforms as
a projective multiplet of weight zero, up to a $\L$-gauge transformation
\begin{align}
\delta V = -(\xi^A \bm D_A + \lambda^{ij} J_{ij}) V + \ri \,(\L - \breve \L)~.
\end{align}

From the discussion of the intrinsic vector multiplet in section \ref{IntrinsicVH}, we know that
$\G$ is given in the AdS frame by $\G = 2 \ve$. Repeating that analysis
in the Minkowski frame, that relation becomes\footnote{Calculating $\G$ is a bit
more technical in the Minkowski frame than in the AdS frame. In particular, there are
additional more complicated terms in \eqref{eq_GammaMinkowski}, which vanish when we project
to $\bm\q_\2 = \bar{\bm\q}^\2=0$. But \eqref{eq_GammaMinkowski} is the only result
we require.}
\begin{align}\label{eq_GammaMinkowski}
\G\vert = - \frac{1}{s} \L~.
\end{align}
This leads to
\begin{subequations}
\begin{align}
\delta \hat\Phi^I
	&= - \xi^\ha \partial_\ha \hat\Phi^I - \xi^\a {\bm D}_\a \hat\Phi^I
	+ \bar\rho_\a \bar {\bm D}^\a \hat\Sigma^I - \xi^z \cD_z \hat\Phi^I
	+ \frac{\L}{s} \ri \hat e \hat\Phi^I \\
\delta \hat\Sigma^I &= -\xi^\ha \partial_\ha \hat\Sigma^I - \xi^\a {\bm D}_\a \hat\Sigma^I
	- \bar \xi_\a \bar {\bm D}^\a \hat\Sigma^I
	- \rho^\a {\bm D}_\a \hat\Phi^I + \bar\rho_\a \bar {\bm D}^\a \hat\Upsilon_2^I
	\eol & \quad
	- \xi^z \cD_z \hat\Sigma^I + 2 \ri \Lambda \hat\Sigma^I
	+ \frac{\L}{s} \ri \hat e \hat\S^I
\end{align}
\end{subequations}
where
\begin{align}
\ri \hat e \hat\Phi^I = X^I(\hat\Phi)~,\qquad \ri \hat e \hat\S^I = \hat\S^J \pa_J X^I(\hat\Phi)~.
\end{align}
The additional terms we have added to $\delta\hat\Phi^I$ lead to
\begin{align}
\delta \hat\Phi^I = - \xi^A D_A \hat\Phi^I
	+ \frac{\bm \s}{s} \hat e \hat\Phi^I
	- \frac{\ri z}{4} \bar{\bm D}^2 (\s \hat\S^I)
\end{align}
which is manifestly chiral. Note the appearance of the new gauged isometry
term. Performing the duality in the usual way leads to a similar modification
for $\delta \hat\Psi_I$, and we find
\begin{subequations}
\begin{align}
\delta \hat\Phi^I 
	&= -\xi^A \cD_A \hat\Phi^I
	+ \frac{\bm\s}{s} \hat e \hat\Phi^I
	- \frac{\ri z}{4} \bar {\bm D}^2 \left(\s\frac{\pa \mathbb K}{\pa \hat\Psi_I}\right)~, \\
\delta \hat\Psi_I
	&= -\xi^A \cD_A \hat\Psi_I
	- 2 \bm\s \hat\Psi_I
	+ \frac{\bm\s}{s} \hat e \hat\Psi_I
	+ \frac{\ri z}{4} \bar {\bm D}^2 \left(\s\frac{\pa \mathbb K}{\pa \hat\Phi^I}\right)~,
\end{align}
\end{subequations}
where $\ri \hat e \hat\Psi_I = -\partial_I X^J(\hat\Phi) \,\hat\Psi_J$. These are the isometries
of the action \eqref{eq_gaugedSigma}. Generalizing the action and the transformation
law to general chiral coordinates, we find \eqref{eq_generalSigmaPot} and \eqref{eq_generalSigmaTrans}.

\section{Hyperk\"ahler geometry for $\s$-models in AdS}\label{HK_geometry}
In the preceding sections, we have mainly been interested in 4D
$\s$-models with $\cN=2$ supersymmetry drawing inspiration from
those which emerge naturally from projective superspace descriptions
of AdS$^{4|8}$. In rewriting these models in terms of a superspace with
only four Grassmann coordinates, there are two possible choices:
the conventional AdS frame corresponding to
\eqref{1.3a}, and the 3D foliated frame corresponding to
\eqref{1.3b}.
However, no matter which  intermediate superspace we choose, it is
clear that the component actions must be identical after
eliminating the remaining auxiliaries and the target spaces must
also be identical.
For that reason, we focus in this section on demonstrating the
\emph{general} features of a hyperk\"ahler target space $\cM$
with a U(1) Killing vector $V^\mu$ which rotates the complex
structures. We collect a number of geometric results and give
the equivalent objects in both the AdS and 
3D foliated pictures. Background information on hyperk\"ahler geometry can be found, e.g., 
in \cite{HitchinKLR,HKLR}.

\subsection{General structure of the $\s$-model target spaces}
Let $\cM$ be a hyperk\"ahler manifold equipped with three complex
structures $(\cJ_A){}^\mu{}_\nu$ along with a U(1) isometry $V^\mu$
which acts as a rotation on them. Without loss of generality, we take
\begin{align}\label{eq_VrotatescJ}
\cL_V \cJ_1 = - \cJ_2~, \quad
\cL_V \cJ_2 = + \cJ_1~, \quad
\cL_V \cJ_3 = 0~.
\end{align}
The three K\"ahler two-forms are
\begin{align}
\Omega_A = \frac{1}{2} (\Omega_A){}_{\mu \nu} \, \rd \phi^\mu \wedge \rd\phi^\nu~, \qquad
(\Omega_A)_{\mu \nu} = g_{\mu \rho} (\cJ_A)^\rho{}_\nu~.
\end{align}
From $\Omega_1$ and $\Omega_2$ we construct the complex $(2,0)$ and $(0,2)$ forms
with respect to $\cJ_3$
\begin{align}
\Omega_\pm = \frac{1}{2} \Omega_1 \pm \frac{\ri}{2} \Omega_2~, \qquad
\cL_V \Omega_{\pm} = \pm \ri\, \Omega_{\pm}.
\end{align}
$\Omega_+$ is holomorphic with respect to $\cJ_3$.

Each of these two-forms is closed by construction. Due to the properties of the Killing
vector $V^\mu$, it turns out that complex structures $\Omega_+$ and $\Omega_-$ (and any linear
combination) are actually
exact. This is easily proven. Consider $\rho_+ := -\ri \, \imath_V\Omega_+$, which is
a holomorphic (1,0) form with respect to $\cJ_3$. It is a simple exercise to show that
$\rd \rho_+ = \Omega_+$. Similarly, $\rho_- := +\ri \, \imath_V\Omega_-$ obeys $\rd \rho_- = \Omega_-$.
It follows that $\rho_1 = \rho_+ + \rho_-$ and $\rho_2=-\ri (\rho_+ - \rho_-)$
are the potentials for $\Omega_1$ and $\Omega_2$, and they are given by
$\rho_1 = \imath_V \Omega_2$ and $\rho_2 =  -\imath_V \Omega_1$.
Note however that the third K\"ahler two-form $\Omega_3$ need not be exact.
Because some of the K\"ahler two-forms are exact, $\cM$ must be a non-compact manifold.

Because $V^\mu$ is holomorphic with respect to $\cJ_3$, we may introduce
a real Killing potential $\cK$ \cite{BWitten,HKLR}
\begin{align}\label{eq_VKillingPot}
V^\mu = \frac{1}{2} (\cJ_3)^\mu{}_\nu \nabla^\nu \cK~
\end{align}
which is defined up to a constant shift. It is straightforward to show that
$\cK$ is the K\"ahler potential with respect to $\cJ_1$ and $\cJ_2$ and indeed
any complex structure $\cJ_\perp$ which is perpendicular to $\cJ_3$. In other words,
\begin{align}
g_{\mu \nu} &= \frac{1}{2} \nabla_\mu \nabla_\nu \cK
	+ \frac{1}{2} (\cJ_\perp)_\mu{}^\rho (\cJ_\perp)_\nu{}^\sigma \nabla_\rho \nabla_\sigma \cK~.
\end{align}
Because 
$\nabla_\m \cK$ is a globally defined one-form, 
this implies that the
$\Omega_1$ and $\Omega_2$ are exact, $\Omega_1 = \rd \rho_1$
and $\Omega_2 = \rd \rho_2$ with
\begin{align}
\rho_1 = \frac{1}{2} \nabla_\mu \cK \,(\cJ_1)^\mu{}_\nu \rd \phi^\nu~, \qquad
\rho_2 = \frac{1}{2} \nabla_\mu \cK \,(\cJ_2)^\mu{}_\nu \rd \phi^\nu~.
\end{align}
These may be alternatively written
\begin{align}\label{eq_K12forms}
\rho_1 = V_\mu\, (\cJ_2)^\mu{}_\nu \rd\phi^\nu~, \qquad
\rho_2 = -V_\mu\, (\cJ_1)^\mu{}_\nu \rd\phi^\nu~.
\end{align}

Let us now suppose the space is equipped with a tri-holomorphic isometry $X^\mu$.
It follows that $X^\mu$ is associated with three distinct Killing potentials $D_{(A)}$,
\begin{align}
X^\mu = \frac{1}{2} (\cJ_A)^\mu{}_\nu \nabla^\nu D_{(A)} \qquad \textrm{(no summation on A)}~.
\end{align}
One can show that if $\cJ_{A\perp}$ is some complex structure orthogonal to $\cJ_A$,
then $D_{(A)}$ is the real part of a holomorphic function with respect to $\cJ_{A\perp}$,
\begin{align}
0 = \nabla_\mu \nabla_\nu D_{(A)} + (\cJ_{A\perp})_\mu{}^\rho (\cJ_{A\perp})_\nu{}^\sigma \nabla_\rho \nabla_\sigma D_{(A)}~.
\end{align}

Some of these Killing potentials possess elegant geometric definitions if
we specialize to the case where $[V,X] = 0$. Consider the set of real functions
\begin{align}
F_{(A)} = V^\mu (\Omega_{A}){}_{\mu \nu} X^\nu = \imath_X \imath_V \Omega_A~.
\end{align}
One can show, using Cartan's formula $\cL_X = \imath_X \, \rd + \rd \, \imath_X$ and
the identity $\cL_X \,  \imath_Y - \imath_Y \, \cL_X =\imath_{[X,Y] } $, 
 that
\begin{align}
\rd F_{(A)} = - \imath_X \cL_V \Omega_{A} \quad \Longleftrightarrow \quad
\nabla_\mu F_{(A)} = -X^\nu \cL_V (\Omega_A)_{\nu \mu}
\end{align}
or equivalently 
\begin{align}
\rd F_{(1)} = \frac{1}{2} \rd D_{(2)} ~, \qquad
\rd F_{(2)} = -\frac{1}{2} \rd D_{(1)}~, \qquad \rd F_{(3)} =0~.
\label{8.12}
\end{align}
The first two relations tell us that $D_{(1)}$ and $D_{(2)}$ can be chosen as 
\begin{align}
D_{(1)} := -2 V^\mu (\Omega_{2}){}_{\mu \nu} X^\nu ~, \qquad
D_{(2)} := 2 V^\mu (\Omega_{1}){}_{\mu \nu} X^\nu ~.
\end{align}
Remarkably, the Killing potentials for $X$ for any complex
structure orthogonal to $\cJ_3$ are purely geometric quantities
(i.e. globally defined scalar fields). 
The third relation in \eqref{8.12} means that 
$F_{(3)}$ must be constant and it is easy to
see that it must be given by\begin{align}\label{eq_F3}
F_{(3)} = \frac{1}{2} \cL_X \cK = \textrm{const}~.
\end{align}

We have seen that in models derived from projective
superspace, the constant in \eqref{eq_F3} is actually zero. 
It will be shown in subsection \ref{subsection8.3} that 
this constant is always zero. 

We may construct a new U(1) vector $V'^\mu = V^\mu + r X^\mu$ for $r \in \mathbb R$,
which also rotates the complex structures. The constant $r$ is determined by our normalization
convention for $X^\mu$. The choices we have made in earlier sections amount to
$r = -1/2s$. It is easy to see that the corresponding Killing potential
is $\cK' = \cK + r D_{(3)}$.
This is just a K\"ahler-like transformation: $\cK$ is the K\"ahler potential and $D_{(3)}$
is the real part of a holomorphic function with respect to any complex
structure orthogonal to $\cJ_3$.
Similarly, the shift in $V'^\mu$ induces gauge transformations in the one-forms
\begin{align}
\rho'_1 = \rho_1 + \frac{r}{2} \rd D_{(2)}~, \qquad
\rho'_2 = \rho_2 - \frac{r}{2} \rd D_{(1)}~.
\end{align}
Of course, $\Omega_1$ and $\Omega_2$ are unchanged by this shift.

\subsection{Geometry in the 3D foliated frame}\label{subsection8.3}
Let us now specialize to 3D foliated frame, where the complex structures are
\begin{align}
\mathbb J_1 = \left(\begin{array}{cc}
0 & \newomega^\ra{}_{\bar \rb} \\
\newomega^{\bar \ra}{}_\rb & 0
\end{array}\right)~, \quad
\mathbb J_2 = \left(\begin{array}{cc}
0 & \ri\, \newomega^\ra{}_{\bar \rb} \\
-\ri\, \newomega^{\bar \ra}{}_\rb & 0
\end{array}\right)~,\quad
\mathbb J_3 = \left(\begin{array}{cc}
\ri \,\delta^\ra{}_\rb & 0 \\
0 & -\ri \,\delta^{\bar \ra}{}_{\bar \rb}
\end{array}\right)~,
\end{align}
where $V^\mu$ acts on them as in \eqref{eq_VrotatescJ}. We may choose
$\cJ_3 = {\mathbb J}_3$. However, it is not completely obvious that
$\mathbb J_1$ and $\mathbb J_2$ should be identified with
$\cJ_1$ and $\cJ_2$. The most we can say is that they
are identified up to some rotation. Let us
assume $\cJ_1 = \mathbb J_1$ and $\cJ_2 = \mathbb J_2$.

Here the K\"ahler two-forms have the conventional form
\begin{subequations}
\begin{align}
\Omega_1 &= \frac{1}{2} \newomega_{\ra \rb} \,\rd\phi^\ra \wedge \rd\phi^\rb
	+ \frac{1}{2}\newomega_{\bar a \bar b} \,\rd\bar\phi^{\bar \ra} \wedge \rd\bar\phi^{\bar \rb}~, \\
\Omega_2 &= -\frac{\ri}{2} \newomega_{\ra \rb} \,\rd\phi^\ra \wedge \rd\phi^\rb
	+ \frac{\ri}{2} \newomega_{\bar \ra \bar \rb} \,\rd\bar\phi^{\bar \ra} \wedge \rd\bar\phi^{\bar \rb}~, \\
\Omega_3 &= -\ri g_{\ra \bar \rb} \,\rd\phi^\ra \wedge \rd\bar\phi^{\bar \rb}~,
\end{align}
\end{subequations}
with
\begin{align}
\Omega_+ = \frac{1}{2} \newomega_{\ra \rb} \,\rd\phi^\ra \wedge \rd\phi^\rb~, \qquad
\Omega_- = \frac{1}{2} \newomega_{\bar\ra \bar\rb} \,\rd\bar\phi^{\bar\ra} \wedge \rd\bar\phi^{\bar\rb}~.
\end{align}
$\Omega_+$ and $\Omega_-$ are exact (and similarly $\Omega_1$ and $\Omega_2$) with
\begin{align}
\rho_+ = H_\ra \rd\phi^\ra~,\qquad
\rho_- = \bar H_{\bar\ra} \rd\bar\phi^{\bar\ra}~.
\end{align}
These are holomorphic and anti-holomorphic with respect to ${\mathbb J}_3$.
We recall that the holomorphic Killing vector field $V^\ra (\f) $ is related to $H_\ra (\f)$ according to eq. 
\eqref{VHeqn}.

Let us now demonstrate that the constant in \eqref{eq_F3} is actually zero. 
As argued at the end of subsection \ref{subsection7.1}, we can always choose local complex coordinates 
$\f^\ra = (\F^I, \J_I)$  on $\cM$ such that $H_\ra$ and $V^\ra$ have the form
\eqref{Dar1} and \eqref{Dar2} respectively. In these Darboux coordinates, 
the hyperk\"ahler potential ${\mathbb K} (\f , \bar \f)$ is given by eqs. 
\eqref{h-master} and \eqref{h}. Moreover, for the Killing potential $\cK(\f, \bar \f)$ associated with 
$V$ we derived the explicit expression \eqref{KP7.15}. Since $\cK $ is at least quadratic in $\f$'s and $\bar \f$'s, 
the function $\cL_X \cK$ is at least linear in the superfield variables (i.e. no constant term is
present in the Taylor series for $\cL_X \cK$). As a result, the only option for the relation 
 \eqref{eq_F3} is 
 \bea
 \cL_X \cK =0~.
 \eea
If we have a tri-holomorphic Killing vector $X^\ra$, we may conventionally choose
\begin{align}
X^\ra = -\newomega^{\ra \rb} W_{\rb}
\end{align}
for some holomorphic function $W$. It follows that
\begin{align}
D_{(1)} = - 2(W + \bar W)~, \qquad
D_{(2)} &= 2\ri (W - \bar W)~,
\end{align}
where
\begin{align}
W = -\ri \,V^\ra \newomega_{\ra \rb} X^\rb~.
\end{align}

\subsection{Geometry in the AdS frame}
The complex structures for this geometry were given in \eqref{complex_structure1}
and \eqref{complex_structure2}.
However, we are free to choose a different basis for the complex
structures. To make contact with the assumptions made in eq. \eqref{eq_VrotatescJ},
let us choose $\cJ_3$ to be the invariant complex structure,
\begin{align}
\cJ_3 &= J_1 \cos\theta + J_2 \sin \theta =
\frac{1}{|\mu|}
\begin{pmatrix}
0 & \mu \omega^a{}_{\bar b} \\
\bar\mu \omega^{\bar a}{}_b & 0
\end{pmatrix}~.
\end{align}
We are free to choose $\cJ_1$ and $\cJ_2$ however we like,
provided we maintain the conditions \eqref{eq_VrotatescJ}.
The simplest choice is to let one of them be the diagonal complex
structure, say
\begin{align}
\cJ_1 = J_3 = \begin{pmatrix}
\ri \delta^a{}_{b} & 0 \\
0 & -\ri \delta^{\bar a}{}_{\bar b}        
\end{pmatrix}~.
\end{align}
The other is then found to be
\begin{align}
\cJ_2 &= J_1 \sin \theta - J_2 \cos\theta =
\frac{1}{|\mu|}
\begin{pmatrix}
0 & -\ri \mu \omega^a{}_{\bar b} \\
\ri \bar\mu \omega^{\bar a}{}_b & 0
\end{pmatrix}~.
\end{align}
It is a simple exercise to check that this choice for the complex
structures respects the quaternionic algebra.

We easily see that $V^\mu$ given by \eqref{eq_VKillingPot} is indeed
\begin{align}
V^a = \frac{\mu}{2|\mu|} \omega^{ab} \cK_b~, \qquad
V^{\bar a} = \frac{\bar\mu}{2|\mu|} \omega^{\bar a \bar b} \cK_{\bar b}~.
\end{align}
By construction, $\cK$ is the K\"ahler potential with respect to $\cJ_1$;
moreover, it must also be the K\"ahler potential with respect to any complex
structure orthogonal to $\cJ_3$. As discussed in the previous subsection,
$\cK$ is also a globally defined function.

The first K\"ahler two-form is the usual K\"ahler form,
\begin{align}
\Omega_1 &= -\ri g_{a \bar b} \rd \varphi^a \wedge \rd \bar\varphi^{\bar b}~.
\end{align}
It is exact, $\Omega_1 = \rd \rho_1$, with
\begin{align}
\rho_1 = \frac{\ri}{2} \cK_a \rd\varphi^a - \frac{\ri}{2} \cK_{\bar b} \rd\bar\varphi^{\bar b}~.
\end{align}
The second is
\begin{align}
\Omega_2 &= \frac{\ri \bar \mu}{|\mu|} \omega_{a b} \,\rd \varphi^a \wedge \rd \varphi^{b}
	- \frac{\ri \mu}{|\mu|} \omega_{\bar a \bar b} \,\rd \bar\varphi^{\bar a} \wedge \rd \bar\varphi^{\bar b}~.
\end{align}
Remarkably, this is also exact, with
\begin{align}
\rho_2 &= \frac{\ri \bar\mu}{2|\mu|} \cK_{\bar a} \omega^{\bar a}{}_{b} \rd\varphi^{b} 
	-\frac{\ri \mu}{2|\mu|} \cK_a \omega^a{}_{\bar b} \rd\bar\varphi^{\bar b}
	= - \ri V_{a} \rd \varphi^a + \ri V_{\bar b}\rd\bar\varphi^{\bar b}~.
\end{align}
However, the third K\"ahler two-form,
\begin{align}
\Omega_3 &= \frac{\bar \mu}{|\mu|} \omega_{a b} \,\rd \varphi^a \wedge \rd \varphi^{b}
	+ \frac{\mu}{|\mu|} \omega_{\bar a \bar b} \,\rd \bar\varphi^{\bar a} \wedge \rd \bar\varphi^{\bar b}~,
\end{align}
is not exact.

Suppose again that $X^a$ is a holomorphic isometry which commutes with $V^a$.
We know that there must be two Killing potentials with purely
geometric definitions. The first is
\begin{align}
D_{(1)} = -\ri X^a \cK_a + \ri X^{\bar a} \cK_{\bar a}~.
\end{align}
This is the usual Killing potential for a holomorphic
isometry since $\cJ_1$ is the diagonal complex structure.
The remaining two Killing potentials, $D_{(2)}$ and
$D_{(3)}$ must be the real part of a holomorphic field.
Because $X^a$ is tri-holomorphic, we may introduce a 
holomorphic function $W$, with
\begin{align}
X^a = -\omega^{ab} W_b~.
\end{align}
It is straightforward to show that (up to a constant)
\begin{align}
D_{(2)} = -2 \ri |\mu| \left(\frac{W}{\mu} - \frac{\bar W}{\bar \mu}\right)~,\qquad
D_{(3)} = -2 |\mu| \left(\frac{W}{\mu} + \frac{\bar W}{\bar \mu}\right)~.
\end{align}
In particular, the shift $V \rightarrow V - X / 2|\mu|$ does indeed correspond to
a redefinition of the AdS Lagrangian by $\cK \rightarrow \cK + W / \mu + \bar W / \bar\mu$.

The second Killing potential with a purely geometric definition is
\begin{align}
D_{(2)} = 2\ri X^a V_a - 2\ri X^{\bar a} V_{\bar a}~.
\end{align}
It follows that
\begin{align}
-2\ri |\mu| \left(\frac{W}{\mu} - \frac{\bar W}{\bar \mu}\right) = 2\ri X^a V_a - 2\ri X^{\bar a} V_{\bar a}~,
\end{align}
which we can interpret as a definition of the imaginary part of $W / \mu$.
There is no corresponding definition for the real part.

\section{$\cN=2$ AdS supersymmetric  $\s$-model on $T^*{\mathbb C}P^n$}\label{section9}
\setcounter{equation}{0}

For the general off-shell $\s$-model \eqref{1.25}, the two schemes to eliminate the auxiliary superfields, 
which we presented in sections 3 and 6, were purely formal. Indeed, in sections 3 and 6
we assumed that the problems of solving 
the auxiliary field equations \eqref{AuxEOMAdS} and \eqref{AuxEOM}, which originate in the AdS frame 
and in the 3D foliated frame respectively, had been solved. 
In practice, however, these equations are very difficult to solve.
In the 3D foliated frame, solutions to the equations \eqref{AuxEOM} are known for a large class
of $\cN=2$ supersymmetric $\s$-models in $\rm AdS_4$ in which the fields take their values in the cotangent 
bundles of Hermitian symmetric spaces. These solutions were briefly described in subsection 
\ref{subsection6.3}. As concerns the AdS frame, no exact solution of the auxiliary field equations 
\eqref{AuxEOMAdS} is known except for the case when $K(\F, \bar \F)$ corresponds to a flat 
K\"ahler space. Here  we consider a specific example, the $\s$-model on the  cotangent bundle of $\mathbb C P^n$,
where the problem of solving  the auxiliary field equations 
\eqref{AuxEOMAdS} can be bypassed,  and the hyperk\"ahler potential, 
$  \cK(\F, \J , \bar \F , \bar \J )$, 
in \eqref{eq_dualActionAdS} can be found by using superconformal techniques 
developed in \cite{KLvU}.  

We recall that the manifestly $\cN=2$ supersymmetric action in $\rm AdS_4$, eq. \eqref{InvarAc1}, 
can be reduced to $\cN=1$ AdS superspace \cite{KT-M-4D-conf-flat}. The result is eq. \eqref{3.1sma},
which we repeat here for convenience,
\bea
S = \int \rd^4 x \, {\rm d}^2\q\,{\rm d}^2{\bar \q}\,{E}\,\cL~, \qquad
\cL = \oint_C \frac{\rd \z}{2\pi \rm i\z}\cL^{[2]} (\z)\Big|~,
\eea
where we have introduced the Lagrangian $\cL^{ [2]} (\z )$ defined\footnote{The inhomogeneous 
complex coordinate for the north chart of ${\mathbb C}P^1$,  $\z$, is defined as usual: $v^i = v^{\1}(1, \z)$.} by
\bea
\cL^{(2)} (v) = \ri (v^{\1})^2 \z \cL^{[2]} (\z )
\eea
and made use of 
the bar-projection \eqref{bar-projection}. 

Let us  consider a special  $\cN=2$ supersymmetric field theory 
in $\rm AdS_4$ describing a tensor multiplet $\cH^{(2)} $, 
\bea
\cH^{(2)}(v) = \breve{\cH}{}^{(2)} (v)
=\cH^{ij} v_i v_j ~, \qquad \cD^{(k}_\a \cH^{ij)} = \bar \cD^{(k}_\ad \cH^{ij)}=0~,
\eea 
coupled to a system of weight-zero arctic multiplets $\U^I$ and their smile-conjugates $\breve{\U}^{\bar I}$.
The Lagrangian is
\bea
\cL^{(2)}=  \cH^{(2)}  {K}({ \U}, \breve{ \U})~,
\label{6.4}
\eea
where ${K}(\F^I, {\bar \F}^{\bar J}) $ is the K\"ahler potential of a real analytic K\"ahler manifold $\cX$.
If we freeze the tensor multiplet, 
\bea
 \cH^{(2)} ~\longrightarrow~ \frac{1}{2s}  \cS^{(2)}
 \label{6.5}
\eea
then the Lagrangian reduces to that describing the  $\s$-model  \eqref{1.25}. 
Note that the action is invariant under the K\"ahler transformations \eqref{1.26}.

Upon projection to $\cN=1$ AdS superspace, $\cH^{[2]} (\z)$ reads\footnote{In what follows, 
we do not indicate the bar-projection.}
\bea
\cH^{[2]} (\z) = \frac{1}{\z}\, \c + G - \z \,{\bar \c}~, \qquad {\bar \cD}_{\ad} \c =0~, 
\qquad ({\bar \cD}^2 -4\m)G =0~, \quad {\bar G}=G~.
\label{tensor-series}
\eea
For the arctic multiplets we get  
\be
 \U^I (\z) = \sum_{n=0}^{\infty}  \, \z^n \U_n^I  = 
\F^I + \z \, \S^I  + O(\z^2) ~,\qquad
{\bar \cD}_{{\ad}} \F^I =0~, \quad ({\bar \cD}^2 -4\m)\S^I = 0 ~.
\ee
The $\cN=1$ AdS superfields $\U^I_2$, $\U^I_3, \dots $, are complex unconstrained.

The theory \eqref{6.4} is $\cN=2$ superconformal. Since  ${\rm AdS^{4|8} } $ is conformally flat, 
the theory can be re-formulated in $\cN=2$ Minkowski superspace where its Lagrangian
has essentially the same form, $L^{(2)}=  H^{(2)}  {K}({ \U}, \breve{ \U})$, but 
the supermultiplets $H^{(2)}$, $\U$ and $\breve \U$ are projective with respect to the
flat covariant derivatives.
In Minkowski superspace, the auxiliary superfields have been eliminated in \cite{KLvU} 
for one particular case of $\cX$ -- the complex projective space ${\mathbb C}P^n  $.
Here we can use the flat-superspace results of \cite{KLvU} and then lift them to AdS 
using the consideration of superconformal invariance. This will allow us  to obtain 
a formulation in terms of $\cN=1$ chiral superfields
for the $\s$-model  \eqref{1.25} in the case $\cX ={\mathbb C}P^n  $. 
In other words, we can use the same superconformal model to derive the hyperk\"ahler 
potential for $T^*{\mathbb C}P^n  $ both in the AdS frame and in the 3D foliated frame 
by making a different choice of $\mathcal S^{[2]}$,
\begin{align}
\cH^{[2]}= \frac{1}{2s} \mathcal S^{[2]} =
\begin{cases}
\dfrac{\ri \mu}{2|\mu|} \dfrac{1}{\zeta} + \dfrac{\ri \bar\mu}{2|\mu|} \zeta~, & \text{AdS frame} ~;\\
1~, & \text{3D foliated frame}~.
\end{cases}
\end{align}

Using standard inhomogeneous  coordinates for ${\mathbb C}P^n  $,
the K\"ahler potential and the metric are
\be 
K (\F, {\bar \F}) = r^2 \ln \left(1 + \frac{1}{r^2}  
\F^L \overline{\F^L} \right)
~,~~~
g_{I {\bar J}} (\F, \bar \F) =  
\frac{ r^2 \d_{I {J}} }{r^2 + \F^L \overline{\F^L} }
- \frac{  r^2   \overline{\F^I} \F^J  }
{(r^2 + \F^L \overline{\F^L})^2 }~,
\label{s2pot}
\ee
where $I,\bar{J}=1,\dots, n$ and $r^2={\rm const}$.
We recall that the Riemann curvature of  ${\mathbb C}P^n$ is 
\bea
R_{I_1 {\bar  J}_1 I_2 {\bar J}_2}  &:=& 
K_{I_1 {\bar  J}_1 I_2 {\bar J}_2}  
- g_{M \bar N} \G^M_{I_1I_2} {\bar \G}^{\bar N}_{{\bar J}_1{\bar J}_2} 
= -\frac{1}{r^2} \Big\{ g_{I_1 {\bar J}_1 } g_{I_2 {\bar J}_2 }
+g_{I_1 {\bar J}_2 } g_{I_2 {\bar J}_1 } \Big\}~,
\eea
and hence
\bea
\S^{I_1} {\bar \S}^{ {\bar J}_1 } \S^{I_2}\,
R_{I_1 {\bar J}_1 I_2 {\bar J}_2}  
=-\frac{2}{r^2}  g_{I_2 {\bar J}_2}   \S^{I_2}  |\S|^2~, 
\qquad |\S|^2 := g_{I \bar J}(\F, {\bar \F})\, \S^{I} {\bar \S}^{ {\bar J} } ~.
\label{curvature}
\eea

Upon elimination of the auxiliary superfields, 
the  Lagrangian becomes\footnote{We view the tensor multiplet as a background field.} 
\cite{KLvU}
\bea
&&\cL(\F, \bar \F , \S, \bar \S)  = G\,K(\Phi, {\bar \Phi})
+ \c K_I (\Phi, {\bar \Phi})\Sigma^I 
+\bar\c K_{\bar J} (\Phi, {\bar \Phi})\bar \Sigma^{\bar J}
\label{L-H} \non \\
&&\qquad+ r^2 \Big\{ 
G\, \ln \frac{ 1 - |\S|^2/r^2 } 
{ \sqrt{ G^2 + 4{\bar \c} \c ( 1 - |\S|^2/r^2 )} +G }
+  \sqrt{ G^2 + 4{\bar \c} \c ( 1 - |\S|^2/r^2 )} 
\Big\} \non \\
&&\qquad- r^2  \Big\{     {\mathbb H} - G \, \ln \big( G+  {\mathbb H}\big) \Big\} ~,
\eea
where
\bea
{\mathbb H}:= \sqrt{ G^2 +4\c \bar \c }~.
\eea

The theory (\ref{L-H})
possesses a dual formulation obtained by dualizing the complex linear tangent 
variables $\S^I$ and their conjugates
${\bar \S}^{\bar I}$ into chiral superfields $\J_I$ and their conjugates 
${\bar \J}_{\bar I}$, $ {\bar \cD}_{ \ad} \J_I =0$.
One first replaces the action with a first order one, 
\bea 
S= \int  {\rm d}^4 x \, {\rm d}^2\q\,{\rm d}^2{\bar \q}\,{E}
\,\Big\{ \cL(  \F , {\bar \F}, \S, {\bar \S})  
+\S^I \J_I +{\bar \S}^{\bar J} {\bar \J}_{\bar J}
\Big\}~,
\label{6.13}
\eea
where $\S^I $ and ${\bar \S}^{\bar J} $ are chosen to be complex unconstrained.
Next, one eliminates these superfields with the aid of their algebraic 
equations of motions,  ending  up with the dual  Lagrangian \cite{KLvU}:
 \bea
&&{\bf K}(\F,  \J , \bar \F, \bar \J) = G\,K(\Phi, {\bar \Phi})
- r^2  \Big\{     {\mathbb H} - G \, \ln \big( G+  {\mathbb H}\big) \Big\} 
\non \\
&&\quad+ r^2\Big\{   \sqrt{ {\mathbb H}^2 + 4 | \J + \c \nabla K |^2/r^2 } 
- G\, \ln \Big(   \sqrt{ {\mathbb H}^2 + 4 | \J + \c \nabla K |^2/r^2 } +G \Big) \Big\}~,~~~~~~~
\label{Calabi}
\eea
where
\bea 
 | \J + \c  \nabla K |^2 := g^{I{\bar J}} \Big( \J_I +\c K_I (\Phi, {\bar \Phi})\Big) 
 \Big( {\bar \J}_{\bar J} +{\bar \c} K_{\bar J}(\Phi, {\bar \Phi}) \Big) ~. 
 \eea
Under the K\"ahler transformation (\ref{1.26}), the chiral  one-form $\J_I$ changes as
\be
\J_I \quad \longrightarrow \quad \J_I - \c \,F_I (\F)~,  
\ee
and this transformation is clearly consistent with the chirality of $\J_I$.
The reason for the non-invariance of $\J_I$ is that the terms 
\bea
\int  {\rm d}^4 x \, {\rm d}^2\q\,{\rm d}^2{\bar \q}\,{E}
\,\Big\{\c K_I (\Phi, {\bar \Phi})\Sigma^I 
+\bar\c K_{\bar J} (\Phi, {\bar \Phi})\bar \Sigma^{\bar J}\Big\}
\eea
in \eqref{6.13} are not invariant under the K\"ahler transformations when $\S^I $ is complex unconstrained.

In the limit $G=1$ and $\c =0$,\footnote{It should be kept in mind that the  limit $G=1$ and $\c =0$
cannot be performed in the AdS frame, but instead only in the 3D foliated frame.}
the Lagrangian (\ref{Calabi})  
reduces to the standard hyperk\"ahler potential for the cotangent bundle of ${\mathbb C}P^n$, 
see e.g. \cite{AKL1}, 
\begin{align}\label{eq_CPn3D}
{\mathbb K}(\F,  \J , \bar \F, \bar \J) &= K(\Phi, {\bar \Phi})
	+ r^2 \Big( \sqrt{1 + 4 | \J|^2/r^2 } -1\Big) 
	- r^2 \ln  \frac{  \sqrt{1 + 4 | \J|^2/r^2 } + 1 }{2}~.
\end{align}
This is the hyperk\"ahler potential appearing in the 3D foliated  action \eqref{SDarboux}
with Darboux coordinates $\Phi^I$ and $\Psi_I$. It is invariant under 
a U(1) Killing vector field $V$ of the standard form \eqref{Dar2}. The complex coordinates $\Phi^I$
and $\Psi_I$ naturally diagonalize the preferred complex structure $\mathbb J = \cJ_3$
with respect to which $V$ is holomorphic.

We are actually interested in a different limit, $G=0$ and $\c = \ri \m /2|\m|$, 
which gives us the formulation in terms of $\cN=1$ chiral superfields
 for the $\s$-model  \eqref{1.25}. Implementing this limit gives 
\bea
\cK (\F,  \J , \bar \F, \bar \J ) = r^2\  \sqrt{  1 +  \frac{4}{r^2}\Big| \J + \ri \frac{\m}{2 | \m |}  \nabla K (\F, \bar \F)
\Big|^2 } -r^2~,
\eea
with a \emph{different} set of Darboux coordinates $\Phi^I$ and $\Psi_I$,
which diagonalize a different complex structure $J_3 = \cJ_1$.
The Lagrangian is globally defined on $T^*{\mathbb C}P^n$.
The corresponding  U(1) Killing
vector field \eqref{eq_VAds} is not holomorphic with respect to the diagonalized $J_3 = \cJ_1$,
but rather with respect to $\mathbb J = \cJ_3$. This should be compared with eq. \eqref{7.18} which is 
the expression for $\cK$ in terms of the complex coordinates which diagonalize $\mathbb J=\cJ_3$.

\section{Conclusions}
\setcounter{equation}{0}
 
One of the  important findings of this work is the observation that 
the most general $\cN=2$ supersymmetric $\s$-model in $\rm AdS_4$ can be described 
in terms of an off-shell $\s$-model in projective superspace given by eqs. \eqref{InvarAc1}
and \eqref{1.25}. This model is associated with a real analytic K\"ahler manifold $\cX$ with 
K\"ahler potential $K(\F^I, \bar \F^{\bar J})$ which appears in  \eqref{1.25}.
As demonstrated above, there are two ways to relate this off-shell formulation, 
realized in terms of covariant weight-zero arctic multiplets, to a formulation in terms of ordinary chiral superfields:
(i) using the AdS frame; and (ii) using the 3D foliated frame. In the AdS frame, one ends up 
with the $\s$-model \eqref{4DN1action} in which the Lagrangian $\cK(\vf^a, \bar\vf{}^{\bar b})$
is a globally defined function over the hyperk\"ahler target space $\cM$ such that $\cK$ is the K\"ahler 
potential, $g_{a\bar b} = \pa_a \pa_{\bar b} \cK$,  
with respect to any complex structure orthogonal to the preferred one $\mathbb J$, 
eq.  \eqref{eq_JAdS}, which is invariant under the  Killing vector $V$ rotating the complex structure. 
The covariantly chiral superfields $\vf^a $ in \eqref{4DN1action} are complex coordinates with respect to 
a certain complex structure,  $ J_3$, orthogonal to $\mathbb J$.
In the 3D foliated frame, one ends up with the $\s$-model \eqref{SDarboux} in which
the Lagrangian is
\bea
{\mathbb K} (\f^\ra , \bar \f^{\bar \rb} ) \equiv
{\mathbb K} (\F^I ,  \J_I, \bar \F^{\bar J},  \bar \J_{\bar J}) = K \big( \F, \bar{\F} \big)+    
\cH \big(\F,  \J ,\bar \F, \bar \J \big)~, 
\label{10.1}
\eea
where
\bea
\cH \big(\F,  \J , \bar \F,  \bar \J \big)&=& 
\sum_{n=1}^{\infty} \cH^{I_1 \cdots I_n {\bar J}_1 \cdots {\bar 
J}_n }  \big( \F, \bar{\F} \big) \J_{I_1} \dots \J_{I_n} 
{\bar \J}_{ {\bar J}_1 } \dots {\bar \J}_{ {\bar J}_n } ~.
\label{10.2}
\eea
Here $\cH^{I {\bar J}}  = g^{I {\bar J}}  $ and 
the coefficients $\cH^{I_1 \cdots I_n {\bar J}_1 \cdots {\bar 
J}_n }$, for  $n>1$, 
are tensor functions of the K\"ahler metric
$g_{I \bar{J}} \big( \F, \bar{\F}  \big) 
= \pa_I 
\pa_ {\bar J}K ( \F , \bar{\F} )$ on $\cX$,  
the Riemann curvature $R_{I {\bar J} K {\bar L}} \big( \F, \bar{\F} \big) $ and its covariant 
derivatives. The superfield Lagrangian ${\mathbb K} (\f^\ra , \bar \f^{\bar \rb} ) $
is the K\"ahler potential of $\cM$ in complex coordinates $\f^\ra$ with respect to the 
preferred complex structure $\mathbb J$. 
Associated with ${\mathbb K} (\f^\ra , \bar \f^{\bar \rb} ) $ is 
the globally defined function of $\cM$
\bea
{\cK} (\f^\ra , \bar \f^{\bar \rb} ) = 2 \J_I \frac{\pa}{\pa \J_I } \cH \big(\F,  \J , \bar \F,  \bar \J \big)
~,
\eea
which is the Killing potential for the Killing vector $V$  (which is holomorphic with respect to $\mathbb J$). 
The function  ${\cK} (\f^\ra , \bar \f^{\bar \rb} ) $ coincides with the superfield Lagrangian 
$\cK(\vf^a, \bar\vf{}^{\bar b})$ in  \eqref{4DN1action}, however they are written down in terms of 
different coordinates for $\cM$. The former is given in terms of the complex coordinates
with respect to $\mathbb J$, while the latter is defined in terms of 
the variables $\vf^a$ which are  complex coordinates 
with respect to the orthogonal complex structure  $J_3$.

In conjunction with the results of \cite{BKsigma1,BKsigma2}, 
our work shows that there is a one-to-one correspondence between $\cN=2$ supersymmetric 
$\s$-models in $\rm AdS_4$ and those hyperk\"ahler manifolds which possess a Killing vector field 
generating an SO(2) group of rotations on the two-sphere of complex structures.
This clearly differs from  $\cN=2$ Poincar\'e supersymmetry
where  arbitrary hyperk\"ahler manifolds can originate
as target spaces of $\cN=2$ supersymmetric $\s$-models \cite{A-GF,HKLR}.
The difference between the $\s$-model target spaces which are allowed by 
$\cN=2$ Poincar\'e and AdS supersymmetries
can nicely be  demonstrated in terms of the most general off-shell 4D $\cN=2$ 
supersymmetric $\s$-model in  {\it flat} projective superspace ${\mathbb R}^{4|8}\times {\mathbb C}P^1$ 
formulated in \cite{LR88}. The action is
\bea
S=\oint_C \frac{\rd \z}{ 2\p \ri \z}
\int \rd^4 x\, {\rm d}^2 \q\,{\rm d}^2\bar \q\,  
{L}({ \U}^I, \breve{ \U}^{\bar J} , \z)~,
\eea
where the Lagrangian is an essentially arbitrary function of  its arguments. 
As shown in \cite{LR2008},  ${L}({ \U}^I, \breve{ \U}^{\bar J} , \z)$ has a geometric origin and can be defined
for any hyperk\"ahler manifold. 
The target space of this $\cN=2$ supersymmetric $\s$-model in Minkowski space 
can at the same time originate as the  target space of some $\cN=2$ supersymmetric $\s$-model 
in $\rm AdS_4$ only if the Lagrangian has no explicit $\z$-dependence, 
\bea
{L}({ \U}^I, \breve{ \U}^{\bar J} , \z) ~\to ~ {K}({ \U}^I, \breve{ \U}^{\bar J} )~.
\eea

In the case of Minkowski space, it is well known \cite{A-GF2,HKLR} that adding a  superpotential to 
an $\cN=2$ supersymmetric $\s$-model requires the target space to possess a tri-holomorphic 
Killing vector field.
We have found an additional restriction in the AdS case: this tri-holomorphic Killing vector 
must commute  with the Killing vector $V$ which rotates the complex structures.

Many results of our work can be naturally extended to five dimensions. Within the projective-superspace setting,
general off-shell $\s$-models in 5D $\cN=1$ AdS superspace were formulated in \cite{KT-M}. 
A 5D analogue of the 3D foliated frame was developed in \cite{KT-M_5D_conf-flat}.
One can repeat the analysis of section 6 for the case of  the off-shell $\cN=1$ $\s$-models in $\rm AdS_5$
proposed in \cite{KT-M}. The results of such an analysis will be the most general 
5D $\cN=1$ supersymmetric $\s$-model in $\rm AdS_5$ realized in terms of 4D $\cN=1$ chiral superfields
\cite{BaggerXiong,BaggerLi}.

There still remain a number of interesting open questions. In particular, in our discussion of gauged
$\cN=2$ supersymmetric $\s$-models in $\rm AdS_4$  the vector multiplet was chosen to be intrinsic, 
since our goal was to derive the
superpotential generated. 
It is of interest to study the general structure of gauged $\cN=2$ supersymmetric $\s$-models in $\rm AdS_4$. 
This will be reported in a separate publication \cite{BKprogress}. 

Another interesting issue, which we only briefly touched upon in section \ref{subsection7.1},
was that the supersymmetry algebra of the general 3D foliated $\s$-model closes off-shell,
which is quite distinct from the Minkowski case \cite{HKLR}. 
It was shown in \cite{BKsigma1, BKsigma2} that the general
$\cN=2$ supersymmetric $\s$-model in the AdS frame also has a closed algebra, with the SO(2) generator of AdS
mimicking the action of a central charge. It is unsurprising that the 3D foliation should
have the same feature, and it would be interesting to develop an off-shell Fayet-Sohnius
$\cN=2$ superfield formulation (as in the AdS frame \cite{BKsigma2}) for the 3D foliated
frame.

One last question regards the two choices of $s^{ij}$ we have made, eq. \eqref{1.3b}, 
which led to the AdS and 3D foliated frames. At the level of the
hyperk\"ahler target space, these two frames are related by a
non-holomorphic coordinate transformation which effects a rotation
on the complex structures. At the same time, this coordinate
transformation acts as a simple SU(2) rotation on the original
projective multiplets which defined the action. The explicit link
between these two operations remains unexplored. We expect that
for the wide class of symmetric spaces studied in
\cite{GK1,GK2,AN,AKL1,AKL2,KN}, it should be possible to construct
this coordinate transformation explicitly.
\\

\noindent
{\bf Acknowledgements:}\\
We are grateful to Martin Ro\v{c}ek for useful discussions and for sharing his insights with us. 
SMK, UL and GT-M are grateful
to the program {\it Geometry of Strings and Fields} at Nordita (November, 2011)
where part of this work was
carried out, for providing a stimulating atmosphere.
SMK is grateful to the  IGA/AMSI Workshop {\it The Mathematical Implications of Gauge-String Dualities}
at Adelaide University (March, 2012) where part of this work was
carried out, for hospitality.
GT-M thanks the School of Physics at the University of Western Australia for the kind hospitality and support during part of this work.
The work of DB and SMK  is supported in part by the Australian Research Council. 
The work of UL was supported by VR-grant 621-2009-4066.
The work of GT-M is supported by the European Commission, Marie Curie Intra-European Fellowships 
under contract No. PIEF-GA-2009-236454.

\appendix

\section{Killing vectors}\label{app_KV}
\setcounter{equation}{0}

Within the formulation of $\cN=2$ conformal supergravity used in this
paper \cite{KLRT-M1}, a superconformal Killing vector consists of a superspace
diffeomorphism and structure group transformation encoded
in the parameter
\begin{align}
\xi = \xi^\cA \cD_\cA + \frac{1}{2}\l^{cd} M_{cd} + \l^{ij} J_{ij}~,
\end{align}
along with a super-Weyl transformation\footnote{Within \cite{KLRT-M1}, the super-Weyl
parameter was denoted $\s$. Here we use $\S$ to avoid confusion with the
finite super-Weyl transformation $\s$ connecting the AdS frame to the Minkowski frame.}
associated with a chiral superfield $\Sigma$, so that the covariant derivatives are invariant,
\begin{align}
\delta_\xi \cD_\cA + \delta_\Sigma \cD_\cA = 0~.
\end{align}
If the covariant derivatives $\cD_\cA$ are associated with an AdS geometry,
then the AdS Killing vectors are those with $\Sigma=0$, i.e.
$\delta_\xi \cD_\cA = 0$. For such a geometry, the properties that
the Killing vector $\xi$ must obey have been worked out in
detail \cite{KT-M-4D-conf-flat}. For our purposes, the relevant
features are that all the parameters can be derived from
$\xi_{\a \ad}$, which must obey the relations
\begin{subequations}\label{AdSKilling}
\begin{gather}\label{eq_AdSMaster}
\cD_{(\beta}^i \xi_{\alpha) \ad} = \bar\cD^{(\bd}_i \xi^{\ad) \alpha} = 0~, \\
\cD_\beta^j \bar\cD_{\bd j} \xi^{\bd \b} = \bar\cD_{\bd j} \cD_\beta^j \xi^{\bd \b} = 0~.
\end{gather}
\end{subequations}
The first condition, known as the master equation, holds for a superconformal Killing vector,
while the second imposes the additional requirement that the vector is AdS Killing.
Together, it is easy to show that these imply the superspace version of the
usual Killing equation
\begin{align}
\cD_a \xi_b + \cD_b \xi_a = 0~.
\end{align}
The remaining parameters $\xi^\a_i$, $\l^{ab}$ and $\l^{ij}$ can be derived
from $\xi_a$ and are given in \cite{KT-M-4D-conf-flat} (see also
\cite{BKsigma2}). For example, one finds that $\l^{ij} = 2 \ve \cS^{ij}$
for some real superfield $\ve$.

Our interest is in the 3D foliated version of AdS, which we constructed
explicitly in section \ref{geometry_N=2_AdS} in terms of
a chiral superfield $\s$ given in \eqref{2.45} with the choice
\eqref{eq_sijvalues} and $\a=-\ri$. In section A.1, we explicitly
construct a solution for the AdS Killing vectors using $\s$.
Then in section A.2, we perform the rotation described in section \ref{section5.2}
and give a new form of the Killing vectors relevant for the 3D $\cN=2$
superspace used in sections \ref{section6} and \ref{section7}.

\subsection{Killing vectors in 3D foliated AdS}
As discussed in section \ref{geometry_N=2_AdS}, the flat Minkowski derivatives $D_\cA$
are related to the AdS covariant derivatives $\cD_\cA$ by a super-Weyl transformation.
Superconformal Killing vectors on both spaces are similarly related.
The easiest way to derive the relation is to consider a superconformal Killing
isometry acting on a scalar function $\cF$ of vanishing super-Weyl weight. Within
AdS, this isometry is
\begin{align}
\delta \cF = -\xi^\cA \cD_\cA = -\xi^\cA E_\cA \cF
\end{align}
while in the Minkowski frame
\begin{align}
\delta \cF = -\widetilde \xi^\cA D_\cA \cF~.
\end{align}
Equating the two results, one easily finds
\begin{align}
\widetilde \xi_{\alpha \ad} = \xi_{\alpha \ad} \,\re^{(\s + \bar\s)/2}~, \qquad
\widetilde \xi^\alpha_i = \xi^\alpha_i e^{\bar\s/2}
	+ e^{(\s + \bar \s)/2} \left(\frac{\ri}{4} \xi^\alpha{}_\bd \bar D^\bd_i \bar\s \right)
\end{align}
where $\s$ is the super-Weyl transformation connecting the AdS
frame to the Minkowski frame. One can show that if $\xi_\cA$ is
a superconformal Killing vector in AdS, $\widetilde\xi_\cA$ must be
a superconformal Killing vector in Minkowski, and vice-versa.

Now we specialize to the case where $\xi_\cA$ is an AdS Killing
vector, obeying both equations \eqref{AdSKilling}. One can show that
$\widetilde\xi_\cA$ obeys
\begin{subequations}
\begin{gather}\label{confKillingMink}
D_{(\beta}^i \widetilde\xi_{\alpha) \ad} = \bar D^{(\bd}_i \widetilde\xi^{\ad) \alpha} = 0~,\\
D_\b^j \bar D_{\bd j} \widetilde \xi^{\bd \b} = 16 \ri \partial_a \bar\sigma \widetilde \xi^a
	- 2 \bar D^\ad_j \bar\sigma D^{\a j} \widetilde \xi_{\a \ad}~. \label{folKillingMink}
\end{gather}
\end{subequations}
Given these equations, it is straightforward to construct explicitly
the solution for $\widetilde\xi_{\alpha \ad}$, from which all the other
parameters can be constructed. The general
solution to the first equation \eqref{confKillingMink} is the general
$\cN=2$ superconformal Killing vector in Minkowski, and is given by
\begin{subequations}
\begin{align}
\widetilde\xi_i^\a &= \e^\a_i + \bar r \theta^a_i
	- \theta_i^\b \omega_\b{}^\a
	- \Lambda_i{}^j \theta^\a_j
	+ \theta_i^\b k_{\b \bd} x_L^{\bd \a}
	- \ri \bar \eta_{i \bd} x_L^{\bd \a}
	- 4 \theta^\b_i \eta_\b^k \theta^\a_k \\
\widetilde\xi^a &= \frac{1}{2} (\widetilde\xi_L^a + \widetilde\xi_R^a)
	- \ri \widetilde\xi_k \sigma^a \bar\theta^k
	- \ri \widetilde{\bar\xi}{}^k \ts^a \theta_k \\
\widetilde\xi_L^{\ad \a} &= p^{\ad \a} + (r + \bar r) x_L^{\ad \a}
	- \bar \omega^{\ad}{}_{\bd} x_L^{\bd \a}
	- x_L^{\ad \b} \omega_{\b}{}^{\a} 
	+ x_L^{\ad \b} k_{\b \bd} x_L^{\bd \a}
	+ 4\ri \bar\e^{\ad k} \theta_k^\a
	- 4 x_L^{\ad \b} \eta_\b^k \theta^\a_k
\end{align}
\end{subequations}
where $x_L^a := x^a + \ri \q_j \sigma^a\bar \q^j$.
The last equation above can be rewritten
\begin{align}
\widetilde\xi_L^{a} &= p^{a} + (r + \bar r) x_L^{a}
	+ \omega^a{}_b x_L^b
	-2 x_L^b k_b x_L^a
	+ x_L^2 k^a
	- 2\ri \bar\e^{k} \ts^a\theta_k
	- 2 x_L^b \eta^k \sigma_b \ts^a \theta_k~.
\end{align}
The constant parameters $\e^\a_i$ and $p_a$ are the
supersymmetry and spacetime translation parameters,
$\omega^a{}_b$ is the constant Lorentz parameter, and
$\Lambda_i{}^j$ is the SU(2) parameter.
The real and imaginary parts of $r$ give constant
dilatation and U(1) transformations. Finally,
$k_a$ and $\eta_\a^i$ are the special conformal
and $S$-supersymmetry parameters.

The second condition \eqref{folKillingMink} imposes
restrictions on some of these constant parameters.
One finds for the bosonic parameters
\begin{align}
p_z = 0~,\qquad \omega^z{}_b = 0~, \qquad \Lambda^{\1\1} = \Lambda^{\2\2} = 0~,\qquad
r = \bar r ~, \qquad k_z = 0~.
\end{align}
The first two constraints eliminate translations in the $z$ direction
as well as rotations mixing $z$ with the 3D coordinates.
The third constraint arises from $\Lambda^{ij} \propto s^{ij}$
as usual in AdS. The fourth eliminates the global U(1)
freedom while the last eliminates special conformal transformations
in the $z$ direction.
The constraints we find on the fermionic parameters are\footnote{Here we 
make the choice $\alpha=-\ri$ as in the main body of the paper.}
\begin{align}\label{KillingFermConstraints}
\e_{\a \1} = \ri \bar \e^\2_\a~,\qquad
\e_{\a \2} = \ri \bar \e^\1_\a~, \qquad
\eta^{\a \1} = \ri \bar \eta_\2^\a~,\qquad
\eta^{\a \2} = \ri \bar \eta_\1^\a~.
\end{align}
In addition to the parameters $\widetilde\xi^A$, we will also need the
explicit form of the SU(2) parameter $\widetilde\lambda_{ij}$, which is given by
\begin{align}
\widetilde \lambda_{ij} = \frac{1}{2} D_{\beta (i} \widetilde\xi^\b_{j)} = -\Lambda_{ij} - 4 \q_{(i} \eta_{j)} + 4 \bar\q_{(i} \bar\eta_{j)} + 4\ri k_\hb \q_{(i} \s^\hb \bar\q_{j)}~.
\end{align}
One could easily derive the Lorentz parameter $\widetilde\l^{ab}$, but we have
no need for it.

\subsection{A new basis for the Killing vectors}
In section \ref{section5.2}, we introduced a new basis for the spinor covariant
derivatives. The Killing vectors in the new basis, which we denote
$\bm\xi^\cA$, can be found by identifying
\begin{align}
\widetilde \xi^a \partial_a + \widetilde \xi^\a_i D_\a^i + \bar{\widetilde\xi}{}_\ad^i \bar D^\ad_i
= \bm\xi^\ha \partial_\ha + \bm\xi^z \partial_z + {\bm\xi}^\a_i {\bm D}_\a^i + \bar{\bm\xi}_\a^i \bar{\bm D}^\a_i
\end{align}
where the index $\ha$ is restricted now to run over the 3D coordinates.
A direct identification of these two formulae leads to\footnote{We take $\l=\ri$
as in the main body of the paper.}
\begin{align}
\bm \xi^\ha := \widetilde\xi^\ha~,\quad \bm\xi^z := \widetilde\xi^2 ~,\quad
{\bm\xi}^\a_i :=\frac{1}{\sqrt{2}}\left(\widetilde\xi^\a_i-\ri \bar{\widetilde\xi}{}^\a_i\right)~, \quad
{\bm{\bar \xi}}_\a^i:=\frac{1}{\sqrt{2}}\left(\bar{\widetilde\xi}{}_\a^i-\ri\widetilde\xi_\a^i\right)~.
\end{align}
In constructing the Killing vectors in this rotated basis, we should also
rotate the constant fermionic parameters. Defining
\begin{align}
\bm \e_i^\a := \frac{1}{\sqrt 2} (\e_i^\a - \ri \bar\e_i^\a)~,\qquad
\bm \eta^{\a i} := \frac{1}{\sqrt 2} (\eta^{\a i} + \ri \bar\eta^{\a i})~,
\end{align}
we find that the constraints \eqref{KillingFermConstraints} amount to
\begin{align}
\bm \e_\2^\a = 0~,\qquad \bm \eta^{\a \2} = 0~.
\end{align}

In this transformed basis, a projective multiplet of weight zero transforms as
\begin{align}\label{deltaQfull}
\delta \cQ = - \left(
	\bm\xi^\ha \partial_\ha
	+ \bm\xi^z \partial_z
	+ {\bm\xi}^\a_i {\bm D}_\a^i
	+ \bar{\bm\xi}_\a^i \bar{\bm D}^\a_i
	+ \left(\widetilde\lambda_{\1\1} + 2 \widetilde\lambda_{\1\2} \zeta + \widetilde\lambda_{\2\2} \zeta^2\right)\partial_\zeta \right)\cQ~.
\end{align}
It is possible to work out the new Killing vectors explicitly,
but we need only their projection to $\bm\q_\2 = \bar{\bm \q}^\2 = 0$.
It turns out that $\widetilde\lambda_{\1\1}| = \widetilde\lambda_{\2\2}| = 0$.
The remaining pieces are defined as
\begin{align}
\xi^\ha := \bm\xi^\ha |~, \qquad
\xi^z := \bm\xi^z |~, \qquad
\xi^\a := {\bm\xi}^\a_\1 |~, \qquad
\rho^\a := {\bm\xi}^\a_\2 |~, \qquad
\Lambda := \ri \widetilde\lambda_{\1\2} |~.
\end{align}
We also use abbreviated notation for the constant parameters
and for the residual $\bm\q_\1$ and $\bar{\bm \q}^\1$ coordinates:
\begin{align}
\e^\a := \bm\e^\a_\1~, \qquad
\eta^\a := \bm\eta^{\a\1}~, \qquad
\bm\q = \bm\q_\1~.
\end{align}
In terms of these, one finds\footnote{The 4D sigma matrices used here can be replaced
with 3D gamma matrices. The result is given in \eqref{3DKilling} and \eqref{3DKillingSecond}.}
\begin{subequations}
\begin{align}
\xi^\ha &= p^\ha + \omega^\ha{}_\hb x^\hb
- \frac{\ri}{2} \ve^{\ha \hb \hc} \omega_{\hb \hc} \bm\q \bar{\bm\q}
- 2\ri\, \bar{\e} \ts^\ha \bm\theta
- 2\ri\, {\e} \sigma^\ha \bar{\bm\theta}
+ 2 r x^\ha 
+ 2\ri \Lambda_\1{}^\1 {\bm\theta} \sigma^\ha \bar{\bm\theta}
\eol & \quad
- 2 x^\hb k_\hb x^\ha + x^2 k^\ha - 2\ri \ve^{\ha \hb \hc} k_\hb x_\hc \bm\theta \bar{\bm\theta}
+ z^2 k^\ha 
-\hf \bm\theta^2\bar{\bm\theta}^2 k^\ha
\eol & \quad
- 2 x^\hb (\eta \sigma_\hb \ts^\ha \bm\theta)
- 2 x^\hb (\bar{\eta} \ts_\hb \sigma^\ha \bar{\bm\theta})
-\ri \,(\eta \sigma^\ha \bar{\bm\q}){\bm\q}^2
+\ri \,(\bm\q \sigma^\ha \bar{\eta})\bar{\bm\q}^2 ~,\\
\xi^\a &=
\e^\a
- \frac{1}{2} \omega^{\hb \hc} (\bm\q \sigma_{\hb \hc})^\a
- \Lambda_\1{}^\1 \bm\q^\a
+ k^\hb x^\hc (\bm\q \sigma_\hb \ts_\hc)^\a
- \frac{\ri}{2} k_\hb {\bm\q}^2{\bm\qb}^\b (\sigma^\hb)_{\b}{}^{ \a} 
\eol & \quad
- \ri x^\hb \bar{\bm \eta}_\b (\ts_\hb)^{\b \a}
+ (\bm\q \bar {\bm \q}) \bar{\eta}^\a
+\eta^\a \bm\q^2 
+ 2 \bm\q^\b \bar{\eta}_\b \bar{\bm\q}^\a~, \\
\xi^z &= 2 r z - 2 x^\hb k_\hb z + 2 (\eta \bm\theta + \bar{\eta} \bar{\bm\theta}) z~, \\
\rho^\a &= \ri z \eta^\a +z k_\hb \bar{\bm\q}^\b (\sigma^\hb)_{\b}{}^{ \a} ~, \\
\Lambda &= -\ri \Lambda_\1{}^\1 - 2 \ri \eta\bm\q + 2 \ri \bar{\eta} \bar{\bm\q}
- 2 k_\ha \bm\q \sigma^\ha \bar{\bm\q} ~.
\end{align}
\end{subequations}

Taking the $\bm\q_\2 = \bar{\bm\q}^\2 = 0$ projection of \eqref{deltaQfull} gives
the transformations \eqref{deltaQ}.

\section{Prepotential for the intrinsic vector multiplet}\label{app_IntrinsicVector}
\setcounter{equation}{0}

An abelian vector multiplet is described in Minkowski superspace by a reduced
chiral superfield $W$ obeying
\begin{align}
\DB^\ad_iW=D_\a^i\bar{W}=0~,~~~D^{ij}W=\DB^{ij}\bar{W}~.
\end{align}
Within projective superspace, these constraints are solved in terms of
a projective prepotential $V$, which is a real weight zero tropical superfield
\begin{align}\label{eq_tropV}
V(z,v)&=\sum_{k=-\infty}^{+\infty}\z^kV_k(z)~,~~~\bar{V}_k=(-1)^kV_{-k}~,
\quad
D_\a^{(1)}V=\DB_\a^{(1)}V=0
~,\quad
\end{align}
so that
\bea
W&=&
-{1\over 8\pi}\oint{(v,\rd v)\over(v,u)^2}\,\DB^{(-2)}V
~,~~~~~~
\bar{W}=
-{1\over 8\pi}\oint{(v,\rd v)\over(v,u)^2}\,D^{(-2)}V
~.
\label{Vector-prepot}
\eea
Here we have introduced the operators
\bsubeq
\bea
&&D_\a^{(-1)}:=\frac{1}{(v,u)}u_i D_\a^i~,~~~
\DB_\ad^{(-1)}:=\frac{1}{(v,u)}u_i \DB_\a^i~,
\\
&&D^{(-2)}:=D^{\a(-1)}D_\a^{(-1)}~,~~
\DB^{(-2)}:=\DB_\ad^{(-1)}\DB^{\ad(-1)}
~.
\eea
\esubeq
The field strength $W$ is invariant under the gauge transformations
\begin{align}
V \rightarrow V + \ri \L - \ri \breve \L
\end{align}
for a weight zero arctic superfield $\L$.

Let us now specialize to the AdS intrinsic vector multiplet.
For the case $\bms^{\1\1}=\bms^{\2\2}=0$ and $\bms^{\1\2}=\a s$,
with $\a = \pm \ri$, we have
\bea
W&=&\frac{1}{s}\frac{1}{\(z_L+2\a \q_{\1\2}\)}
~.
\eea
In the north chart of $\mathbb{C}P^1$, the prepotential $V$ can be chosen
\bea
V&=&
\frac{1}{s z_A(\z)}\frac{\ri}{\z}\(\q^{(2)}(\z)+\qb^{(2)}(\z)\)
-\frac{1}{s (z_A(\z))^2}\frac{\ri\a}{\z^2}\q^{(2)}(\z)\qb^{(2)}(\z)
~,
\label{prep-intr-VM}
\eea
where the analytic coordinates in the north chart are
\bsubeq
\bea
&\q_\a^{(1)}(\z):=\z_i\q_{\a}^{i}~,~~~~~~
\qb_\a^{(1)}(\z):=\z_i\qb_\a^i~, \quad \z^i := (1, \z)~,
\\
&
\q^{(2)}(\z)=\q^{\a(1)}(\z)\q_\a^{(1)}(\z)
~,~~~~~~
\qb^{(2)}(\z)=\qb_\a^{(1)}(\z)\qb^{\a(1)}(\z)
~,
\\
&z_A(\z)=z+\q^{\a (1)}(\z)\qb_{\a}^{\1}+\q^{\a \1}\qb_\a^{(1)}(\z)
~.
\eea
\esubeq
One can check that this expression for $V$ is indeed real under smile
conjugation.
Upon reduction to 3D $\cN=2$ superspace, we find
\bea
V\big|&=&
\frac{\ri}{sz}\Big(\frac{1}{\z}\bmq^2
+\z\bmqb^2
\Big)
~.
\eea
This prepotential corresponds to a frozen vector multiplet, which
necessarily has vanishing component field strength $F_{mn}$. The specific
choice of $V$ made above turns out to correspond to a vanishing
component gauge connection $A_m$ as well. It remains possible to perform
a gauge transformation on $V$ to yield a non-vanishing (but pure
gauge) $A_m$. We will demonstrate this for the special case of
an arbitrary $z$-dependent gauge connection $A_z$.

Let us perform a gauge transformation with
$\L = \L(z_A)$ where $\L$ is a real function and
\begin{subequations}
\begin{align}
z_A &:= z + \q^{(1)} \bar\q^{(-1)} + \q^{(-1)} \bar\q^{(1)}~, \\
\q_\a^{(-1)} &:= \frac{\q^i u_i}{\z^i u_i}~,\qquad
\bar\q_\a^{(-1)} := \frac{\bar\q^i u_i}{\z^i u_i}~,\qquad u_i = (1,0)~.
\end{align}
\end{subequations}
The quantity $z_A$ has already been introduced, and is well-defined in
the north chart. Its smile conjugate $\breve z_A$ is given by
\begin{align}
\breve z_A = z + \q^{(1)} \bar\q'^{(-1)} + \q'^{(-1)} \bar\q^{(1)}~,\quad
u_i' = (0, 1)~,
\end{align}
and is well-defined in the south chart. The antarctic gauge parameter is
then given by $\breve \L = \L(\breve z_A)$.
Now rotating the theta coordinates and projecting to the 3D $\cN=2$
variables, we find
$z_A \vert = z + \ri \zeta \bar{\bm\q}^2$
which leads to
\begin{align}
\ri \L\vert - \ri \breve \L\vert = \ri \L(z + \ri \zeta \bar{\bm\q}^2) - \ri \L(z + \ri {\bm\q}^2 / \zeta)
	= \L'(z) \left(\frac{1}{\zeta} \bm\q^2 -\zeta \bar{\bm\q}^2 \right)~.
\end{align}
The intrinsic vector multiplet with arbitrary pure gauge $A_z$ is then given by
\begin{align}
V \vert = \frac{1}{\zeta} \bm\q^2 \left(\frac{\ri}{mz} + \L'(z) \right)
	+ \zeta \bar {\bm\q}^2 \left(\frac{\ri}{mz} - \L'(z) \right)~.
\end{align}

It remains to verify our claim that $V$ possesses a non-vanishing connection
$A_z$. We begin by observing that the 3D $\cN=4$ spinor derivatives must obey
\begin{align}
v_i \bm \cD_\a^i = \bm\cD_\a^{(1)} = \exp\left(V_{(+)} \hat e \right) \bm D_\a^{(1)} \exp\left(- V_{(+)} \hat e \right)
	= \bm D_\a^{(1)} - \bm D_\a^{(1)} V_{(+)} \hat e~.
\end{align}
This identifies the
spinor U(1) connection as
$\bm A_\a^{(1)} = \ri \bm D_\a^{(1)} V_{(+)}$.
Note that because $V = V_{(+)} + V_{(-)}$ is analytic, we can just as well denote
$\bm A_\a^{(1)} = -\ri \bm D_\a^{(1)} V_{(-)}$,
which implies that $\bm A_\a^{(1)} = v_i \bm A_\a^i$ is defined everywhere on
$\mathbb CP^1$. This allows the construction of all the connections.
Using \eqref{eq_tropV}, we note that the analyticity
condition $\bm D_\a^{(1)} V = 0$ amounts to
\begin{align}
{\bm D}_\a^\2 V_n = {\bm D}_\a^\1 V_{n-1}~,\qquad \bar {\bm D}^\a_\2 V_n = -\bar {\bm D}^\a_\1 V_{n+1}~.
\end{align}
Defining 
$V_{(+)} := \frac{1}{2} V_0 + \sum_{n=1}^\infty V_n \z^n$
it is straightforward to calculate that
\begin{align}
v_i \bm A_\a^i &= \ri \bm D_\a^{(1)} V_{(+)}
	= \frac{\ri}{2} \z D_\a V_0 + \frac{\ri}{2} D_\a V_{(-1)} \implies \eol
\bm A_\a^\1 &= -\frac{\ri}{2} \bm D_\a^\1 V_0~,\qquad
\bm A_\a^\2 = \frac{\ri}{2} \bm D_\a^\1 V_{(-1)}~.
\end{align}
The vector connections can be easily derived from the spinor connections using
\begin{subequations}
\begin{align}
\{\bm \cD_\a^\1, \bm \cD_{\b}^\2\} + \{\bar{\bm \cD}_{\a \1}, \bar{\bm \cD}_{\b \2}\}
	&= -4 \ri \ve_{\a \b} \bm\cD_z~, \\
\{\bm \cD_\a^\1, \bar{\bm \cD}_{\b \1}\} + \{\bm \cD_\b^\1, \bar{\bm \cD}_{\a \1}\} &= -4 \ri \bm\cD_{\a \b}~.
\end{align}
\end{subequations}

Now let us specialize to the intrinsic vector multiplet and consider
the projection to $\bm\q_\2 = \bar{\bm\q}^\2 = 0$. It is immediately evident
that the zero mode $V_0\vert$ vanishes, and so $\bm A_\a^\1 \vert = \bar{\bm A}_{\a \1} \vert = 0$.
In other words, from the point of view of the 3D $\cN=2$ superspace, there is
no U(1) connection. The remaining connection we need to determine is
$\bm A_z$. It is straightforward to calculate
\begin{align}
\bm A_z
	= -\frac{1}{8} \Big((\bm D^\1)^2 V_{(-1)} - (\bar {\bm D_\1})^2 V_{(1)}\Big)
\implies \bm A_z\vert &= \L'(z)~.
\end{align}

%%%%%%%%%%%%%%%%%%%%%%%%%%%%%%%%%%%%%%%%%%%%%%%
%%%%%%%%%%%%%%%%%%%%%%%%%%%%%%%%%%%%%%%%%%%%%%%
%%%%%%%%%%%%%%%%%%%%%%%%%%%%%%%%%%%%%%%%%%%%%%%

%%%%%%%%%%%%%%%%
%%%%%%%%%%%%%%%%
\end{document}